\documentclass[journal ]{new-aiaa}

\usepackage[utf8]{inputenc}
\usepackage{textcomp}

\usepackage{graphicx}
\usepackage{amsmath}
\usepackage[version=4]{mhchem}
\usepackage{siunitx}
\usepackage{longtable,tabularx}
\newcolumntype{K}[1]{>{\centering\arraybackslash}p{#1}}
\newcolumntype{L}[1]{>{\arraybackslash}p{#1}}
\newcolumntype{P}[1]{>{\raggedright\arraybackslash}p{#1}}

\usepackage{xcolor} % For making notes to myself in color
\usepackage{float} % For H option in figures
\usepackage{subcaption} % For ... subfigures. 
\usepackage{multirow} %deal with merged cells in tables
\usepackage{mathtools} % For matrix* 
\usepackage{booktabs} % For table options like toprule 
\usepackage{arydshln} % For hdashline cdashline

\definecolor{red}{RGB}{231, 76, 60}
\definecolor{orange}{RGB}{243, 156, 18}
\definecolor{yellow}{RGB}{241, 196, 15}
\definecolor{green}{RGB}{39, 174, 96}
\definecolor{blue}{RGB}{46, 134, 193}
\definecolor{purple}{RGB}{155, 89, 182}
\definecolor{pink}{RGB}{236, 64, 122}
\definecolor{turquoise}{RGB}{0, 172, 193}

\newcommand\tab[1][1cm]{\hspace*{#1}} % Tabs 
\newcommand\smalltab[1][.5cm]{\hspace*{#1}} % Smaller tabs

\title{Closed-Form Optimal Impulsive Control of Spacecraft Relative Motion Using Reachable Set Theory}

\author{Michelle Chernick \footnote{Ph.D. Candidate, Space Rendezvous Laboratory, Aeronautics and Astronautics, 496 Lomita Mall} and Simone D'Amico \footnote{Assistant Professor, Space Rendezvous Laboratory, Aeronautics and
Astronautics, 496 Lomita Mall}}
\affil{Stanford University, Stanford, California, 94305}
\begin{document}

\maketitle

\begin{abstract}
This paper addresses the spacecraft relative orbit reconfiguration problem of minimizing the delta-v cost of impulsive control actions while achieving a desired state in fixed time. The problem is posed in relative orbit element (ROE) space, which yields insight into relative motion geometry and allows for the straightforward inclusion of perturbations in linear time-variant form. Reachable set theory is used to translate the cost-minimization problem into a geometric path-planning problem and formulate the reachable delta-v minimum, a new metric to assess optimality and quantify reachability of a maneuver scheme. Next, this paper presents a methodology to compute maneuver schemes that meet this new optimality criteria and achieve a prescribed reconfiguration. Though the methodology is applicable to any linear time-variant system, this paper leverages a state representation in ROE to derive new globally optimal maneuver schemes in orbits of arbitrary eccentricity. The methodology is also used to generate quantifiably sub-optimal solutions when the optimal solutions are unreachable. Further, this paper determines the mathematical impact of uncertainties on achieving the desired end state and provides a geometric visualization of those effects on the reachable set. The proposed algorithms are tested in realistic reconfiguration scenarios and validated in a high-fidelity simulation environment.
\end{abstract}

\section{Introduction}
\lettrine{D}{istributed} space systems enable advanced missions in fields such as astronomy and astrophysics, planetary science, and space infrastructure by employing the collective usage of two or more cooperative spacecraft.  Proposed formation flying missions in eccentric orbits introduce new challenges for spaceborne control architectures, such as efficiency (reduced on-board processing power and propellant), autonomy (no ground-in-the-loop), and operational constraints (interference with payload, and predictability) \cite{bib:DamicoMDSS}. To address these challenges, this work develops globally-optimal closed-form solutions to the problem of minimizing the delta-v cost of a set of impulsive control actions that accomplish a fixed desired satellite relative orbit configuration in fixed time.

In  literature, most approaches to finding impulsive maneuver schemes that accomplish the fixed-time, fixed-end condition reconfiguration fall into three categories: direct optimization techniques, indirect optimization techniques, and closed-form solutions. For spaceborne applications, closed-form solutions are frequently favored because they are simple, predictable, and computationally efficient. Numerous authors have developed closed-form solutions in Hill's coordinates \cite{bib:Ichi,bib:Khalil} using the Hill-Clohessy-Wiltshire equations (HCW), which are valid for near-circular orbits, small separations as compared with the orbit radius, and unperturbed Keplerian motion. However, these limitations are overcome by using a different state representation, called relative orbit elements (ROE). 

As will be discussed in the following section, use of the ROE state representation allows for the derivation of simple, high-fidelity solutions to the spacecraft formation control problem. The ROE state inherently enables the linearization of the equations of relative motion with minimal loss of accuracy \cite{bib:Damico_ROE_as_Integration_Constants}, and it is this linearization that allows for application of reachable set theory to enable a geometric solution of the resulting optimal control problem. Despite the benefits of using the ROE state representation, few authors have solved the reconfiguration problem in closed-form using ROE. Gaias and D'Amico \cite{bib:GaiasDamico1,bib:GaiasDamico2} proposed a set of closed-form guidance and control algorithms for optimal control in near-circular orbits. Chernick and D'Amico \cite{bib:ChernickDamico} extended these algorithms to include perturbations due to Earth's oblateness ($J_2$) in near-circular orbits and developed control solutions in eccentric unperturbed orbits, but optimality was limited to specific cases. In addition, it was assumed that the in-plane and out-of-plane control problems were decoupled. Zhang and Mortari  \cite{bib:zhang_mortari} removed the need for the  transcendental equation in Chernick’s closed-form solution and used it as initial guess to derive both single- and multi-impulse minimum norm solutions through a nonlinear least-squares iterative method using the second order Gauss Variational Equations. Vaddi et al. \cite{bib:Vaddi} proposed analytical solutions to the optimal reconfiguration problem, but only for control of the in-plane orbital elements in near-circular orbits. For eccentric chief orbits, Schaub and Alfriend \cite{bib:SchaubAlf} developed an impulsive feedback controller to establish mean orbit element differences. These solutions (and their optimality) are case-specific and have not yet been advanced to include other perturbations.  

Direct optimization techniques allow for a greater degree of generality because the optimal control problem can be formulated with the times, magnitudes, and directions as variables \cite{bib:Betts}. Though widely applicable, direct optimization techniques are not guaranteed to converge to a global optimum because the minimum cost is generally a non-convex function of the control action times \cite{bib:SobiesiakDamaren}. In addition, direct optimization techniques do not provide insight into the maneuver optimality. Therefore, the majority of numerical approaches in literature utilize indirect optimization techniques, taking advantage of the characteristics of the so-called primer and dual pair of the optimal control problem. For example, Roscoe et al. \cite{bib:Roscoe} designed an optimal algorithm for eccentric perturbed orbits based on Lawden's primer vector theory and Pontryagin’s optimal control \cite{bib:ConwayPrussing}. This algorithm involves an iterative process that depends on a good initial guess and yields large computational loads. Gilbert and Harasty \cite{bib:Gilbert} proposed a different approach to indirect optimization based on reachable set theory, which converges to a globally optimal sequence of impulsive control inputs for problems with norm-like constant cost functionals. Koenig \cite{bib:KDAutomatica} generalized Gilbert's algorithm to time-variant cost functions with the only restriction that the cost function be represented as the integral of a time-variant norm-like function of the control input vector. In optimal control and robotics applications, reachable set theory is commonly used to assess cost-reachability and safety \cite{bib:Allen}. Vinh et al. characterized the surface of the reachable domain for a given fuel cost in a fixed time interval, and applied the concept to assess the feasibility of an interceptor capturing a target missile \cite{bib:Vinh}. Recently, Zagaris et al. used reachability theory to determine initial conditions for which rendezvous maneuvers were possible in a binary formation with a tumbling chief in a near-circular orbit \cite{bib:Zagaris_Romano}. 

This paper shows that reachable set theory can be used not just to assess the cost of a maneuver scheme, but as a geometric tool to derive the closed-form maneuver scheme itself. This work improves upon current literature through four main contributions to the state of the art. First, by leveraging domain specific knowledge and the linear scaling properties of the reachable set, this paper develops the reachable delta-v minimum, a new metric for assessing maneuver scheme optimality. Second, this paper presents a general methodology to derive impulsive control solutions that meet the new optimality criteria. This general methodology is applied to a new, quasi-nonsingular relative orbit element state representation to obtain closed-form maneuver schemes that are applicable to orbits of arbitrary eccentricity. The same methodology can be used to generate quantifiably sub-optimal solutions when the optimal solutions are unreachable. Third, this paper presents a visual and mathematical analysis of the effect of maneuver execution and navigation errors on the reachable sets. Finally, the new algorithms are validated by comparison to Koenig's numerical optimization algorithm \cite{bib:KDAutomatica} and by numerical integration of the Gauss Variational Equations including a full-force dynamics model. What follows in Sec. \ref{sec:prob_statement} is the formal statement of the energy-optimal formation control problem.

\section{Problem Statement}\label{sec:prob_statement}
Without loss of generality, a formation here consists of two satellites: the chief, which defines the reference orbit and is uncontrolled, and the deputy, which is controlled by a 3D thrust input. The relative motion between two satellites in a formation is commonly defined in terms of relative position and velocity in the Hill's coordinate frame (also called radial/along-track/cross-track coordinates, or RTN), whose origin is at the chief's center of mass. The RTN frame is defined by the basis $[o_r, o_t, o_n]$, where $o_r$ is aligned with the radial direction and positive outward, $o_n$ is aligned with the chief angular momentum vector and positive in the orbit normal direction, and $o_t$ completes the right-handed triad. However, the relative motion can be equivalently described using combinations of non-dimensional orbit elements of the chief and deputy, called relative orbit elements (ROE), $\delta \pmb{\alpha} = \delta\pmb{\alpha}(\pmb{\alpha}_c, \pmb{\alpha}_d)$. The benefits of using the ROE state representation are numerous. First, the homogeneous, unperturbed solution to the ROE dynamics equations is the trivial solution of the Keplerian two-body problem, $\delta\pmb{\alpha}=\text{const.}$, whereas there is no available unperturbed solution to the equations of relative motion in Hill's coordinates \cite{bib:ChernickDamico}. In addition, in the presence of perturbations, a state based on ROE slowly varies in time, whereas Hill's coordinates vary rapidly. Secular and long-period effects of perturbing forces are simply included in the equations of relative motion for the ROE state using a state transition matrix (STM), $\pmb{\Phi}(t_j,t_i)$. The STM propagates the state forward in time, while the control input matrix, $\pmb{\Gamma}(t_k)$, represents the effect of a 3D control input $\delta \pmb{v}_k$ at time $t_k$. With this notation in mind, the linearized dynamics that govern the reconfiguration from an initial set of ROE, $\delta \pmb{\alpha}_0$, to a final desired set of ROE, $\delta\pmb{\alpha}_f$, under the influence of $p$ impulsive maneuvers can be written as
\fontsize{8.5}{11}\selectfont\begin{equation}\label{eqn:prob_statement}
 \delta \pmb{\alpha}_f = \pmb{\Phi}(t_f,t_0)\delta\pmb{\alpha}_0 + \sum_{k=1}^p\pmb{\Phi}(t_f,t_k)\pmb{\Gamma}(t_k)\delta \pmb{v}_k.
 \end{equation}\normalsize
Given an initial set of chief orbital elements (OE), $\pmb{\alpha}_{c,0}$, and an initial set of ROE, $\delta\pmb{\alpha}_0$, the fixed-time, fixed-end conditions relative orbit reconfiguration problem is defined by a desired final set of ROE, $\delta\pmb{\alpha}_f \in \mathbb{R}^6$, and a reconfiguration time span $T$. With these reconfiguration parameters, the optimal control problem to be solved in this paper is
\fontsize{8.5}{11}\selectfont\begin{equation}\label{eqn:optcontrolproblem}
\text{Minimize } \sum_{k=1}^p ||\delta\pmb{v}_k||_2 \text{ \tab subject to } \Delta\delta{\pmb{\alpha}} = \sum_{k=1}^p \pmb{\Phi}(t_f,t_k)\pmb{\Gamma}(t_k)\delta\pmb{v}_k\text{, } t_k\in T\end{equation}\normalsize
where a pseudo-state $\Delta\delta {\pmb{\alpha}}$ is introduced as $\Delta\delta {\pmb{\alpha}} = \delta\pmb{\alpha}_f - \pmb{\Phi}(t_f,t_0)\delta\pmb{\alpha}_0 $ to simplify notation. The next section outlines the dynamics of relative motion as they apply to the adopted ROE state definition.

\section{Background}\label{sec:background}
\subsection{Astrodynamics of Relative Motion}
The state representation of choice is the 6D quasi-nonsingular ROE, defined \cite{bib:DamicoEI,bib:ChernickDamico} as \fontsize{8.5}{10.5}\selectfont\begin{equation}\label{eqn:ROE}
    \delta\pmb{\alpha} = \begin{bmatrix}\delta a \\ \delta \lambda_e \\ \delta e_x \\ \delta e_y \\ \delta i_x \\ \delta i_y \end{bmatrix} = 
    \begin{bmatrix}\frac{a_d-a_c}{a_c} \\ M_d - M_c + \eta(\omega_d - \omega_c + (\Omega_d - \Omega_c)\cos i_c) \\ e_d \cos\omega_d - e_c \cos\omega_c \\ e_d\sin\omega_d - e_c\sin\omega_c\\ i_d-i_c\\ (\Omega_d - \Omega_c)\sin i_c\end{bmatrix},
    \end{equation}\normalsize
which consists of the relative semi-major axis $\delta a$, the eccentric relative mean longitude $\delta\lambda_e$, and the $x$ and $y$ components of the relative eccentricity vector $\delta\pmb{e}$ and the relative inclination vector $\delta\pmb{i}$. They are nonlinear combinations of the classical absolute orbital elements (OE) $[a, e, i, \Omega, \omega, M]$ of the chief and deputy spacecraft, denoted with subscripts $c$, $d$, respectively. The state representation is termed quasi-nonsingular because the state is valid for circular chief orbits ($e_c = 0$) but becomes singular for strictly equatorial chief orbits ($i_c = 0$). Apart from the eccentric relative mean longitude $\delta\lambda_e$, it is the same state used for Guidance, Navigation, and Control (GNC) and orbit design on multiple formation flying missions such as PRISMA \cite{bib:DamicoFlorio} and TanDEM-X \cite{bib:ArdaensFischer}. 

To derive closed-form maneuver schemes in the ROE state representation, it is necessary to first find the STM and control input matrix that propagate the dynamics. The STM from $t_0$ to $t_f$ is derived using Koenig et al.'s approach \cite{bib:KoenigDamico} for the quasi-nonsingular ROE (provided in Appendix A, Eq. \eqref{eqn:STMJ2}).
The change in the mean ROE due to an impulsive maneuver $\delta\pmb{v}_k$ is described by the control input matrix $\pmb{\Gamma}(t_k)$ (also written as $\pmb{\Gamma}_k$) and derived from the Gauss Variational Equations (GVE). The GVE describe the rate of change of the osculating OE $\pmb{\alpha}_{osc}$ as a function of perturbing accelerations in the RTN frame. The effect of an impulsive maneuver $\delta\pmb{v}_k$ on the osculating ROE is found by integrating the GVE over the duration of the maneuver with the assumption that the OE are constant and can be written in matrix form using the chain rule as $\Delta\delta\pmb{\alpha}_{osc} = \frac{\partial \delta \pmb{\alpha}(\pmb{\alpha}_{c,osc},\pmb{\alpha}_{d,osc})}{\partial \pmb{\alpha}_{d,osc}}\frac{\partial\pmb{\alpha}_{d,osc}}{\partial \pmb{v}_{RTN}}\delta \pmb{v}_k$. To employ the same approach for the mean ROE state representation used in this paper, Brouwer's transformation from osculating to mean OE $f_{mean}$ is applied \cite{bib:Brouwer}. The partial derivatives of $f_{mean}$ with respect to the osculating OE form a Jacobian matrix that is near identity. For the mean ROE state, the control input matrix is then \fontsize{9}{11}\selectfont\begin{equation}\label{eqn:Gamma_chain}
    \pmb{\Gamma}(\pmb{\alpha}_c) \approx  \left.\frac{\partial \delta\pmb{\alpha}(\pmb{\alpha}_c,\pmb{\alpha}_d)}{\partial \pmb{\alpha}_d}\right\vert_{\pmb{\alpha}_c=\pmb{\alpha}_d}\frac{\partial f_{mean}(\pmb{\alpha}_{d,osc})}{\partial \pmb{\alpha}_{d,osc}}\frac{\partial\pmb{\alpha}_{d}}{\partial \pmb{v}_{RTN}},
    \end{equation}\normalsize where the subscript $osc$ denotes the osculating OE and the effect of an impulsive maneuver on the mean ROE is given by $\Delta\delta\pmb{\alpha} = \pmb{\Gamma}(\pmb{\alpha}_c)\delta\pmb{v}$ \cite{bib:Vaddi}. For the quasi-nonsingular ROE state in particular, the control input matrix for an impulsive maneuver at true argument of latitude $\theta_k = \nu_k + \omega_k$ is given by Eq. (13) in Ref. \cite{bib:ChernickDamico} for eccentric orbits. By virtue of the form of $\pmb{\Gamma}$, when the eccentricity is close to zero, the control input matrix reduces so ROE control is effectively decoupled; Specifically, in-plane maneuvers (radial, tangential) affect only the in-plane ROE, $\delta a$, $\delta \lambda$, and $\delta \pmb{e}$, and out-of-plane maneuvers (normal) affect only the out-of-plane ROE, $\delta\pmb{i}$.
    In contrast, in the eccentric case, the relative eccentricity vector is affected by maneuvers in the normal direction, as is evident by the nonzero third column entries in the control input matrix in Eq. (13) of Ref. \cite{bib:ChernickDamico}, so an inherent decoupling cannot be claimed. To avoid this issue, the ROE state can be redefined to include a modified relative eccentricity vector as 
    \fontsize{8.5}{11}\selectfont\begin{equation}\label{eqn:ROE_modified}\begin{bmatrix} \delta e_x' \\ \delta e_y' \end{bmatrix} = \begin{bmatrix}e_d - e_c \\ \omega_d - \omega_c + (\Omega_d - \Omega_c)\cos{i_c}\end{bmatrix},
    \end{equation}\normalsize
which replaces the relative eccentricity vector $\delta e_x, \delta e_y$ in Eq. \eqref{eqn:ROE}. The control input matrix for the new ROE state is derived in the same way as for the quasi-nonsingular ROE state: using Eq. \eqref{eqn:Gamma_chain}. The corresponding control input matrix for the ROE including the modified relative eccentricity vector and eccentric relative mean longitude is given by
    
    \fontsize{9}{11}\selectfont\begin{equation}\label{eqn:control_input_matrix}
        \Delta\delta\pmb{\alpha}_k = \pmb{\Gamma}_k\delta\pmb{v}_k = \frac{1}{na} \begin{bmatrix}\frac{2}{\eta} e \sin\nu_k & \frac{2}{\eta}(1+e\cos\nu_k) & 0 \\
-\frac{2\eta^2}{1+e\cos\nu_k} & 0 &0 \\
\eta\sin\nu_k & \eta \frac{e+\cos\nu_k(2+e\cos\nu_k)}{1+e\cos\nu_k} & 0 \\
-\frac{\eta}{e}\cos\nu_k & \frac{\eta}{e}\sin\nu_k\frac{2+e\cos\nu_k}{1+e\cos\nu_k} & 0\\
0 & 0 & \eta \frac{\cos(\theta)}{1+e\cos\nu_k} \\
0 & 0 & \eta \frac{\sin(\theta)}{1+e\cos\nu_k}\end{bmatrix}\begin{bmatrix}\delta v_R \\ \delta v_T \\ \delta v_N\end{bmatrix}.\end{equation}\normalsize
As shown in Eq. \eqref{eqn:control_input_matrix}, control of the in-plane and out-of-plane ROE is fully decoupled. The relationship between the $\delta \pmb{e}$ and $\delta \pmb{e}'$ states in Eqs. \eqref{eqn:ROE} and \eqref{eqn:ROE_modified} itself is nonlinear, but the relationship between the expressions for the pseudo-states is actually very simple. This will be further discussed and used to derive closed-form optimal impulsive control solutions in Sec. \ref{sec:cf_maneuver_schemes}.

For unperturbed orbits, the STM for the new ROE state is the same as the STM for the quasi-nonsingular ROE, given in Eq. (12) in Ref. \cite{bib:ChernickDamico}. The unperturbed STM is used in the calculation of optimal maneuver schemes because the effect of perturbations due to maneuivers is small. However, in calculating the pseudo-state in Eq. \eqref{eqn:optcontrolproblem}, where the desired change in ROE is the difference between the final desired ROE and the initial ROE propagated by the dynamics over the entire reconfiguration span, a more accurate method can be used to accurately represent the free dynamics, such as numerical propagation with the full-force nonlinear dynamics. This is not a problem addressed in this paper, as intermediate pseudo-state calculation is an optimal guidance problem.

\subsection{Reachable Set Theory}\label{sec:RST}
This section provides an introduction to reachable set theory, which will be used to derive a simple geometric optimality criteria and closed-form optimal reconfiguration schemes.
Let $U(c)$ be the set of all control actions in RTN whose magnitude (two-norm, as in Eq. \eqref{eqn:optcontrolproblem}) is less than or equal to $c$. $S(c,t_j)$ is the set of pseudo-states $\Delta\delta {\pmb{\alpha}}$ that can be reached at the end of the reconfiguration $T$ given a single control action $\pmb{u}$ of magnitude less than or equal to $c$ at time $t_j$,
\fontsize{9}{11}\selectfont\begin{equation}\label{eqn:Sdvt}
S(c,t_j) = \{ \Delta\delta {\pmb{\alpha}} : \Delta\delta {\pmb{\alpha}}=\pmb{\Phi}(t_f,t_j)\pmb{\Gamma}(t_j)\pmb{u}, \pmb{u}\in U(c) \}, t_j \in T.
\end{equation}\normalsize
$S(c,T)$ is the set of pseudo-states $\Delta\delta {\pmb{\alpha}}$ that can be reached at the end of the reconfiguration time $T$ given a single control action of magnitude less than or equal to $c$ at any time in $T$, given by
\fontsize{9}{11}\selectfont\begin{equation}\label{eqn:Sdv}
S(c,T) = \bigcup_{t_j\in T} S(c,t_j).
\end{equation}\normalsize
Finally, $S^*(c,T)$ is the set of pseudo-states $\Delta\delta {\pmb{\alpha}}$ that can be reached at the end of the reconfiguration $T$ given $p\geq 1$ control actions of total magnitude less than or equal to $c$ at any time in $T$. For any $\Delta\delta {\pmb{\alpha}}^* \in S^*(c,T)$, there must exist a set of $p$ control actions $\{\pmb{u}_1, ..., \pmb{u}_p\}$ executed at times $\{t_1, ..., t_p\}$ that satisfies
$\Delta\delta {\pmb{\alpha}}^* = \sum_{k=1}^{p} \pmb{\Phi}(t_f,t_k)\pmb{\Gamma}(t_k)\pmb{u}_k\text{, } \sum_{k=1}^{p} ||\pmb{u}_k||_2 \leq c, t_k \in T.$ Because the cost of a control action scales linearly with its magnitude and $\pmb{\Phi}(t_f,t_k)\pmb{\Gamma}(t_k)\pmb{u}_k$ is an element of $S(c,T) \text{ }\forall t_k$, $S^*(c,T)$ can be defined as
\fontsize{9}{11}\selectfont\begin{equation}\label{eqn:Sstar}
S^*(c,T) = \{\Delta\delta {\pmb{\alpha}}^* : \Delta\delta {\pmb{\alpha}}^* = \sum_{k=1}^p c_k \Delta\delta {\pmb{\alpha}}_k, \Delta\delta {\pmb{\alpha}}_k \in S(c,T), c_i \geq 0, \sum_{k=1}^p c_k = 1 \},
\end{equation}\normalsize
which is the definition of the convex hull of $S(c,T)$. Based on this definition, a pseudo-state that lies on the boundary of the convex hull $S^*(c,T)$ is reachable with total delta-v equal to $c$. 

There are three important properties of the reachable set that will be leveraged to derive reachable delta-v minima and energy-optimal maneuver schemes. First, the reachable set is a linear function of a control action for linear time-variant (LTV) systems, so the reachable set scales linearly with cost. This allows for the development of a very simple optimality condition; if $\delta v_{min}$ is the optimal cost for a given reconfiguration, then $\Delta\delta {\pmb{\alpha}}$ will lie on the boundary of $S^*(\delta v_{min},T)$. Second, it is clear from the definitions above that if and only if a desired endpoint is on the boundary of both $S$ and $S^*$ for a given reconfiguration, the reconfiguration can be achieved with a single maneuver. Also, it has been previously demonstrated that any state can be reached by a maximum number of impulses equal to the dimension of the state $n$ \cite{bib:KDAutomatica}. These facts prove useful in categorizing and deriving closed-form maneuver schemes because they define both the minimum and maximum number of maneuvers required. Third, a $2n$-dimensional ($2n$-D) state can be projected into $n$ 2D planes and analyzed separately without loss of generality. The minimum delta-v required to achieve the entire desired $2n$-D reconfiguration cannot be less than the maximum delta-v required to achieve any of the $n$ separate 2D reconfigurations. To achieve a $2n$-D reconfiguration given by $
    \Delta\delta {\pmb{x}}_{des} = \begin{bmatrix}\Delta\delta  {x}_{1,des}, \Delta\delta  {x}_{2,des}, ..., \Delta\delta  {x}_{2k-1,des}, \Delta\delta  {x}_{2k,des},...,\Delta\delta  {x}_{2n-1,des},\Delta\delta  {x}_{2n,des}\end{bmatrix} \text{\smalltab for } k = 1,...,n$,  
the minimum delta-v required is given by
\fontsize{9}{11}\selectfont
\begin{equation}\label{eqn:dvminproof}
    \delta v_{min} \geq \max \{ \delta v_{min,\delta x_{1,2}}, ..., \delta v_{min,\delta x_{2k-1,2k}}, ... \delta v_{min,\delta x_{2n-1,2n}}  \},
\end{equation}\normalsize
where $\delta v_{min,\delta x_{2k-1,2k}}$ is the optimal delta-v required to achieve the reconfiguration in only the $\delta x_{2k-1,2k}$ plane. A proof of this claim is given in Appendix B and shows that the total cost of the entire reconfiguration is driven by one of the 2D planes used to decompose the state. 

The inequality in Eq. \eqref{eqn:dvminproof} stems from the fact that projecting a higher dimensional reachable set onto a lower dimensional space inherently loses information about the shape of the original reachable set. For example, it is not possible to know whether a circle was projected from a cone or a cylinder. Nonetheless, it is possible to quantify when the expression in Eq. \eqref{eqn:dvminproof} is an equality. The minimum delta-v of the entire reconfiguration \textit{equals} $\delta v_{min,\delta x_{2k-1,2k}}$ when the desired pseudo-state is contained in $S^*_n(\delta v_{min,\delta x_{2k-1,2k}}, T_{opt,\delta x_{2k-1,2k}})$. $S_n^*$ is computed using Eq. \eqref{eqn:Sstar} in all of the $n$ 2D planes, where $T_{opt,\delta x_{2k-1,2k}}$ is the set of optimal maneuver times. The subscript $n$ in $S^*_n$ denotes a nested reachable set, which is generated using a subset of times in the total reconfiguration time and is therefore itself a subset of the reachable set $S^*$. If $S^*_n$ includes the desired pseudo-state when mapped onto the other 2D plane(s), the expression in Eq. \eqref{eqn:dvminproof} is an equality. A more in depth discussion of how to find those optimal times for the quasi-nonsingular ROE follows in Sec. \ref{sec:cf_maneuver_schemes}.

In ROE space, the plane $k$ whose total required delta-v defines the maximum in Eq. \eqref{eqn:dvminproof} is referred to as the dominant plane, and the individual ROE whose total required delta-v defines the maximum within the plane is known as the dominant ROE. For the 6D ROE state, three 2D planes can be defined: the $(\Delta\delta\lambda,\Delta\delta a)$ plane (also called the $\Delta\delta\pmb{a}$ plane), the $\Delta\delta \pmb{e}$ plane, and the $\Delta\delta\pmb{i}$ plane. Control actions in the $\Delta\delta\pmb{i}$ plane are decoupled from the other two planes using the new state representation. Cost optimality is reduced from a 6D problem to three 2D problems that can be analyzed separately. The next section categorizes reconfigurations in each 2D plane according to the number of required maneuvers. This analysis drives the derivation of the reachable delta-v minima. 

\subsection{Analysis of Reachable Sets in ROE Space}
\begin{figure}[H]
    \centering
    \begin{subfigure}[h]{0.24\textwidth}
        \includegraphics[width=\textwidth]{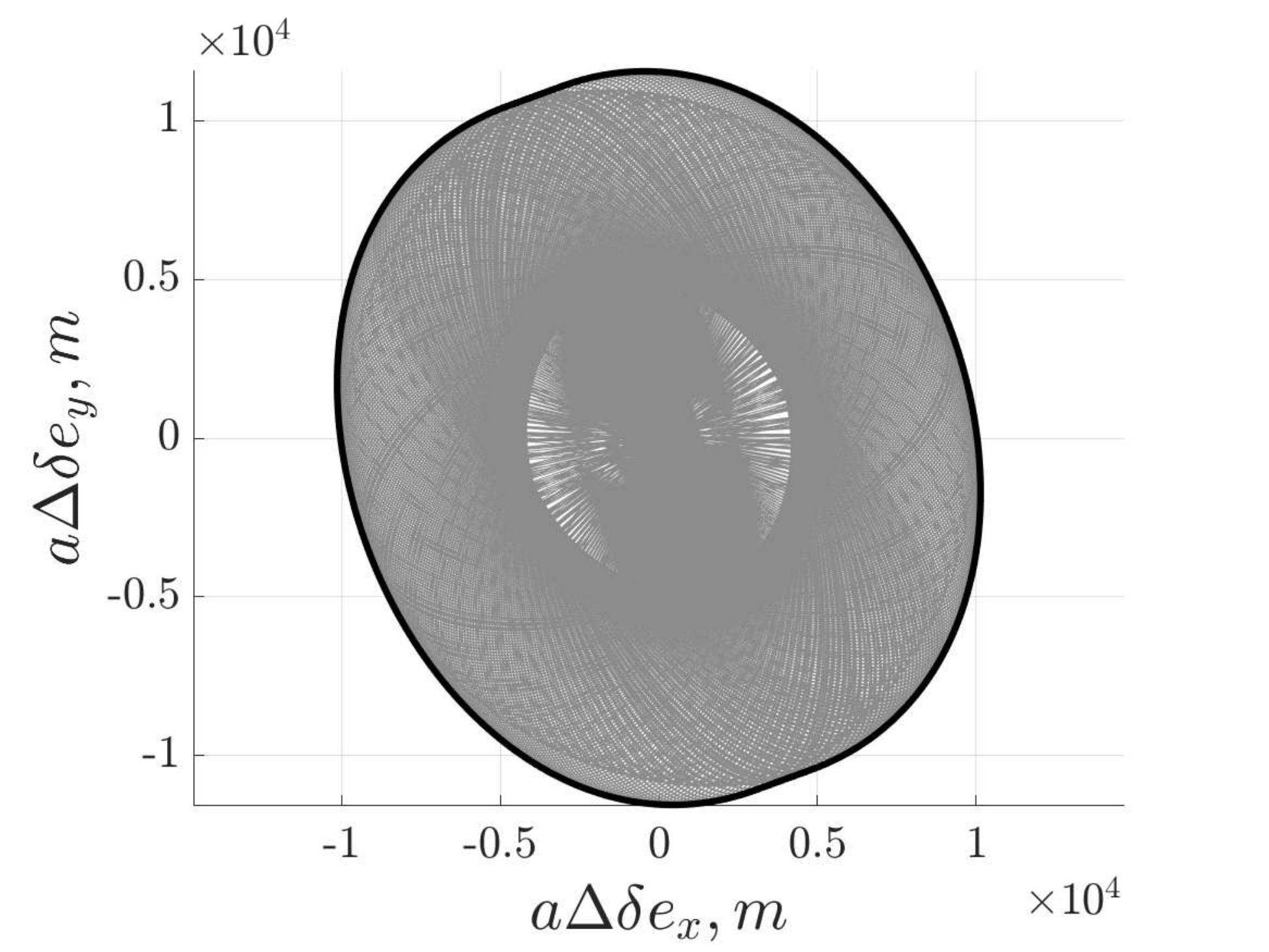}
        \caption{$\Delta\delta\pmb{e}$ plane}
        \label{fig:Sdv_ecc_de}
    \end{subfigure}
    \begin{subfigure}[h]{0.24\textwidth}
        \includegraphics[width=\textwidth]{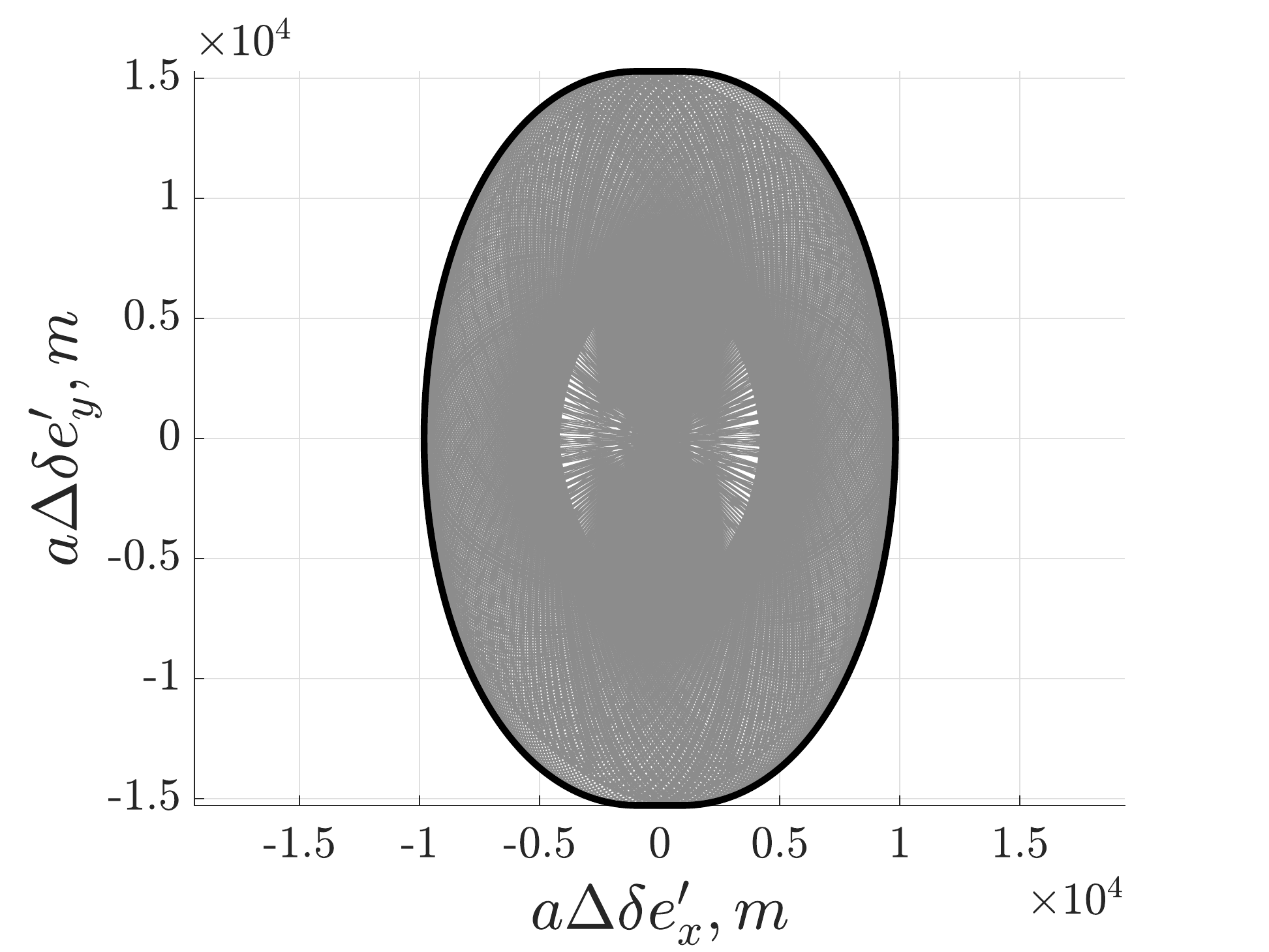}
        \caption{$\Delta\delta\pmb{e}'$ plane}
        \label{fig:Sdv_ecc_deprime}
    \end{subfigure}
        \begin{subfigure}[h]{0.24\textwidth}
        \includegraphics[width=\textwidth]{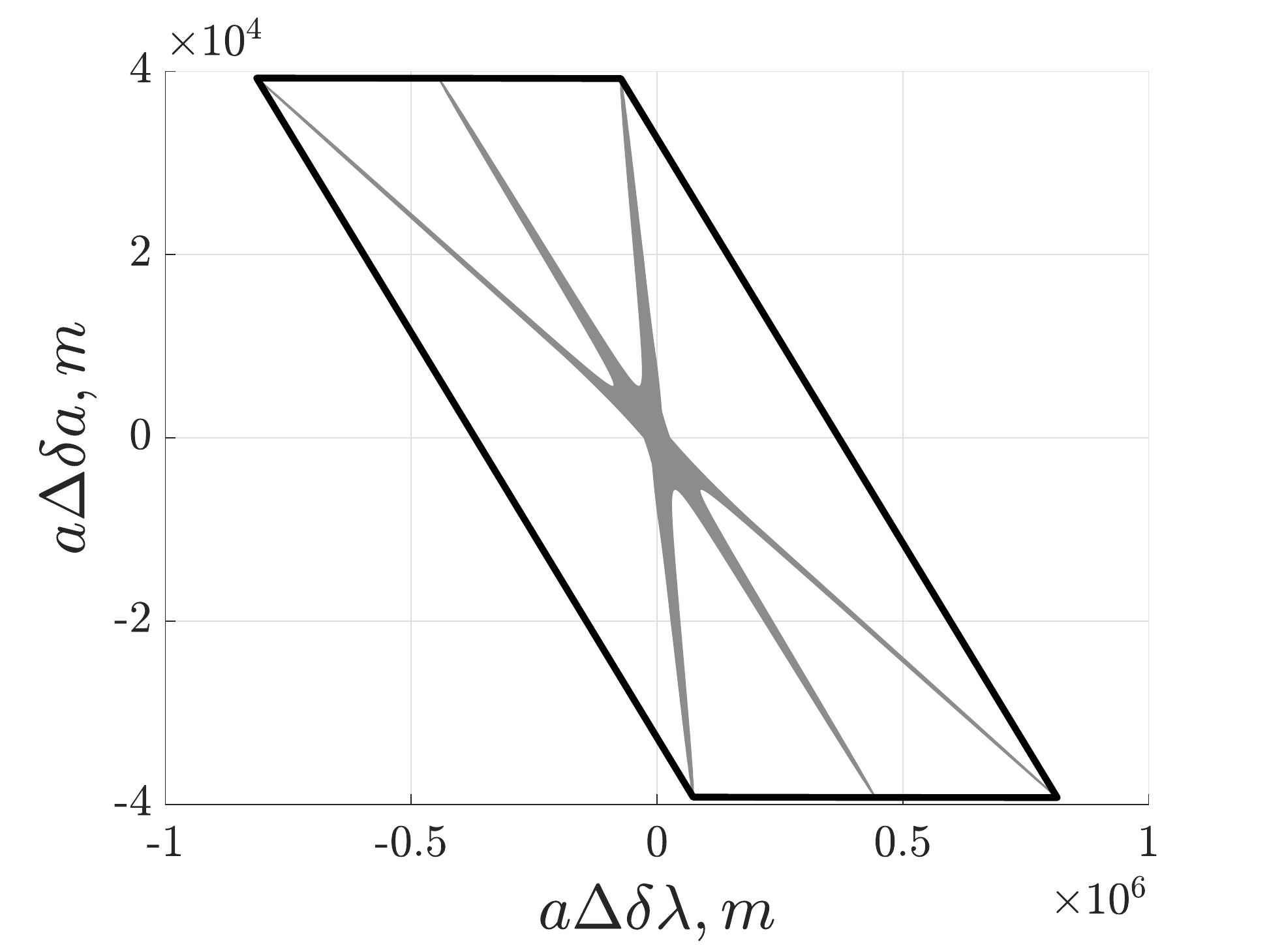}
        \caption{$\Delta\delta \pmb{a}$ plane}
        \label{fig:Sdv_ecc_dadl}
    \end{subfigure}
        \begin{subfigure}[h]{0.24\textwidth}
        \includegraphics[width=\textwidth]{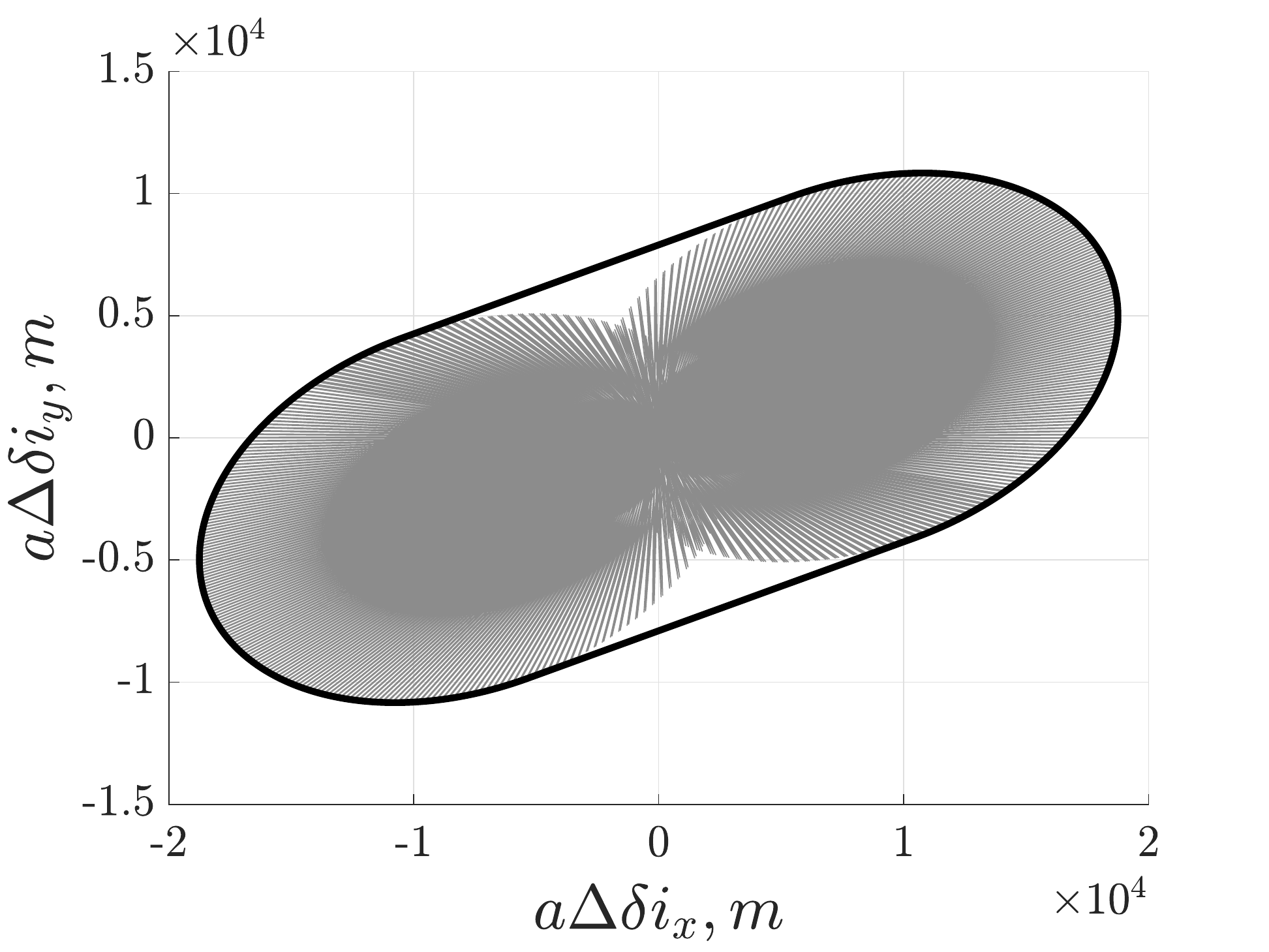}
        \caption{$\Delta\delta \pmb{i}$ plane}
        \label{fig:Sdv_ecc_di}
    \end{subfigure}
    \caption{$S(1,T)$ (gray) and convex hull $S^*(1,T)$ (solid line) boundary in eccentric orbits, generated using Eq. (12)-(13) in Ref. \cite{bib:ChernickDamico}}
    \label{fig:Sdv_ecc}
    \end{figure}

Recall, if a desired pseudo-state $\Delta\delta {\pmb{\alpha}}_{des}$ lies on the boundary of $S(c,T)$ and $S^*(c,T)$ for some cost $c$, the reconfiguration can be achieved with a single maneuver. If the desired pseudo-state lies only on $S^*(c,T)$'s boundary, then the number of maneuvers required is no more than the size of the state $n$ \cite{bib:KDAutomatica}. Looking at the three 2D planes separately, reconfigurations can be categorized based on whether they can be achieved with one or two maneuvers by determining the times $\in T$ where $S(c,t)$ intersects the boundary of $S^*(c,T)$, where $S$ and $S^*$ are computed with Eqs. \eqref{eqn:Sdv} and \eqref{eqn:Sstar}, respectively.  
The reachable sets for eccentric orbits (Fig. \ref{fig:Sdv_ecc}) are found by substituting $\pmb{\Phi}$ (Eq. (12) in Ref. \cite{bib:ChernickDamico}) and $\pmb{\Gamma}$ (Eq. (13) in Ref. \cite{bib:ChernickDamico}) into the equations for $S$ and $S^*$. In the $\Delta\delta\pmb{e}$ plane (Fig. \ref{fig:Sdv_ecc_de}), $S$ and $S^*$ are equal except for a very small region near the axis of symmetry of the reachable set which appears only for very high eccentricities (i.e. $e>0.85$). As long as the desired pseudo-state does not lie within this region, the point can be reached with a single maneuver. Consistent with Eq. \eqref{eqn:Ddeprime_to_Dde}, the reachable set in the $\Delta\delta\pmb{e}'$ plane (Fig. \ref{fig:Sdv_ecc_deprime}) has the same attributes as the reachable set in the $\Delta\delta\pmb{e}$ plane, but it is unaffected by the chief argument of perigee and is scaled along the y-axis by the eccentricity of the chief orbit. In the $\Delta\delta \pmb{a}$ plane (Fig. \ref{fig:Sdv_ecc_dadl}), there are only a few locations on the top/bottom surface of the parallelogram that are reachable with a single maneuver. The $\Delta\delta\pmb{i}$ plane is similar to the $\Delta\delta\pmb{e}$ plane, but with a slightly larger region where $S$ and $S^*$ disconnect (Fig. \ref{fig:Sdv_ecc_di}). 
 Remarkably, because of the periodicity of the control input matrix in Eq. \eqref{eqn:Gamma_chain}, the shape of $S^*$ in the $\Delta\delta\pmb{e}$ and $\Delta\delta\pmb{i}$ planes is independent of the reconfiguration time if it is greater than one orbit. In the $\Delta\delta\pmb{a}$ plane, there is no periodicity so $S^*$ is virtually unbounded for time growing to infinity. 
 
\begin{figure}[H]
    \centering 
    \begin{subfigure}[t]{0.24\textwidth}
        \includegraphics[width=\textwidth]{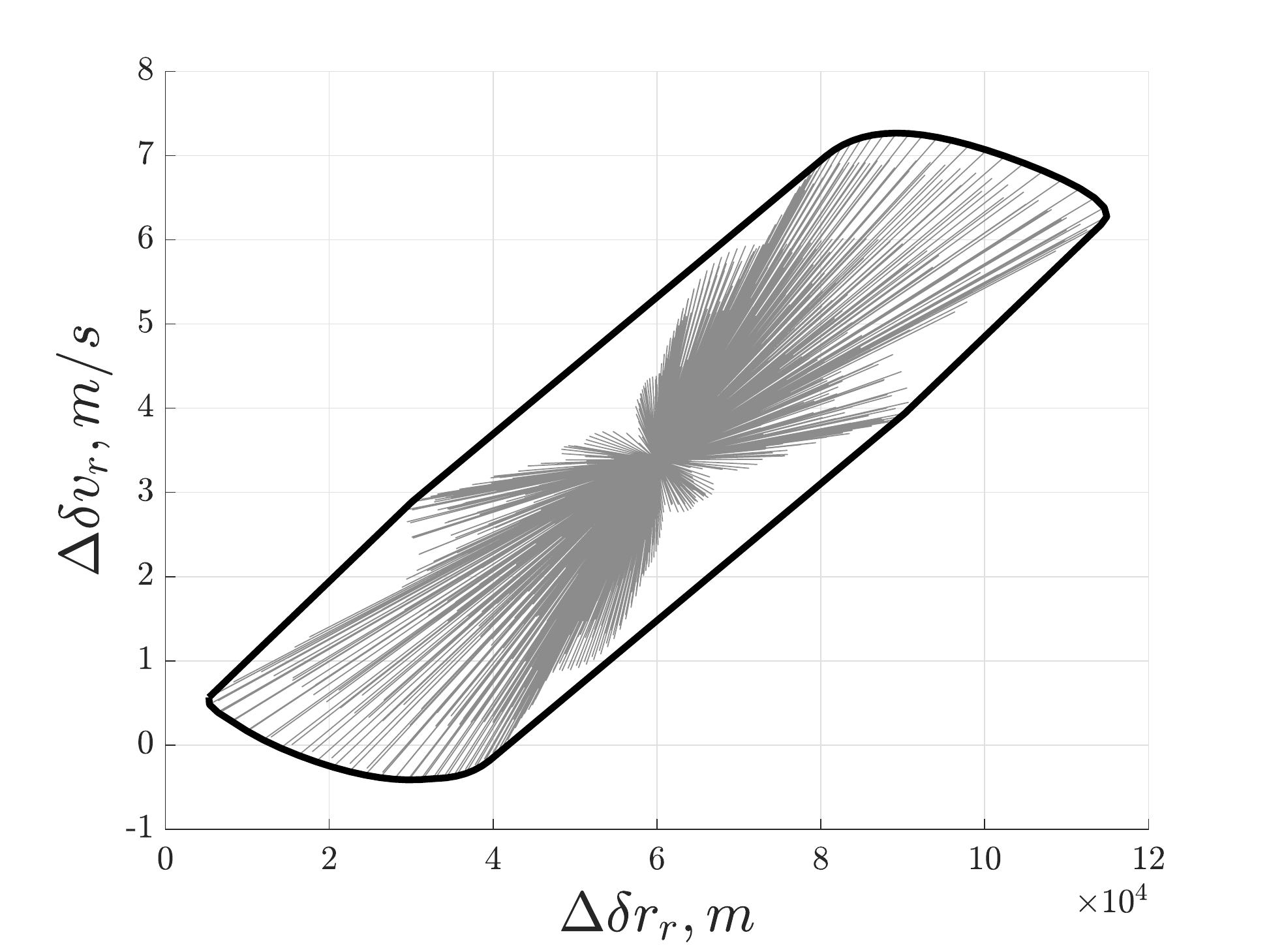}
        \caption{$\delta r, \delta v_r$ plane}
        \label{fig:YA_RR}
    \end{subfigure}\hfill
        \begin{subfigure}[t]{0.24\textwidth}
        \includegraphics[width=\textwidth]{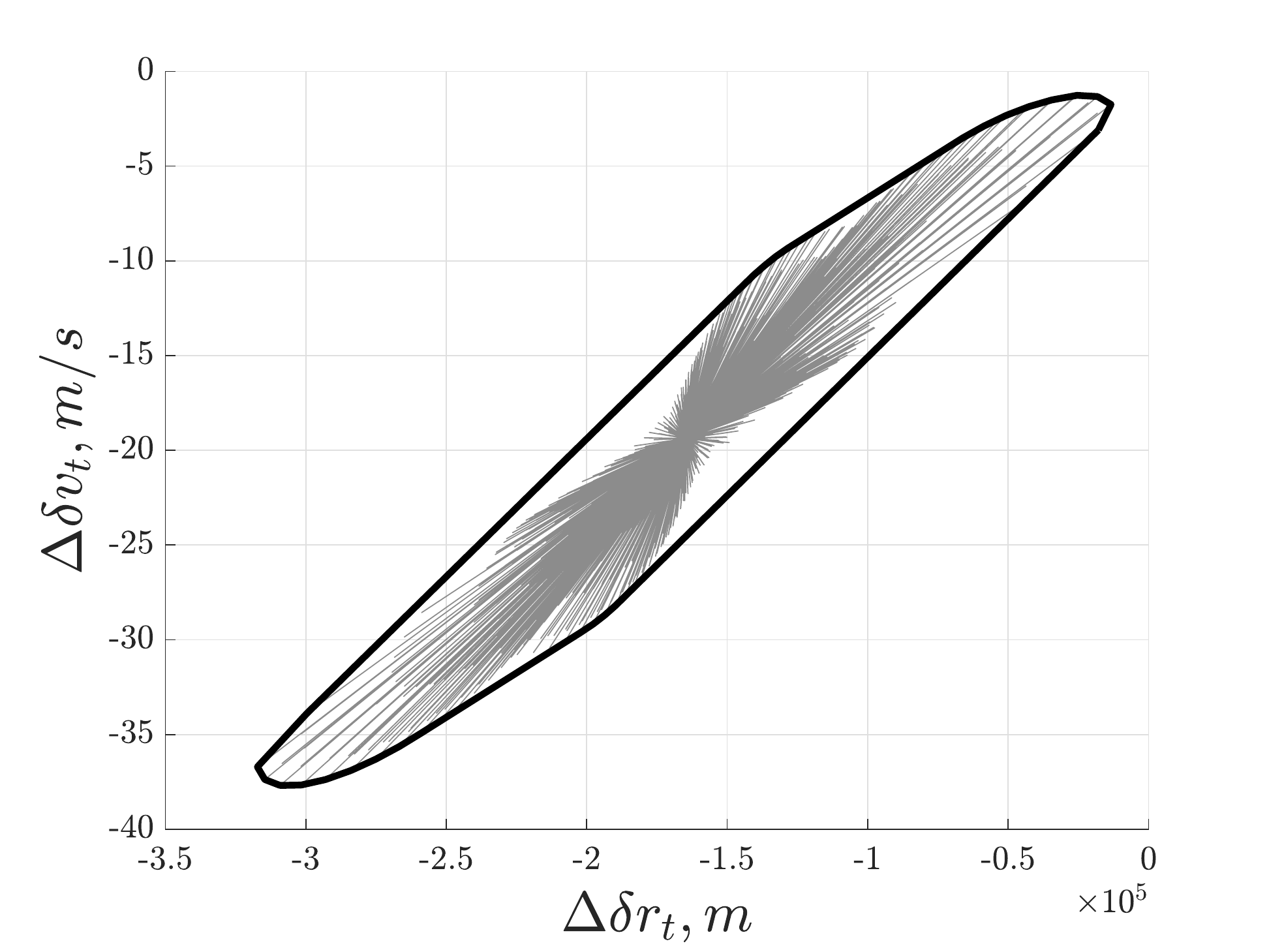}
        \caption{$\delta t, \delta v_t$ plane}
        \label{fig:YA_TT}
    \end{subfigure}\hfill
    \begin{subfigure}[t]{0.24\textwidth}
        \includegraphics[width=\textwidth]{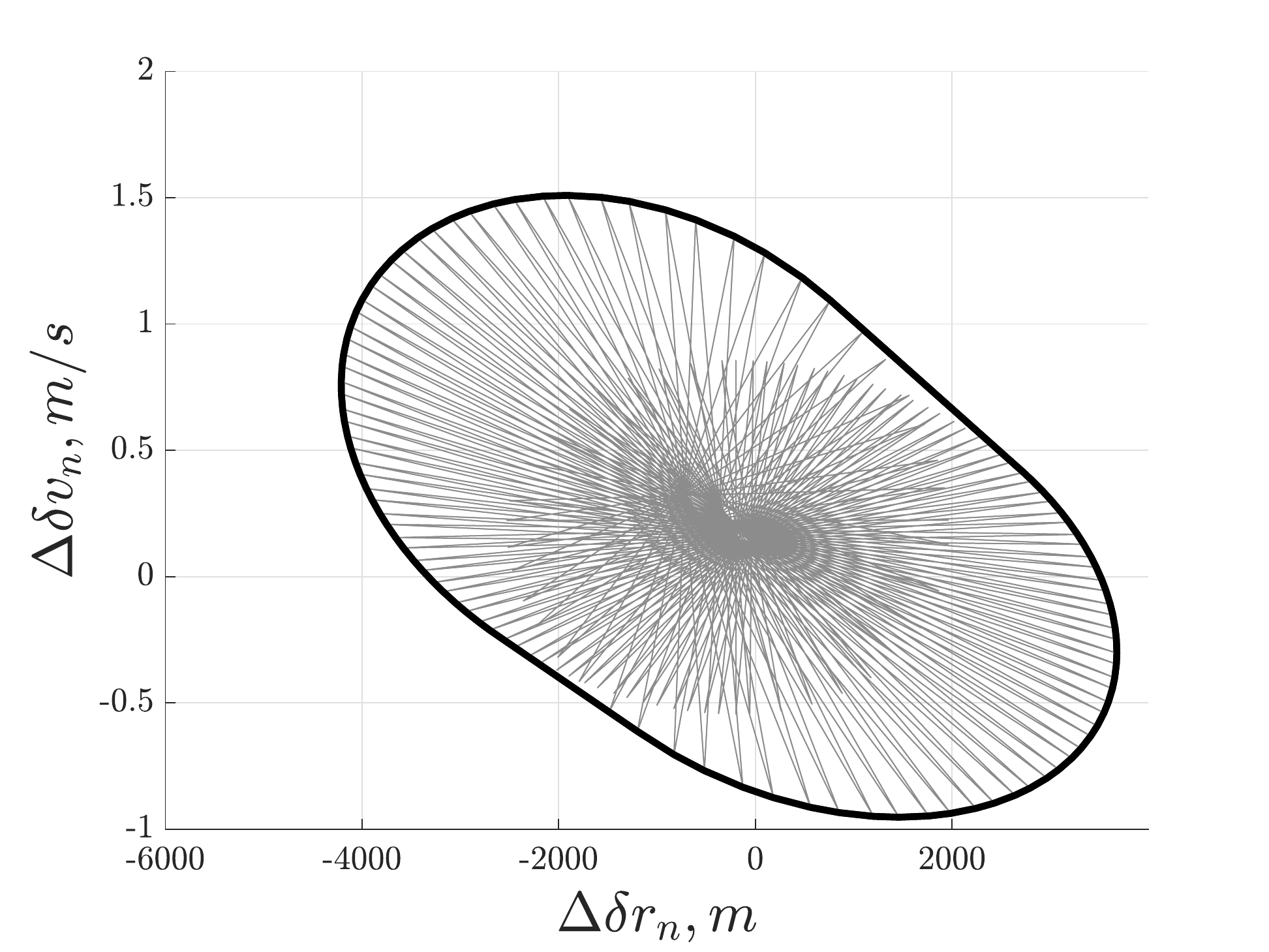}
        \caption{$\delta n, \delta v_n$ plane}
        \label{fig:YA_NN}
    \end{subfigure}
    \caption{$S(1,T)$ (gray) and convex hull $S^*(1,T)$ boundary (solid black line) in $\delta r, \delta v_r$, $\delta t, \delta v_t$, and $\delta n, \delta v_n$ planes in eccentric orbits, generated using YA STM.}
    \label{fig:YA_RS}
    \end{figure}

 The simple geometry of the reachable sets in ROE space above is the basis for derivation of the reachable delta-v minima and closed-form maneuver schemes in this paper. In fact, comparing reachable sets across state representations provides further justification for using ROE. Figure \ref{fig:YA_RS} shows the reachable sets in Hill's coordinates, the most common relative state representation. The reachable sets in the $(\Delta\delta r_r, \Delta\delta v_r)$, $(\Delta\delta r_t, \Delta\delta v_t)$, and $(\Delta\delta r_n, \Delta\delta v_n)$ planes are generated using the Yamanaka-Ankerson (YA) STM \cite{bib:YamanakaAnk} for eccentric orbits. The YA reachable sets are not polygons and cannot be described by explicit linear or polynomial expressions just by inspection. This comparison furthers the claim that using ROE simplifies the relative motion reconfiguration problem. 
Note that to plot these reachable sets, the reconfiguration time is discretized using a small timestep. As the timestep approaches zero, the graph of $S(c,T)$ becomes more dense. The intuition gained from analyzing the simplicity of the ROE reachable sets will be used to define dominance cases, derive closed-form expressions for the reachable delta-v minima, and develop globally optimal closed-form maneuver schemes in the sections that follow.
\section{Reachable Delta-v Minima}\label{sec:reachabledvmin}
In previous work, a metric called the delta-v lower bound, $\delta v_{lb}$, was used to quantify optimality \cite{bib:GaiasDamico1,bib:ChernickDamico}. As the name suggests, $\delta v_{lb}$ is a lower bound on the delta-v required for a given reconfiguration. It was derived assuming the use of only tangential maneuvers because of their inherent efficiency, which restricts the definition to reconfigurations that can be reached using only tangential maneuvers. It is therefore appropriate to define a new metric, called the reachable minimum delta-v, $\delta v_{min}$, which is the minimum delta-v required to achieve a desired reconfiguration. Recall, this expression was used in the proof that an $2n$-dimensional ($2n$-D) state can be decoupled into $n$ 2D planes. This section will derive closed-form expressions and their applicability for the reachable delta-v minima in orbits of arbitrary eccentricity by leveraging the state decoupling proof, domain specific knowledge, and the linear scaling properties of the reachable set. First, this section lays out the general methodology used to derive the reachable delta-v minimum. Second, the general methodology is applied to a specific example. Finally, the explicit closed-form expressions of the reachable delta-v minima are provided in tables for all dominance cases. The general methodology follows, given a desired pseudo-state, $\Delta\delta {\pmb{\alpha}}_{des}$ (see Eq. \eqref{eqn:optcontrolproblem}) and reconfiguration time, $T$. 

\subsection{General Methodology}\label{sec:gen_meth_reachable_dv}
\begin{figure}[H]
    \centering
    \begin{subfigure}[t]{0.23\textwidth}
        \includegraphics[width=\textwidth]{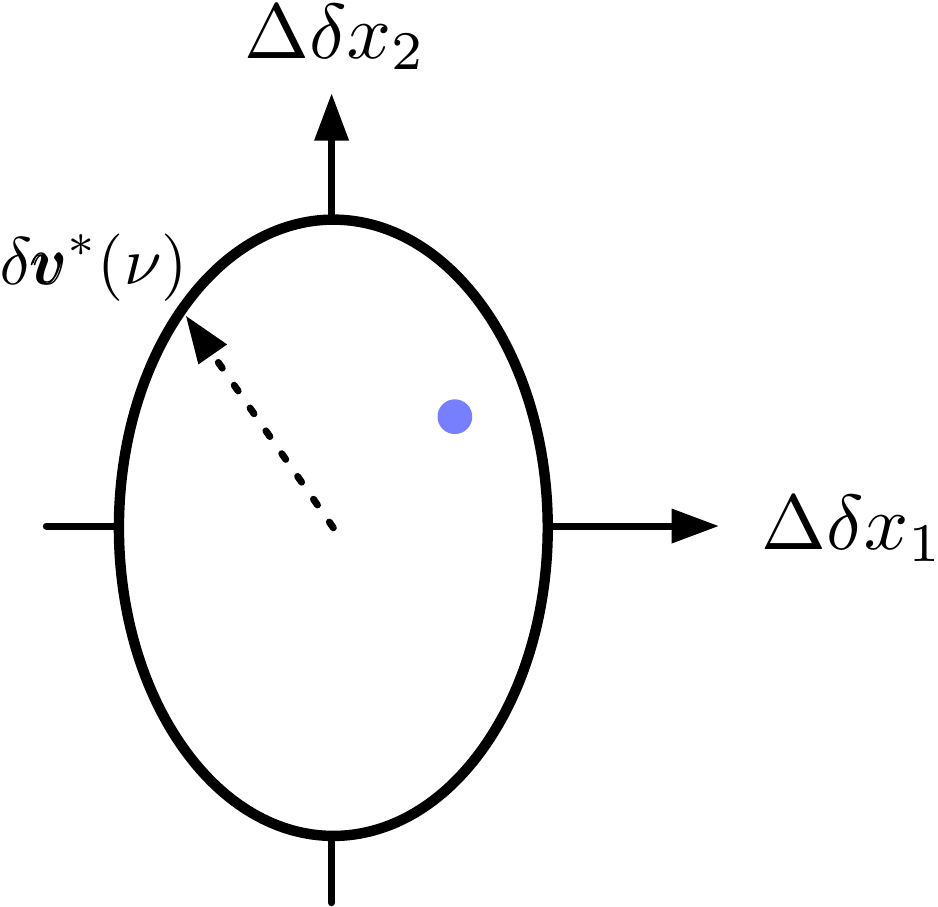}
        \caption{Find the ``best'' unitary delta-v as a function of maneuver location $\delta \pmb{v}^*(\nu)$}
        \label{fig:rdvmin_step1}
    \end{subfigure}\hfill
    \begin{subfigure}[t]{0.23\textwidth}
        \includegraphics[width=\textwidth]{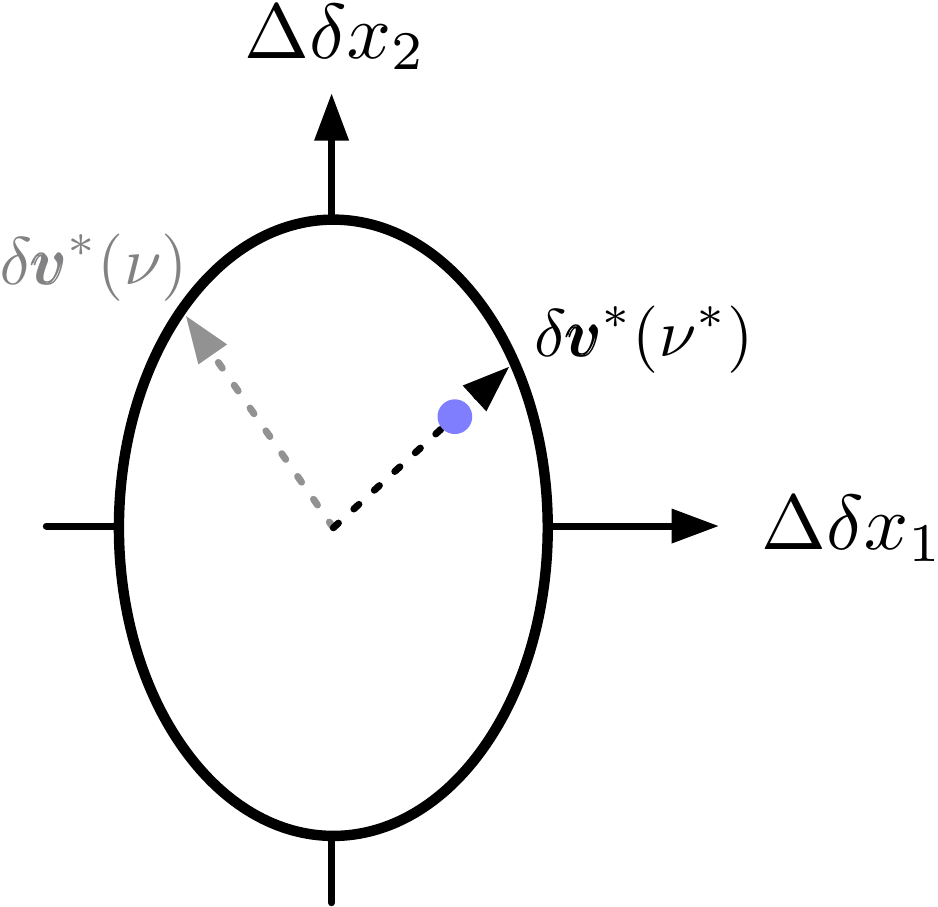}
        \caption{Find the maneuver location $\nu^*$ that aligns with the phase of the desired pseudo-state }
        \label{fig:rdvmin_step2}
    \end{subfigure}\hfill
    \begin{subfigure}[t]{0.23\textwidth}
        \includegraphics[width=\textwidth]{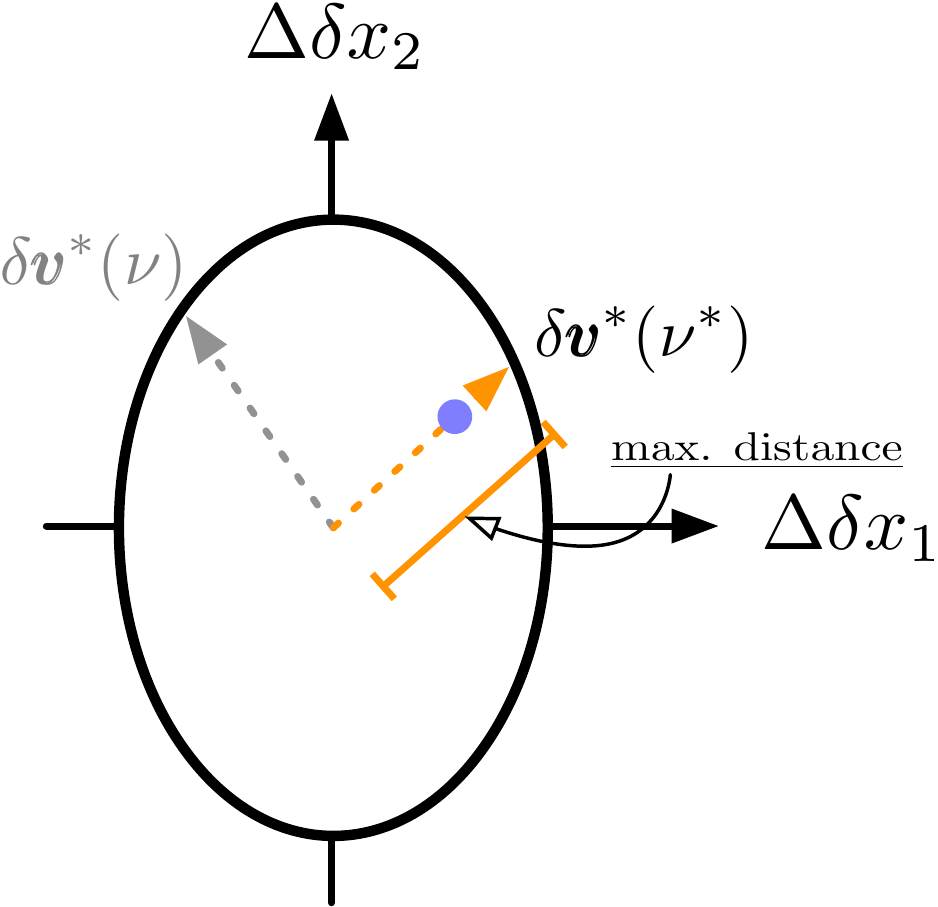}
        \caption{Calculate the maximum distance along the phase direction}
        \label{fig:rdvmin_step3}
    \end{subfigure}\hfill
    \begin{subfigure}[t]{0.23\textwidth}
        \includegraphics[width=\textwidth]{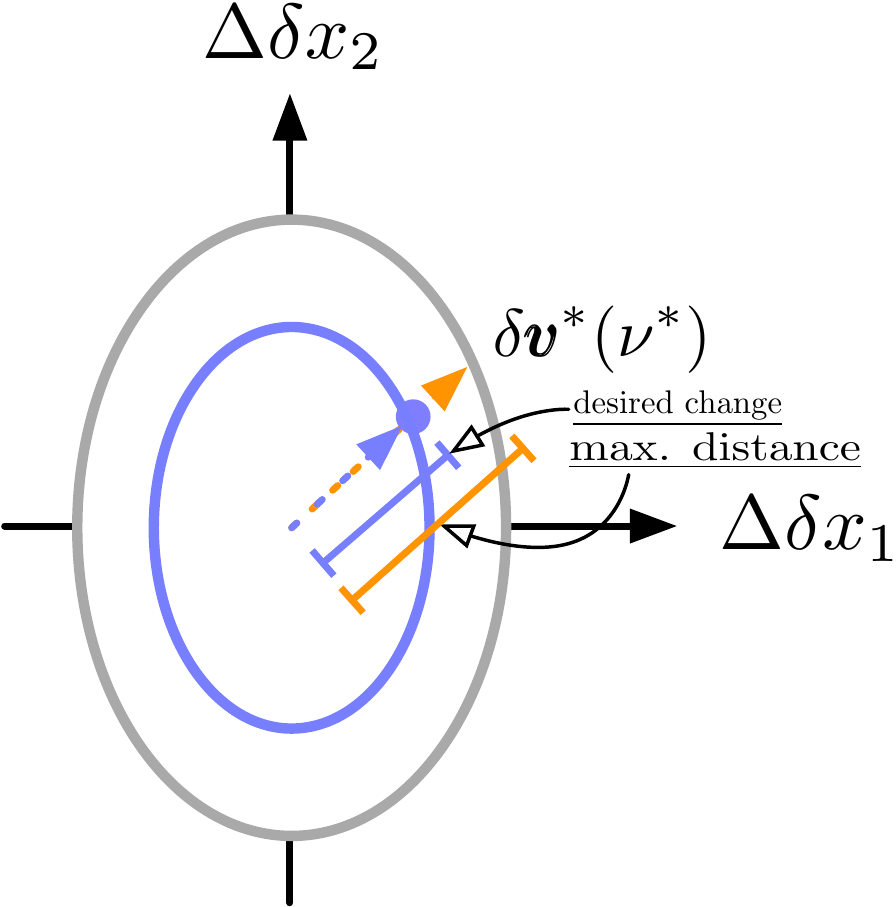}
        \caption{Solve for $\delta v_{min}$ as a ratio of the magnitude of the desired pseudo-state over the maximum distance}
        \label{fig:rdvmin_step4}
    \end{subfigure}    
    \caption{Illustration of the general methodology of the reachable delta-v minimum derivation process. Figures a-d correspond to Steps 1-4.}\label{fig:rdvmin_explain}
\end{figure}

The general methodology to derive closed-form expressions for the reachable delta-v minimum for in-plane reconfigurations is based on the four steps below and visualized in Fig. \ref{fig:rdvmin_explain}. A parameter with notation $(.)^*$ is a specific instance of the variable $(.)$.

1. Find the expression for the maneuver of magnitude one, $\delta \pmb{v}^*$, that achieves the largest change in ROE. $\delta\pmb{v}^*$ is a function of the true anomaly, $\nu$. The tangential component of $\delta \pmb{v}^*$, $\delta v_t^*$ is found by solving for the critical points of $\Delta\delta( {.})_x^2 + \Delta\delta( {.})_y^2$, where $\Delta\delta( {.})_x$ and $\Delta\delta( {.})_y$ are a set of parameterized equations that define the boundary of the convex hull in a given 2D plane $(x,y)$. The parametric equations are found by propagating the effect of a single maneuver at a given $\nu$ (or equivalent time $t$) using the STM and control input matrix and are given by  
\fontsize{8.5}{11}\selectfont\begin{equation}\label{eqn:param_eqns}
\begin{bmatrix}\Delta\delta ( {.})_x (\nu,\delta v_t^*(\nu)) \\ \Delta\delta ( {.})_y (\nu,\delta v_t^*(\nu))\end{bmatrix} = \pmb{\Phi}(t_f,t)\pmb{\Gamma}(t) = \left \{
\begin{matrix*}[l]
\begin{bmatrix}\Delta\delta  {\lambda}(\nu,\delta v_t^*(\nu))  \\ \Delta\delta  {a}(\nu,\delta v_t^*(\nu))\end{bmatrix} \text{ in the } \Delta \delta a, \Delta \delta\lambda \text{ plane, or} \\
\begin{bmatrix}\Delta\delta  {e}_x(\nu,\delta v_t^*(\nu)) \\  \Delta\delta  {e}_y(\nu,\delta v_t^*(\nu)) \end{bmatrix} \text{ in the } \Delta\delta \pmb{e} \text{ plane, or}  \\
\begin{bmatrix}\Delta\delta  {i}_x(\nu,\delta v_n^*(\nu)) \\  \Delta\delta  {i}_y(\nu,\delta v_n^*(\nu)) \end{bmatrix} \text{ in the } \Delta\delta \pmb{i} \text{ plane.}
\end{matrix*}\right.
\end{equation}\normalsize Eq. \eqref{eqn:param_eqns} are functions of the maneuver application time $t$ or equivalent angle $\nu$. The explicit form of Eq. \eqref{eqn:param_eqns} is given in the subsections that follow. Given $\delta v_t^*$, the radial component, $\delta v_r^*$, is found by solving the constraint of unitary delta-v, $\delta v_r^2 + \delta v_t^2 = 1$. For out-of-plane reconfigurations, the unitary maneuver magnitude constraint is $\delta v_n^* = 1$. Figure \ref{fig:rdvmin_step1} illustrates Step 1, where the solid black line represents $S^*(c,T)$ for an arbitrary $c$. The effect of the maneuver is represented by the dashed arrow and is a function of the maneuver location along the reference orbit. The desired pseudo-state does not lie on the boundary, so this is not the optimal delta-v, as expected.

2. \smalltab Solve for the maneuver location, $\nu^*$ or $u^*$, that corresponds to the direction of the desired pseudo-state. The maneuver location is found by equating the phase of the desired pseudo-state to the phase of the $x$, $y$ components of the appropriate set of parameterized functions in Eq. \eqref{eqn:param_eqns} as
\fontsize{8.5}{11}\selectfont
\begin{equation}\label{eqn:ratio}
\tan^{-1}\left(\frac{\Delta\delta ( {.})_y \left(\nu,\delta v_t^*(\nu)\right)}{\Delta\delta ( {.})_x \left(\nu,\delta v_t^*(\nu)\right)}\right) = \tan^{-1}\left(\frac{\Delta\delta ( {.})_{y,des}}{\Delta\delta ( {.})_{x,des}}\right).
\end{equation}\normalsize
Step 2 is illustrated in Fig. \ref{fig:rdvmin_step2}, where a maneuver at $\nu^*$ changes the state in the direction of the desired pseudo-state, indicated by the darker dashed arrow. The explicit values of $\nu^*$ are given in the last column of Table \ref{table:dvstars_ecc}.

3. \smalltab Substitute $\delta\pmb{v}^*$, $\nu^*$ found in Steps 1 and 2 into the parameterized functions in Eq. \eqref{eqn:param_eqns}. This yields $\Delta\delta( {.})_x^* = \Delta\delta( {.})_x(\nu^*,\delta v_t^*(\nu^*))$ and $\Delta\delta( {.})_y^* = \Delta\delta( {.})_y(\nu^*,\delta v_t^*(\nu^*))$, which are the $x$ and $y$ components of the vector that defines the maximum reachable distance $\Delta\delta( {.})_{max} = (\Delta\delta( {.})_x^{*2} + \Delta\delta( {.})_y^{*2})^{1/2}$ in the direction of the desired pseudo-state $\Delta\delta {(.)}_{des} = (\Delta\delta  {(.)}_{x,des},\Delta\delta  {(.)}_{y,des})$. Equivalently, the vector points to the pseudo-state on the boundary of the convex hull defined by $||\delta \pmb{v}^*||_2=1$ in the desired direction, as illustrated in Fig. \ref{fig:rdvmin_step3}. 

4. \smalltab Find $\delta v_{min}$ by dividing the norm of the desired pseudo-state $||\Delta\delta {(.)}_{des}||_2$ by the maximum distance $\Delta\delta( {.})_{max}$, found in Step 3. This scales the reachable set so that the desired pseudo-state lies on the boundary of the convex hull, as illustrated in Fig. \ref{fig:rdvmin_step4}.

\subsection{Example: Application of Methodology to Eccentric Chief Orbit} From Sec. \ref{sec:RST}, the dominant 2D plane drives the delta-v cost of the $2n$-D reconfiguration.
This example demonstrates how the general methodology above is used to derive the expression for $\delta v_{min}$ for one of the 2D planes, $\delta \pmb{e}'$, in eccentric chief orbits. In Ref. \cite{bib:RST_conf}, recall the algorithms were formulated using the state given in Eq. \eqref{eqn:ROE}. For any reconfiguration where the chief orbit has a nonzero (and defined) argument of perigee,the reachable set $S^*$ is oriented at an angle of $\omega$ from the vertical in the $\Delta\delta \pmb{e}$ plane. The pseudo-state $\Delta\delta  {\pmb{e}}$ was corrected for the rotation due to the argument of perigee as $\Delta\delta \tilde{\pmb{e}} = \pmb{R}^{-1}\Delta\delta {\pmb{e}}$, where $\pmb{R}$ represents a counterclockwise rotation through $\omega$ about the origin. The parametric equations in Eq. \eqref{eqn:param_eqns} that defined the boundary of $S(c,T)$ in the $\Delta\delta\tilde{\pmb{e}}$ plane were the $\delta \pmb{e}$ components of $\pmb{\Phi}(t_f,t_k)\pmb{\Gamma}(t_k)$ rotated into the $\Delta\delta\tilde{\pmb{e}}$ plane and expressed in terms of $\nu$, and were given by
 \fontsize{9}{11}\selectfont\begin{equation}\label{eqn:parametric_de}
 \begin{bmatrix} a\Delta\delta \tilde{e}_x(\nu,\delta v_t(\nu)) \\  a\Delta\delta \tilde{e}_y(\nu,\delta v_t(\nu)) \end{bmatrix} = \frac{1}{n}\begin{bmatrix} \eta\sin(\nu)\sqrt{1-\delta v_t^2(\nu)} + \eta\frac{(2+e\cos(\nu))\cos(\nu) + e}{1+e\cos(\nu)}\delta v_t(\nu) \\
     -\eta\cos(\nu)\sqrt{1-\delta v_t^2(\nu)} + \eta\frac{(2+e\cos(\nu))\sin(\nu)}{1+e\cos(\nu)}\delta v_t(\nu)\end{bmatrix}.
 \end{equation}\normalsize
However, in this paper, the state of choice is the modified ROE given in Eq. \eqref{eqn:ROE_modified}, which, as discussed in Sec. \ref{sec:background}, allows for complete decoupling of the in-plane and out-of-plane control solutions. Though the transformation between the $\delta \pmb{e}$ state in Eq. \eqref{eqn:ROE} to the $\delta\pmb{e}'$ state in Eq. \eqref{eqn:ROE_modified} is nonlinear, the transformation between the pseudo-states, $\Delta\delta\tilde{\pmb{e}}$ and $\Delta\delta {\pmb{e}}'$, is very simple. In fact, a desired change in $\delta \pmb{e}'$ relates to a desired change in $\delta\pmb{e}$ as
\fontsize{9}{11}\selectfont\begin{equation}\label{eqn:Ddeprime_to_Dde}
    \begin{bmatrix}
            \Delta\delta \tilde{e}_{x,des} \\
            \Delta\delta \tilde{e}_{y,des}
    \end{bmatrix} =
    \begin{bmatrix}
            \Delta\delta  {e}'_{x,des} \\
            e_c\Delta\delta  {e}'_{y,des}
    \end{bmatrix}.
\end{equation}\normalsize
Therefore, a desired pseudo-state in $\delta\pmb{e}'$, $\Delta\delta {\pmb{e}}'_{des}$, can be simply transformed to the equivalent pseudo-state $\Delta\delta\tilde{\pmb{e}}_{des}$ through Eq. \eqref{eqn:Ddeprime_to_Dde}. Thus, the algorithms that follow in this section, which were derived for the $\delta\tilde{\pmb{e}}$ state, can be used.
The equation for $\delta v_t^*$ is derived according to Step 1 in the general methodology in Sec. \ref{sec:gen_meth_reachable_dv} as \fontsize{8.5}{11}\selectfont\begin{equation}\label{eqn:dvtstar_example}\delta v_t^*(\nu) = \left\{\begin{matrix*}[l]+\sqrt{\frac{1}{2} + \frac{f_1(\nu)}{2\sqrt{4+f_1(\nu)^2}}} \text{\smalltab for } \pm\Delta\delta \tilde{e}_{x,des},\mp\Delta\delta \tilde{e}_{y, des}\\
 -\sqrt{\frac{1}{2} - \frac{f_1(\nu)}{2\sqrt{4+f_1(\nu)^2}}}\text{\smalltab for } \pm\Delta\delta \tilde{e}_{x,des},\pm\Delta\delta \tilde{e}_{y,des}\end{matrix*}\right.\text{, where } \begin{matrix*}[l]f_1(\nu) = \frac{f_2(\nu)}{(1+e\cos(\nu))e\sin(\nu)}\\ f_2(\nu) = 2e^2\cos^2(\nu) + 6e\cos(\nu) + e^2 + 3\end{matrix*}.\end{equation}\normalsize  
To solve for $\nu^*$, Eqs. \eqref{eqn:parametric_de}  and \eqref{eqn:dvtstar_example} are substituted into Eq. \eqref{eqn:ratio}. However, in this case, solving Eq. \eqref{eqn:ratio} yields two values of $\nu^*$ per orbit, so some extra considerations must be made. First, it can be shown that the arbitrary pseudo-states $\pm\Delta\delta  {e}_x, \mp\Delta\delta  {e}_y$ are achieved by the same $\delta v_t^*$ and $\nu^*$.  The same is true for arbitrary pseudo-states $\pm\Delta\delta  {e}_x, \pm\Delta\delta  {e}_y$. Therefore, in place of $\Delta\delta ( {.})_{des}$ in Eq. \eqref{eqn:ratio}, $\Delta\delta\hat{(.)}_{des}$ is used, given by 
\fontsize{9}{11}\selectfont\begin{equation}\label{eqn:changeDdes}(\Delta\delta \hat{(.)}_{x,des},\Delta\delta \hat{(.)}_{y,des}) = 
\left\{\begin{matrix*}[l](\Delta\delta \tilde{e}_{x,des}, \Delta\delta \tilde{e}_{y,des}) &\text{ if } +\Delta\delta \tilde{e}_{y,des}\\
(-\Delta\delta \tilde{e}_{x,des}, -\Delta\delta \tilde{e}_{y,des})& \text{ if } -\Delta\delta \tilde{e}_{y,des}\end{matrix*}\right.
\end{equation}\normalsize
Second, it can be shown that $S(c,t)$ briefly disconnects from the boundary of the convex hull during each orbit. By solving for the $\nu$ values that maximize the parametric $\Delta\delta\tilde{e}_y$  expression in Eq. \eqref{eqn:parametric_de}, it is found that disconnection occurs at $\nu_{dis.} = \pi + \cos^{-1}(e)$ and reconnection occurs at $\nu_{re.} = \pi - \cos^{-1}(e)$. These points define the regions that contain $\nu^*$, the maneuver locations at which $\delta v_t^*(\nu^*)$ aligns with the phase of the desired pseudo-state in each orbit. Table \ref{table:nuboundaries_nuopt} gives the range containing $\nu^*$, based on the sign of the desired pseudo-state. If the phase of $a\Delta\delta\tilde{e}_{des}$ (denoted $\angle a\Delta\delta\tilde{e}_{des}$ in the table) lies in the region bounded by $\nu_{re}$ or $\nu_{dis}$ as 
\fontsize{9}{11}\selectfont
\begin{equation}\label{eqn:condition_on_phase_of_des_change}
    \angle \Delta\delta\tilde{e}_{des}\in [\angle \nu_{dis}, \pi - \angle\nu_{dis}] \text{\tab or \tab} \angle \Delta\delta\tilde{e}_{des}\in [\pi - \angle\nu_{re}, \angle\nu_{re}],
\end{equation}\normalsize
then the desired pseudo-state lies in the disconnected region. Note that the phase in the plane denoted with $\angle$ is not necessarily equivalent to the true anomaly $\nu$. 
\fontsize{9}{11}\selectfont
\begin{longtable}{K{3.3cm} K{3cm} K{1.2cm} K{1.2cm} K{1.2cm} K{1.2cm}} 
	\caption{Left and right boundaries of the range containing the optimal $\nu$ values for a dominant relative eccentricity vector reconfiguration}\label{table:nuboundaries_nuopt}
	\\ \toprule\toprule
	\textbf{Desired change} & \textbf{Disconnected region?} & \multicolumn{2}{c}{$\nu_{opt,1}$ \textbf{range}} & \multicolumn{2}{c}{$\nu_{opt,2}$ \textbf{range}} \\ \midrule
	{} & Eq. \eqref{eqn:condition_on_phase_of_des_change} true? & $\nu_{\text{left}}$ & $\nu_{\text{right}}$ & $\nu_{\text{left}}$ & $\nu_{\text{right}}$\\
	$\pm\Delta\delta \tilde{e}_{x,des},\pm\Delta\delta\tilde{e}_{y,des}$ & No & $\pi$ & $\nu_{dis}$ & 0 & $\nu_{re}$  \\
	{} & Yes & $\nu_{dis}$ & $2\pi$ & 0 & $\nu_{re}$\\
	$\pm\Delta\delta \tilde{e}_{x,des},\mp\Delta\delta\tilde{e}_{y,des}$ & No & $\nu_{re}$ & $\pi$ & $\nu_{dis}$ & $2\pi$ \\
	{} & Yes & $0$ & $\nu_{re}$ & $\nu_{dis}$ & $2\pi$ \\
    \bottomrule
\end{longtable} \normalsize 
There is a single, unambiguous value of $\nu^*$ in the range that satisfies Eq. \eqref{eqn:ratio}, and the Newton-Raphson (NR) method can be employed to efficiently find it. Because there is only one value of $\nu^*$ in the ranges given in Table \ref{table:nuboundaries_nuopt}, it is enough to use the average of the left and right boundary values as a very good initial guess for the NR method. The method requires the evaluation of the function $f$ and its derivative $df/d\nu$ at each step. $f$ comes from rearranging Eq. \eqref{eqn:ratio} and is given by
\fontsize{9}{11}\selectfont\begin{equation}\label{eqn:function_in_NR}
    f = a\Delta\delta \tilde{e}_y(\nu) - \left(a\Delta\delta\tilde{e}_{y,des}/a\Delta\delta\tilde{e}_{x,des}\right)a\Delta\delta \tilde{e}_x(\nu)
\end{equation}\normalsize
$df/d\nu$ is an analytic expression that can be found by simply applying the derivative quotient rule to Eq. \eqref{eqn:function_in_NR}. The bisection method can also be used, and is guaranteed to converge as long as Eq. \eqref{eqn:function_in_NR} evaluated at the left and right boundaries from the third and fourth columns of Table \ref{table:nuboundaries_nuopt} are of opposite signs \cite{bib:Boyd_CVX}. In the case that this criteria fails, columns two and three of Table \ref{table:nuboundaries_guaranteed_nuopt} contains ``relaxed'' boundaries that are guaranteed to evaluate to values of opposite sign when substituted into Eq. \eqref{eqn:function_in_NR}. Note that the boundaries in Table \ref{table:nuboundaries_nuopt} are a subset of the ``relaxed'' boundaries.
\fontsize{9}{11}\selectfont
\begin{longtable}{K{3.3cm} K{1.2cm} K{1.2cm} K{1.2cm} K{1.2cm}}
    \caption{Left and right boundaries of the range containing the optimal $\nu$ values for a dominant relative eccentricity vector reconfiguration with guaranteed convergence}\label{table:nuboundaries_guaranteed_nuopt} \\ \toprule\toprule
    \textbf{Desired change} & \multicolumn{2}{c}{$\begin{matrix*}\nu_{opt,1} \text{\textbf{ range,}} \\ \text{\textbf{guaranteed convergence}}\end{matrix*}$} & \multicolumn{2}{c}{$\begin{matrix*}\nu_{opt,2} \text{\textbf{ range,}} \\ \text{\textbf{guaranteed convergence}}\end{matrix*}$} \\ \midrule
    {} & $\nu_{\text{left}}$ & $\nu_{\text{right}}$  & $\nu_{\text{left}}$ & $\nu_{\text{right}}$ \\
    $\pm\Delta\delta \tilde{e}_{x,des},\pm\Delta\delta\tilde{e}_{y,des}$ &  $\pi^+$ & $^-2\pi$ & $0^+$ & $^-\pi$\\
    $\pm\Delta\delta \tilde{e}_{x,des},\mp\Delta\delta\tilde{e}_{y,des}$ & $0^+$ & $^-\pi$ & $\pi^+$ & $^-2\pi$\\
    \bottomrule
\end{longtable}    \normalsize
Then, following Steps 3 and 4 in Sec. \ref{sec:gen_meth_reachable_dv}, the general expression for reachable delta-v minimum for dominant $\delta\pmb{e}$ is found in closed-form and given in Table \ref{table:dvmin_de}. Derivation of other dominance cases follows the same procedure. The results are summarized in Tables \ref{table:dvmin_de}, \ref{table:dvmin_da}, and \ref{table:dvmin_ecc_oop}. 
\subsection{Reachable Delta-v Minimum in Eccentric Orbits}
Recall from Eq. \eqref{eqn:dvminproof}, the maximum $\delta v_{min}$ in Eq. \eqref{eqn:dvminproof} is the dominant ROE. Table \ref{table:dvstars_ecc} lists the values of $\delta \pmb{v}^*$ and $\nu^*$ found by applying Steps 1-2 in the general methodology in Sec. \ref{sec:gen_meth_reachable_dv} to each dominance case in eccentric orbits. 
The first column gives the name of the potential dominant ROE in the case that it is dominant by Eq. \eqref{eqn:dvminproof}. The second column gives the maneuver that achieves the maximum distance in the 2D plane as a function of the $\nu^*$ values given in the third column.
\fontsize{8.5}{10}\selectfont
\begin{longtable}{l K{7cm} K{4cm}}
		\caption{Optimal maneuver vectors $\delta \pmb{v}^*$ and optimal maneuver location $\nu^*$ for eccentric orbits}\label{table:dvstars_ecc} \\ \toprule\toprule
		
		{\textbf{Dominant...}} & {$\delta \pmb{v}^*$, \textbf{(m/s)}} & {$\nu^*$\textbf{, (rad)}}\\ \midrule
         \textit{... $\delta \pmb{e}$, $|\Delta\delta \tilde{e}_{(.),des}|>0$} & See Eq. \eqref{eqn:dvtstar_example} & Solve Eq. \eqref{eqn:ratio} with Table \ref{table:nuboundaries_nuopt}, \eqref{eqn:changeDdes} \\ \midrule
         
         \textit{... $\delta \pmb{e}$, $\Delta\delta \tilde{e}_{x,des} = 0$} & $\begin{bmatrix}0 & 1 & 0\end{bmatrix}^{\text{T}}$ & 0\\ \midrule
         
         \textit{... $\delta \pmb{e}$, $\Delta\delta \tilde{e}_{y,des} = 0$} & $\begin{bmatrix}0 & 1 & 0\end{bmatrix}^{\text{T}}$ & $\cos^{-1}((\eta - 1)/e_c)$\\ \midrule
    
        \textit{... $\delta a$} & $\begin{bmatrix}0 & 1 & 0\end{bmatrix}^{\text{T}}$ & 0 \\ \midrule
\textit{... $\delta \lambda$} & $\begin{matrix*}[c]\begin{bmatrix}\frac{e\sin(\nu_t)}{\sqrt{e^2 + 2e\cos(\nu_t) +1}} & \frac{1+e\cos(\nu_t)}{\sqrt{e^2 + 2e\cos(\nu_t) +1}} &0\end{bmatrix}^{\text{T}} \\ \begin{bmatrix}0 & 1 & 0\end{bmatrix}^{\text{T}}
\end{matrix*}$ & $\begin{matrix*}[c]\nu_t \\ 0\end{matrix*}$\\ \midrule 
\textit{... $\delta \pmb{i}$} & $\begin{bmatrix}0 & 0 & 1\end{bmatrix}^{\text{T}}$ & $\tan^{-1}\left(\frac{\Delta\delta\tilde{i}_{y,des}}{\Delta\delta \tilde{i}_{x,des}}\right)$\\ \bottomrule
        	\end{longtable}\normalsize             
Note that the case of dominant $\delta \lambda$ lists two values for $\delta \pmb{v}^*$, $\nu^*$. The other maneuver location, $\nu_t$, is the point at which the derivative of the parameterized curve (see Eq. \eqref{eqn:param_eqns}) is equal to the slope from the desired pseudo-state itself to the pseudo-state achieved at $\nu=0$ with a purely tangential maneuver. $\nu_t$ lies between $\text{floor}(\frac{\nu_f}{2\pi})2\pi$ and $\nu_f$ and is found by solving
\fontsize{9}{11}\selectfont
\begin{equation}\label{eqn:nut}\left.\frac{d}{d\nu}\Delta\delta  {a}\left(\nu,\delta v_t(\nu)\right)\Big\rvert_{\nu_t}\middle/\frac{d}{d\nu}\Delta\delta  {\lambda}\left(\nu,\delta v_t(\nu)\right)\Big\rvert_{\nu_t}\right. = \frac{a\Delta\delta {a}\left(\nu_t,\delta v_t(\nu_t)\right) +a\Delta\delta a_0}{a\Delta\delta {\lambda}\left(\nu_t,\delta v_t(\nu_t)\right) +a\Delta\delta\lambda_0}.
\end{equation}\normalsize
\fontsize{10}{11}\selectfont
\begin{equation}\label{eqn:Dd0}
\text{where }    a\Delta\delta  {a}_0, a\Delta\delta {\lambda}_0 \text{ is the pseudo-state achieved with }  \delta\pmb{v} = \begin{bmatrix} 0 & 1 & 0\end{bmatrix}^{\text{T}}\text{ at }\nu = 0.
\end{equation}\normalsize
As discussed in Sec. \ref{sec:cf_maneuver_schemes}, dominant $\delta\lambda$ and $\delta a$ reconfigurations can only be achieved optimally in the dominant plane, so to determine if a given 6D reconfiguration can be achieved, it suffices to prove that it is \textit{not} dominant $\delta\lambda$ or $\delta a$, which is true if the following conditions are met: 1) $\delta v_{min,\delta \pmb{e}}>\delta v_{min,\delta a}$, and 2) The desired pseudo-state in $\delta\lambda$, $\Delta\delta {\lambda}_{des}$, is inside the convex hull in the $\delta \pmb{a}$ plane. The details of the categorized $\delta v_{min,\delta\lambda}$ cases are given in Appendix C.
Using Table \ref{table:dvstars_ecc} and following Steps 3-4, Tables \ref{table:dvmin_de}, \ref{table:dvmin_da}, and \ref{table:dvmin_ecc_oop} present the remaining reachable delta-v minima in all dominance cases in eccentric orbits. The second column gives the conditions on the desired ROE pseudo-state for which the expression of the reachable delta-v minimum in the fourth column applies.

\fontsize{8.5}{11}\selectfont\begin{longtable}{l K{3cm} K{5cm} K{4.2cm}}
		\caption{Reachable $\delta v_{min}$, eccentric chief orbits, in-plane}\label{table:dvmin_de} \\ \toprule\toprule
		
		{\textbf{Dominant...}}& {\textbf{Region definition}} & {\textbf{Max normalized effect}} & {$\delta v_{min}$\textbf{, (m/s)}}\\ \midrule
        
        \multirow{3}{*}{\textit{... $\delta\pmb{e}$}} 
		& $|\Delta\delta\tilde{e}_{(.),des}| >0$ & $\begin{matrix*}[l]||a\Delta\delta\tilde{\pmb{e}}_{max}||^2 = (a\Delta\delta \tilde{e}_x^*)^2 + (a\Delta\delta \tilde{e}_y^*)^2 \\ 
        \text{where } a\Delta\delta \tilde{e}_x^* = a\Delta\delta \tilde{e}_x\left(\nu^*,\delta v_t^*[\nu^*]\right)\end{matrix*}$ & $ \delta v_{min,\delta\pmb{e}} = \frac{||a\Delta\delta \tilde{\pmb{e}}_{des}||}{||a\Delta\delta \tilde{\pmb{e}}_{max}||}$\\ \cline{2-4}
         & $\Delta\delta\tilde{e}_{x,des} = 0$ & 
         $\begin{matrix*}[l]
            a\Delta\delta\tilde{e}_{y,max} = \frac{1}{n}|\frac{ \eta
            \sin\nu_1(2+e_c\cos\nu_1)}{(1+e_c\cos\nu_1)}| \\
            \text{where } \nu_1 = \cos^{-1}((\eta - 1)/e_c)
            \end{matrix*}$ & $\delta v_{min,\delta\pmb{e}} = \frac{|a\Delta\delta\tilde{e}_{y,des}|}{a\Delta\delta\tilde{e}_{y,max}}$\\\cline{2-4}
         & $\Delta\delta\tilde{e}_{y,des} = 0$ & $a\Delta\delta\tilde{e}_{x,max} = \frac{2\eta}{ n}$ & $\delta v_{min,\delta\pmb{e}} =  \frac{|a\Delta\delta\tilde{e}_{x,des}|}{a\Delta\delta\tilde{e}_{x,max}}$\\
        
        \bottomrule
        	\end{longtable}    \normalsize   

\begin{figure}[H]
    \centering
    \begin{subfigure}[t]{0.49\textwidth}
    \centering 
        \includegraphics[height = 1.5 in]{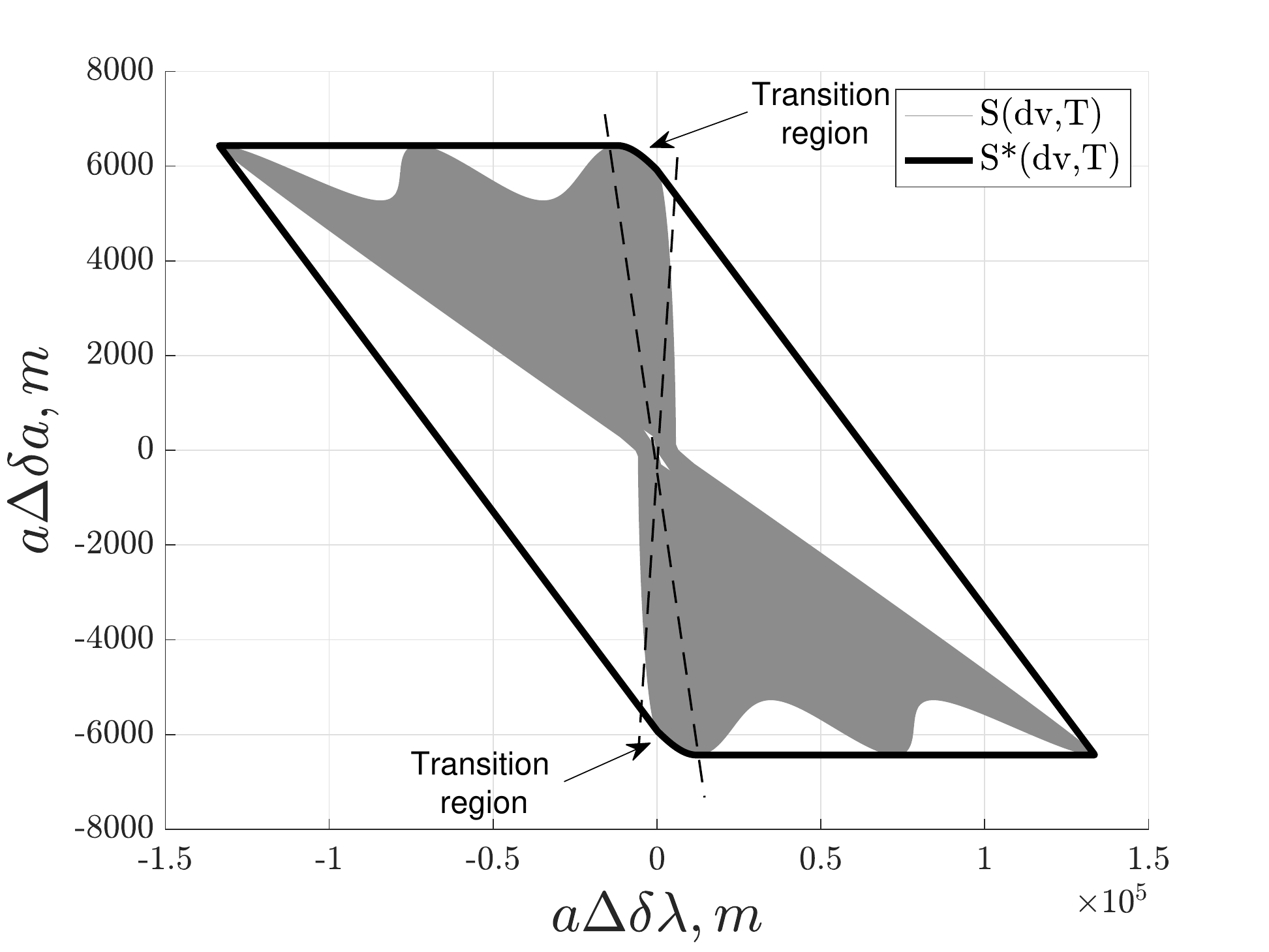}
        \caption{Transition region in the $\Delta\delta\pmb{a}$ plane for low ecc.}
        \label{fig:transition_region}
    \end{subfigure}\hfill~\begin{subfigure}[t]{0.49\textwidth}
    \centering 
        \includegraphics[height = 1.5 in]{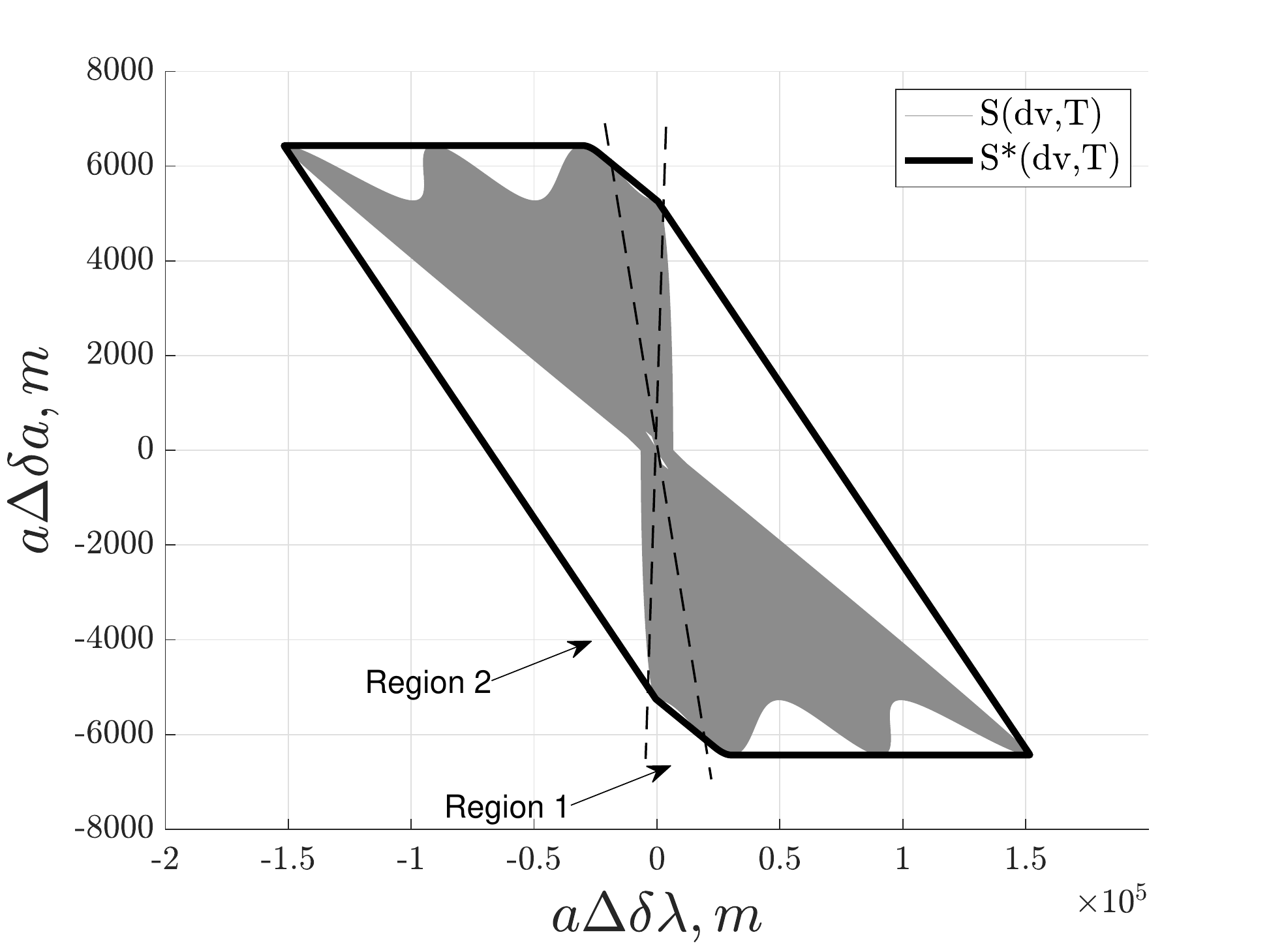}
        \caption{Extended region appears when Eq. \eqref{eqn:extended_regions} is true}
        \label{fig:extended_region}
    \end{subfigure}
    \caption{The different dominance subcases in the $\Delta\delta\pmb{a}$ plane}
    \label{fig:other_dl_regions}
\end{figure}

The reachable delta-v minimum in the other dominance cases in eccentric orbits are derived in the same way. As in Eq. \eqref{eqn:parametric_de}, the parametric equations in the $\Delta\delta a, \Delta\delta\lambda$ plane  in Eq. \eqref{eqn:param_eqns} are the $\delta a,\delta\lambda$ components of $\pmb{\Phi}(t_f,t_k)\pmb{\Gamma}(t_k)$, given by \fontsize{9}{11}\selectfont
\begin{equation}\label{eqn:parametric_dadl}
\begin{bmatrix}a\Delta\delta  {\lambda}(\nu,\delta v_t(\nu))\\ a\Delta\delta  {a}(\nu,\delta v_t(\nu)) \end{bmatrix} = \frac{1}{n}\begin{bmatrix*}[l] 
    \left(-\frac{2\eta^2}{1+e\cos(\nu)} - \frac{3}{\eta}\Delta M e\sin(\nu)\right)\sqrt{1-\delta v_t^2(\nu)} -\frac{3}{\eta}\Delta M \left(1+e\cos(\nu)\right)\delta v_t(\nu) \\ \frac{2}{\eta}e\sin(\nu)\sqrt{1-\delta v_t^2(\nu)} + \frac{2}{\eta}\left(1+e\cos(\nu)\right)\delta v_t(\nu). \end{bmatrix*}\end{equation}\normalsize
The $\Delta\delta\pmb{a}$ plane can be conveniently split into two dominance cases, $\delta a$ and $\delta \lambda$, based on the shape of the reachable set $S^*$. The reconfiguration is denoted dominant $\delta a$ if the desired pseudo-state lies on the horizontal edges of $S^*$ (see Fig. \ref{fig:Sdv_ecc_dadl}), and otherwise is denoted dominant $\delta\lambda$. The pseudo-state that denotes the end of the $\delta a$ region is given by \fontsize{10}{11}\selectfont\begin{equation}\label{eqn:Ddk2pi}
    a\Delta\delta  {a}_{k2\pi},a\Delta\delta {\lambda}_{k2\pi}  \text{, the pseudo-state achieved with }  \delta \pmb{v} = \begin{bmatrix}0 & 1 & 0\end{bmatrix}^{\text{T}}\text{ at }\nu = \text{floor}(\frac{\nu_f}{2\pi})2\pi.
\end{equation}
\normalsize For nonzero but low eccentricity, the transition from the dominant $\delta a$ region to the dominant $\delta \lambda$ region is not a corner, but a curved region known as the transition region (see Fig. \ref{fig:transition_region}). The start of the region is given by Eq. \eqref{eqn:Ddk2pi}, and the end of the region is given by \fontsize{9}{11}\selectfont\begin{equation}\label{eqn:Ddt}
a\Delta\delta {a}_t,a\Delta\delta  {\lambda}_t,  \text{ the pseudo-state achieved with }  \delta\pmb{v} = \frac{1}{\sqrt{e^2 + 2e\cos(\nu_t) +1}} \begin{bmatrix}e\sin(\nu_t) & 1+e\cos(\nu_t) &0\end{bmatrix}^{\text{T}}\text{ at }\nu = \nu_t \text{ (Eq. \eqref{eqn:nut})}.
\end{equation}
\normalsize The size of the transition region is a function of the chief eccentricity, and it exists for any reconfiguration times that are not integer multiples of the chief orbit period. 
For some reconfiguration times, the dominant $\delta\lambda$ region of the boundary of $S^*$ warps into two separate, linear regions, as shown in Fig. \ref{fig:extended_region}. The separate regions, called the extended regions, exist if\fontsize{10}{11}\selectfont
\begin{equation}\label{eqn:extended_regions}
\left|\frac{a\Delta\delta a_t - a\Delta\delta a_f}{a\Delta\delta\lambda_t - a\Delta\delta \lambda_f}\right| < \left|\frac{a\Delta\delta a_t +a\Delta\delta a_0}{a\Delta\delta \lambda_t +a\Delta\delta\lambda_0}\right|, \text{ where $a\Delta\delta {(.)}_0$ is defined in Eq. \eqref{eqn:Dd0} and }
\end{equation}
\normalsize
\fontsize{9}{11}\selectfont
\begin{equation}\label{eqn:Ddf}
a\Delta\delta  {a}_f, a\Delta\delta {\lambda}_f  \text{ is the pseudo-state achieved with }  \delta\pmb{v} = \frac{1}{\sqrt{e^2 + 2e\cos(\nu_f) +1}} \begin{bmatrix}e\sin(\nu_f) & 1+e\cos(\nu_f) &0\end{bmatrix}^{\text{T}}\text{ at }\nu = \nu_f.
\end{equation}
\normalsize
 The dominance cases in the $\Delta\delta\pmb{a}$ plane are given in Table \ref{table:dvmin_da} and in Appendix C, Table \ref{table:dvmin_dl}.     
\fontsize{8}{11}\selectfont
   \begin{longtable}{P{2.1cm} K{7cm} K{2.8cm} K{3.1cm}}
		\caption{Reachable $\delta v_{min}$, eccentric chief orbits, in-plane}\label{table:dvmin_da} \\ \toprule\toprule
		
		{\textbf{Dominant...}}& {\textbf{Region definition}} & \textbf{{Max normalized effect}} & {$\delta v_{min}$, \textbf{(m/s)}}\\ \midrule
        \multirow{2}{*}{\textit{... $\delta a$}} 
		& $\Delta\delta  {a}_{des}>0$ and $\Delta\delta {\lambda}_{des} < -\delta v_{min,\delta a}|\Delta\delta {\lambda}_{k2\pi}|$ 
 & $a\Delta\delta  {a}_{max} = \frac{2(e+1)}{\eta n}$ & $\delta v_{min,\delta a} = \frac{|a\Delta\delta  {a}_{des}|}{|a\Delta\delta  {a}_{max}|}$ \\ 
		
& or $\Delta\delta  {a}_{des}< 0$ and $\Delta\delta {\lambda}_{des} > \delta v_{min,\delta a}|\Delta\delta {\lambda}_{k2\pi}|$ & & \\ \midrule

        %%%%%%%%%%%%%
\multirow{2}{*}{\textit{... $\delta\lambda$, Transition region}} & $\Delta\delta  {a}_{des}>0$, $-|\Delta\delta {\lambda}_t| > \frac{\Delta\delta {\lambda}_{des}}{\delta v_{min,\delta a}} > -|\Delta\delta {\lambda}_{k2\pi}|$ & \multicolumn{2}{c}{See Table \ref{table:dvmin_dl}}\\

& or $\Delta\delta  {a}_{des}<0$, $|\Delta\delta {\lambda}_{k2\pi}| > \frac{\Delta\delta {\lambda}_{des}}{\delta v_{min,\delta a}} > |\Delta\delta {\lambda}_t|$ & & \\ \midrule
        %%%%%%%%% 
        
    \multirow{3}{*}{\textit{... $\delta \lambda$}} & $\Delta\delta  {a}_{des}>0$ and $\Delta\delta {\lambda}_{des} > -\delta v_{min,\delta a}|\Delta\delta {\lambda}_t|$ & \multicolumn{2}{c}{See Table \ref{table:dvmin_dl}}\\
        
    & or $\Delta\delta  {a}_{des}<0$ and $\Delta\delta {\lambda}_{des} < \delta v_{min,\delta a}|\Delta\delta {\lambda}_t|$  \\
        
        & or large $\Delta\delta {\lambda}_{des}$ & & \\  \midrule
    
    ... $\delta\lambda$, \textit{Extended region} & Exists if Eq. \eqref{eqn:extended_regions} is satisfied & \multicolumn{2}{c}{See Table \ref{table:dvmin_dl}} \\ \midrule 
	\end{longtable}\normalsize  
Table \ref{table:dvmin_ecc_oop} presents the reachable delta-v minimum for out-of-plane reconfigurations in eccentric orbits. Just as in the $\Delta\delta\pmb{e}$ plane, the reachable set $S^*$ is oriented at an angle of $\omega$ from the vertical for any reconfiguration where the chief has a nonzero (and defined) argument of perigee. The parametric equations in the $\Delta\delta \pmb{i}$ plane in Eq. \eqref{eqn:param_eqns} are rotated into the $\Delta\delta \tilde{\pmb{i}}$ plane using the same rotation matrix as in the $\Delta\delta\tilde{\pmb{e}}$ case, and are given by

\fontsize{9}{11}\selectfont
\begin{equation}\label{eqn:parametric_di}
\begin{bmatrix}a\Delta\delta \tilde{i}_x(\nu,\delta v_n(\nu))\\ a\Delta\delta \tilde{i}_y(\nu,\delta v_n(\nu)) \end{bmatrix} = \frac{1}{n}\frac{\eta}{1+e\cos(\nu)}\begin{bmatrix*}[l]\cos(\nu) \\ \sin(\nu)\end{bmatrix*}\delta v_n(\nu).\end{equation}\normalsize

As in the dominant $\delta\pmb{e}$ case and shown in Fig. \ref{fig:Sdv_ecc_di}, the reachable set $S(c,t)$ briefly disconnects from the boundary of the convex hull during each orbit. By solving for the $\nu$ values that maximize the parametric $\Delta\delta\tilde{i}_y$  expression in Eq. \eqref{eqn:parametric_di}, it is found that disconnection occurs at $\nu_{dis.} = \pi + \cos^{-1}(e)$ and reconnection occurs at $\nu_{re.} = \pi - \cos^{-1}(e)$, the same $\nu$ locations as in the $\Delta\delta\tilde{\pmb{e}}$ case. Reconfigurations whose desired pseudo-state lies in the disconnected region require a slightly different formulation of the reachable delta-v, given in the second row of Table \ref{table:dvmin_ecc_oop}. In Table \ref{table:dvmin_ecc_oop}, $\nu^*$ is equal to the phase of the rotated desired pseudo-state, $\tan^{-1}(\frac{\Delta\delta\tilde{i}_y}{\Delta\delta\tilde{i}_x})$.
\fontsize{8.5}{11}\selectfont\begin{longtable}{ l K{5cm} K{4cm} K{3cm}}
		\caption{Reachable $\delta v_{min}$, eccentric chief orbits, out-of-plane}\label{table:dvmin_ecc_oop} \\ \toprule\toprule
		
		{\textbf{Dominant...}}& {\textbf{Region definition}} & {\textbf{Max normalized effect}} & {$\delta v_{min}$, \textbf{(m/s)}}\\ \midrule
        
        \multirow{3}{*}{\textit{... $\delta\pmb{i}$}}
		& $\nu^*$ $\in (\nu_{re},\nu_{dis})$ & $||a\Delta\delta\tilde{\pmb{i}}_{max}|| = \frac{\eta}{1+e\cos(\nu^*)}\frac{1}{n}$ & $\frac{||a\Delta\delta\tilde{\pmb{i}}_{des}||}{||a\Delta\delta\tilde{\pmb{i}}_{max}||}$
		\\  
		& $\nu^*+\pi$ $\in (\nu_{re},\nu_{dis})$ & $||a\Delta\delta\tilde{\pmb{i}}_{max}|| = \frac{\eta}{1-e\cos(\nu^*)}\frac{1}{n}$ & $\frac{||a\Delta\delta\tilde{\pmb{i}}_{des}||}{||a\Delta\delta\tilde{\pmb{i}}_{max}||}$
		\\  
        & Both $\nu^*$ and $\nu^*+\pi$ $\not\in (\nu_{re},\nu_{dis})$ & $|a\Delta\delta\tilde{i}_{y,max}| = \frac{\sin(\nu_{re.})\eta }{ 1 + e\cos(\nu_{re.})}\frac{1}{n}$ & $\frac{|a\Delta\delta\tilde{i}_{y,des}|}{|a\Delta\delta\tilde{i}_{y,max}|}$\\
        \bottomrule
        	\end{longtable}    \normalsize  

\section{Closed-form Maneuver Schemes}\label{sec:cf_maneuver_schemes}
This section first presents the general methodology to calculate an optimal maneuver scheme, then gives the explicit closed-form solutions for multiple relevant cases. 

\subsection{General Methodology}\label{sec:gen_meth_cf_solutions}
Given an initial chief orbit $\pmb{\alpha}_{c,0}$, a desired pseudo-state $a\Delta\delta {\pmb{\alpha}}_{des}$, and a reconfiguration time $T$, the general methodology to calculate closed-form, optimal maneuver schemes is given step-by-step as follows.

1. \smalltab  Determine the dominance case and reachable delta-v minimum by computing all independent $\delta v_{min}$ using Tables \ref{table:dvstars_ecc}-\ref{table:dvmin_ecc_oop} and substituting them into Eq. \eqref{eqn:dvminproof}.

2. \smalltab Find the set of optimal maneuver locations $\nu_{opt}$ or $u_{opt}$ (or, equivalently, $T_{opt}$). The optimization problem in Eq. \eqref{eqn:optcontrolproblem} is computationally intractable for many linear time-variant systems. As Koenig \cite{bib:KDAutomatica} details, this problem is remedied by solving the dual of Eq. \eqref{eqn:optcontrolproblem}, which is to maximize $\delta v$ subject to: $\max_{\pmb{z} \in S^*(\delta v,T)} \pmb{\lambda}^{\text{T}}\pmb{z} \leq \pmb{\lambda}^{\text{T}}\Delta\delta {\pmb{\alpha}}_{des}$
where $\pmb{\lambda}$ is the normal vector of a supporting hyperplane of $S^*(\delta v,T)$, called $L(\Delta\delta {\pmb{\alpha}}_{des},\pmb{\lambda})$ (or $L$), which contains the desired pseudo-state $\Delta\delta {\pmb{\alpha}}_{des}$. It is not necessary to solve the optimization problem itself in this paper because the ROE state representation yields geometrically simple reachable sets. However, Koenig's analysis yields properties that help provide some geometric intuition in solving Eq. \eqref{eqn:optcontrolproblem}. Once the reachable delta-v minimum is known for a given reconfiguration, the projection of $L$ in the dominant 2D plane is the line that passes through the desired pseudo-state and is tangent to $S^*(\delta v_{min},T)$. Therefore, every point in the intersection between $L$ and $S^*(\delta v_{min},T)$ is in the boundary of $S^*(\delta v_{min},T)$, and the desired pseudo-state must be in $L\cap S^*(\delta v_{min},T)$ as well. By definition of the convex hull in Eq. \eqref{eqn:Sstar}, it is possible to find a convex combination of points in $L\cap S^*(\delta v_{min},T)$ that achieves the desired reconfiguration with cost equal to $\delta v_{min}$. Therefore, a set of optimal maneuver times $T_{opt}$ must be found so that the optimality condition in the dominant plane is satisfied. In other words, $T_{opt}$ is the set of all times in $T$ at which $L$ can be reached with a single maneuver and is found by determining where $S(c,t)$ (Eq. \eqref{eqn:Sdvt}) intersects the boundary of $S^*(c,T)$.

\begin{figure}[H]
    \centering
    \begin{subfigure}[t]{0.24\textwidth}
        \includegraphics[width=\textwidth]{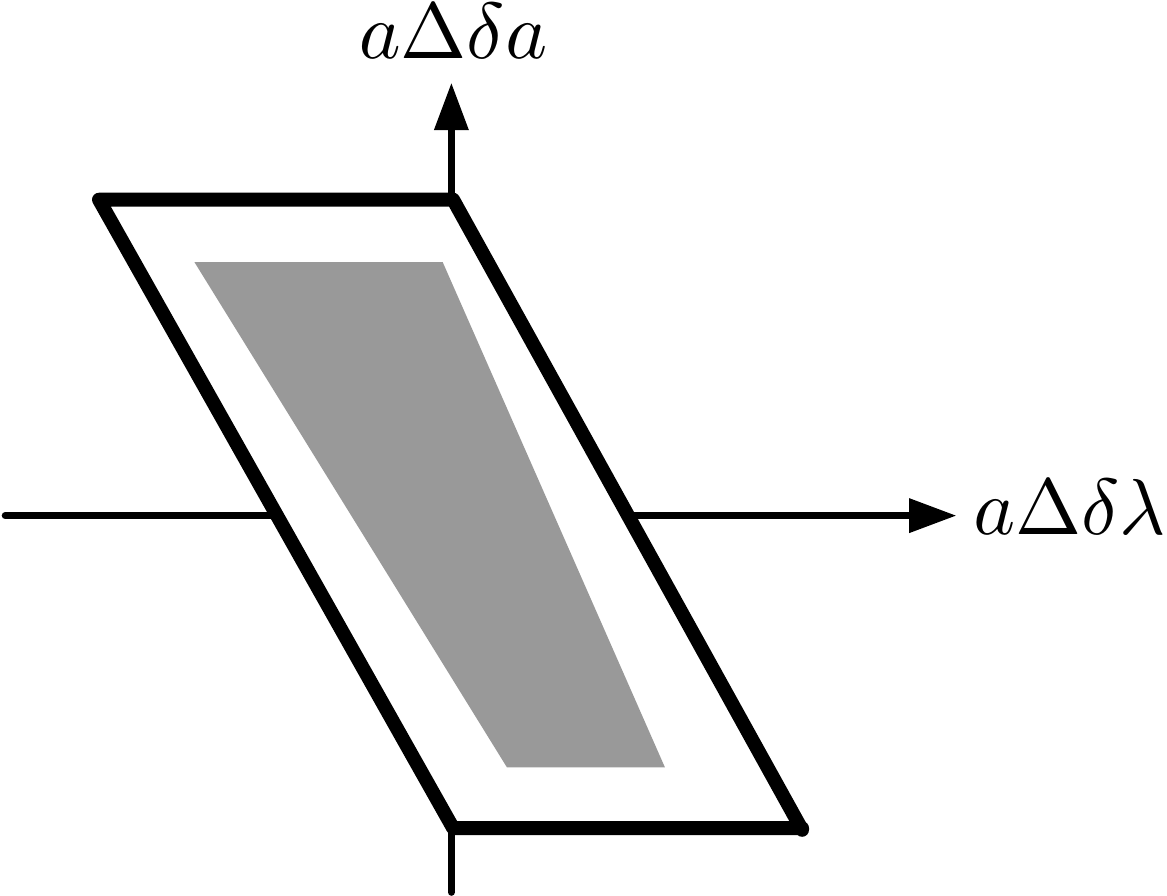}
        \caption{Dominant $\delta\pmb{e}$}
        \label{fig:Sn_Ecc_de}
    \end{subfigure}\hfill
        \begin{subfigure}[t]{0.24\textwidth}
        \includegraphics[width=\textwidth]{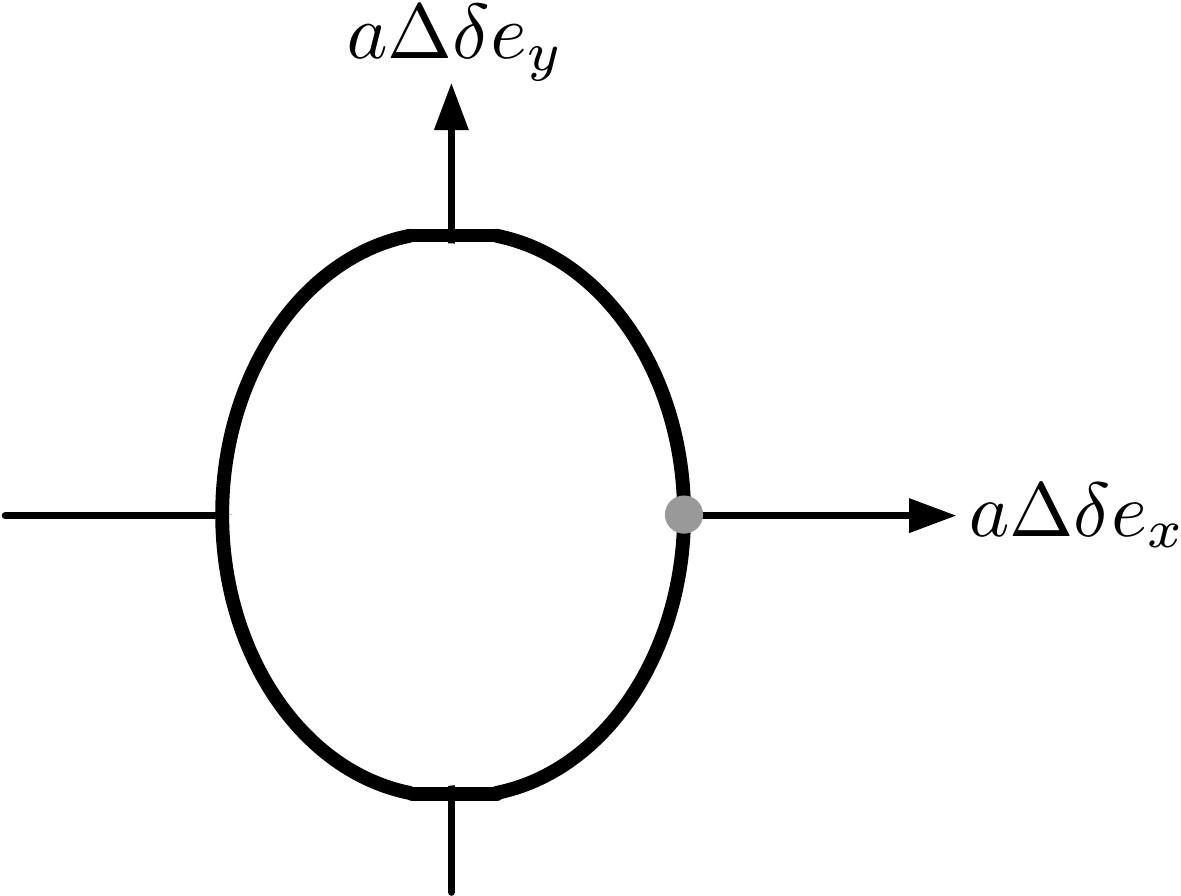}
        \caption{Dominant $\delta a$}
        \label{fig:Sn_Ecc_da}
    \end{subfigure}\hfill
        \begin{subfigure}[t]{0.24\textwidth}
        \includegraphics[width=\textwidth]{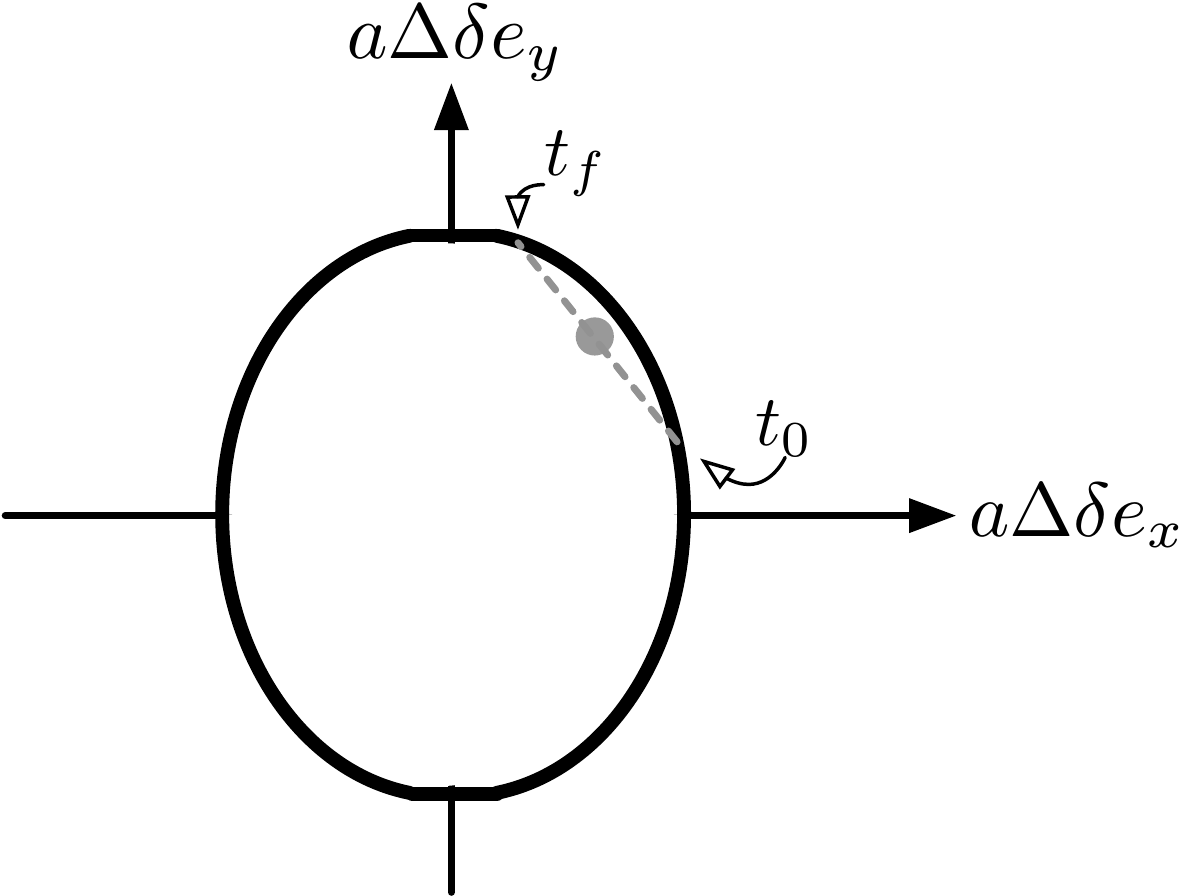}
        \caption{Dominant $\delta \lambda$}
        \label{fig:Sn_Ecc_dl}
    \end{subfigure}
    \caption{In eccentric orbits, the desired pseudo-state must lie in the nested reachable set $S_n^*$ (shaded or gray point) for the full reconfiguration to be achievable with delta-v driven by the dominant plane.}
    \label{fig:Sn_Ecc}
    \end{figure}

3. \smalltab Generate nested reachable sets using $T_{opt}$ from the previous step. The nested reachable set $S_n^*(\delta v_{min},T_{opt})$ is calculated using Eq. \eqref{eqn:Sstar} for only the times in $T_{opt}$, with $\delta \pmb{v}$ equal to $\delta \pmb{v}^*$ (see Table \ref{table:dvstars_ecc}) scaled by $\delta v_{min}$. 
Because $T_{opt}$ is a subset of $T$, $S_n^*(\delta v_{min},T_{opt})$ is a subset of $S^*(\delta v_{min},T)$. As noted in the previous section, the desired pseudo-state projected in the non-dominant plane must lie in the nested reachable set in order for the reachable delta-v minimum to be equal to the values in Tables \ref{table:dvmin_de}-\ref{table:dvmin_ecc_oop}. Figure \ref{fig:Sn_Ecc} illustrates $S_n$ as a shaded region or point. 
$\Delta\delta a$ is dependent on the maneuver location in eccentric orbits, given by multiples of $2\pi$, which means that the set of times in $T_{opt}$ is finite. Therefore, the only pseudo-states that are reachable occur at maneuver locations that are multiples of $2\pi$, so $S_n^*$ is a single point at the boundary of $S^*$ on the $\Delta\delta e_x$-axis (see Fig. \ref{fig:Sn_Ecc_da}). In addition, as shown in Fig. \ref{fig:Sn_Ecc_dl}, for dominant $\delta\lambda$, $S_n^*$ is a single  pseudo-state that lies on the segment (dashed line) connecting the pseudo-states at the two times in $T_{opt}$, $t_0$ at the start of the reconfiguration, and $t_f$ at the end.

4. \smalltab Satisfy constraints. Each dominance case has a set of constraints that must be satisfied in order for 1) a closed-form solution to exist and 2) the closed-form solution to be energy-optimal. A Type 1 constraint requires the desired pseudo-state to be inside or on the boundary of the nested reachable set, and a Type 2 constraint ensures that the total cost of the maneuver scheme is equal to $\delta v_{min}$, i.e., $\sum_{k=1}^p||\delta \pmb{v}_k||_2 = \delta v_{min}$. The third column of Table \ref{table:CFconstraints_ecc} summarizes the Type 1 constraints for all cases. The set of constraints translates to a linear system to be solved for the maneuver magnitudes, $c_1, c_2,...,c_p$. Note that the number of required maneuvers in Column 3 of Table \ref{table:CFconstraints_ecc} is not arbitrary. The size of the state and the location of the desired pseudo-state drive the number of required maneuvers. To control a single 2D state, such as the out-of-plane ROE, one maneuver is required if the desired pseudo-state lies in both S and S*, and two maneuvers are required if the desired pseudo-states lies just in S*. To control a 4D state, such as the in-plane ROE, more maneuvers are required to satisfy the constraints of achieving the desired pseudo-state in each of the four dimensions.

5. \smalltab Solve linear system. At this point, only the maneuver magnitudes are unknown. Sets of $p$ maneuver times  (given in the second column of Table \ref{table:CFconstraints_ecc}) and corresponding pseudo-states in $S_n$ ($\Delta\delta  {a}_k$, $\Delta\delta {\lambda}_k$, $\Delta\delta  {e}_{x,k}$, $\Delta\delta  {e}_{y,k}$ for $k=1,..,p$) must be chosen so that the Type 1 constraints in column three of Table \ref{table:CFconstraints_ecc} are satisfied. The Type 2 constraint always translates to 
$\sum_{k=1}^p c_k = 1$ in the linear system. 

The resulting linear system is of the form $\pmb{A}\pmb{c} = \pmb{b}$ with unknown coefficients $\pmb{c}$. Since the determinant is always non-null based on the choice of maneuvers in $S_n$, the linear system can be solved by $\pmb{c} = \pmb{A}^{-1}\pmb{b}$.
%%
%% end methodology
%%

\subsection{Closed-form Solutions in Eccentric Orbits}
This section gives explicit closed-form maneuver schemes in eccentric orbits, which result from application of the general methodology described in the previous section. As the eccentricity approaches zero, the $\Delta\delta \pmb{e}'$ state is not necessary because control of the quasi-nonsingular ROE in Eq. \eqref{eqn:ROE} is inherently decoupled. The algorithms reduce to the near-circular algorithms given in Ref. \cite{bib:RST_conf}.
Table \ref{table:dvstars_CF_ecc} gives the optimal maneuver $\delta \pmb{v}^*$ (column 2) and set of optimal maneuver locations $\nu_{opt}$ (column 3) for relevant reconfiguration cases in eccentric orbits. The difference between column 2 in Table \ref{table:dvstars_CF_ecc} and column 2 of Table \ref{table:dvstars_ecc} is that Table \ref{table:dvstars_ecc} was only concerned with finding the scaling factor ($\delta v_{min}$) on the reachable set so that the desired pseudo-state was on the boundary of the convex hull. In contrast, Table \ref{table:dvstars_CF_ecc} lists the specific maneuver locations required to achieve the desired reconfiguration with delta-v equal to $\delta v_{min}$.

In the $\Delta\delta\pmb{e}$ plane, $L$ is only reachable at the time at which $S(c,t)$ aligns with the phase of the desired pseudo-state because $L$ is tangent to the reachable set only at the desired pseudo-state. Because $S(c,t)$ is periodic in the $\Delta\delta\pmb{e}$ plane, it will align with the desired direction every half orbit, at so-called $\nu_{opt,1}$ and $\nu_{opt,2}$.  The first value, $\nu_{opt,1}$, is equal to $\nu^*$, which was found by substituting Eqs. \eqref{eqn:parametric_de}-\eqref{eqn:changeDdes} into Eq. \eqref{eqn:ratio} and solving in the range of $\nu$ values given by columns three and four of Table \ref{table:nuboundaries_nuopt}. The second value, $\nu_{opt,2}$, is found using the same solving methods as the previous section, but using an initial guess in the range given in last two columns of Table \ref{table:nuboundaries_nuopt}.
As in Sec. \ref{sec:reachabledvmin}, $\nu_{opt,2}$ can also be found using the bisection method, which is guaranteed to converge as long as substituting the left and right boundaries from columns five and six of Table \ref{table:nuboundaries_nuopt} into Eq. \eqref{eqn:function_in_NR} yields values of opposite sign. If that criteria fails, the ``relaxed'' boundaries in columns four and five of Table \ref{table:nuboundaries_guaranteed_nuopt} can be used.

The explicit values in $\nu_{opt}$ are the union of integer multiples of $2\pi$ away from $\nu_{opt,1}$ and $\nu_{opt,2}$. This is summarized in the last column of row 1 in Table \ref{table:dvstars_CF_ecc}. 
\fontsize{8.5}{11}\selectfont
\begin{longtable}{ l K{5.5cm} P{5.1cm}}
		\caption{Optimal maneuver vectors $\delta \pmb{v}^*$ and set of optimal maneuver times $T_{opt}$ for eccentric orbits}\label{table:dvstars_CF_ecc} \\ \toprule\toprule
		
		\textbf{{Dominant ...}} & {$\delta \pmb{v}^*$\textbf{, (m/s)}} & {$\nu_{opt}$\textbf{, (rad)}}\\ \midrule
        \textit{... $\delta \pmb{e}$}\\ 
         \textit{$|a\Delta\delta\tilde{e}_{(.),des}| > 0$} & See Eq. \eqref{eqn:dvtstar_example} & $\begin{matrix*}[l]\cup\{\nu_{opt,1} + k2\pi\text{, } k = \text{floor}(\frac{\nu_f - \nu_{opt,1}}{2\pi}), \\
        \nu_{opt,2} + k2\pi\text{, } k = \text{floor}(\frac{\nu_f - \nu_{opt,2}}{2\pi})\}\end{matrix*}$\\ 
        \hdashline
        
         \textit{$a\Delta\delta\tilde{e}_{x,des} = 0$, $a\Delta\delta\tilde{e}_{y,des} \gtrless 0$} & $\begin{matrix*}[l]\begin{bmatrix}0 & \pm\delta v_{min,\delta\pmb{e}} & 0\end{bmatrix}^{\text{T}} \\ 
        \begin{bmatrix}0 & \mp\delta v_{min,\delta\pmb{e}} & 0\end{bmatrix}^{\text{T}}
        \end{matrix*}$ & $\begin{matrix*}[l]\cup\{\nu^* + k2\pi\text{, } k = \text{floor}(\frac{\nu_f - \nu^*}{2\pi}), \\
        -\nu^* + (k+1)2\pi\text{, } k = \text{floor}(\frac{\nu_f- \nu^*-2\pi}{2\pi})\}\end{matrix*}$\\ 
        \hdashline
        
         \textit{$a\Delta\delta\tilde{e}_{x,des} \gtrless 0$, $a\Delta\delta\tilde{e}_{y,des} = 0$} & $\begin{matrix*}[l]\begin{bmatrix}0 & \pm\delta v_{min,\delta\pmb{e}} & 0\end{bmatrix}^{\text{T}} \\ 
        \begin{bmatrix}0 & \mp\delta v_{min,\delta\pmb{e}} & 0\end{bmatrix}^{\text{T}}
        \end{matrix*}$ & $\begin{matrix*}[l]\cup\{0 + k2\pi\text{, } k = \text{floor}(\frac{\nu_f}{2\pi}), \\
        \pi + k2\pi\text{, } k = \text{floor}(\frac{\nu_f- \pi}{2\pi})\}\end{matrix*}$\\  
        \midrule
       \textit{... $\delta \pmb{i}$}\\ 
      $\nu^*$ $\in (\nu_{re},\nu_{dis})$ & $\begin{bmatrix} 0 & 0 & +\delta v_{min,\delta\pmb{i}}\end{bmatrix}^{\text{T}}$ & $\nu^* + k2\pi$, $k = \text{floor}(\frac{\nu_f - \nu^*}{2\pi})$ \\ \hdashline
      $\nu^*+\pi$ $\in (\nu_{re},\nu_{dis})$ & $\begin{bmatrix} 0 & 0 & -\delta v_{min,\delta\pmb{i}}\end{bmatrix}^{\text{T}}$
 & $(\nu^*+\pi) + k2\pi$, $k = \text{floor}(\frac{\nu_f - (\nu^*+\pi)}{2\pi})$\\  \hdashline
       Else & $\begin{matrix*}[l]\begin{bmatrix} 0 & 0 & +\delta v_{min,\delta\pmb{i}}\end{bmatrix}^{\text{T}}\\ \begin{bmatrix} 0 & 0 & -\delta v_{min,\delta\pmb{i}}\end{bmatrix}^{\text{T}}\end{matrix*}$
      & $\begin{matrix*}[l]\cup\{\nu_{re} + k2\pi\\\nu_{dis}+k2\pi\}\end{matrix*}$, $k = \text{floor}(\frac{\nu_f - \nu_{dis}}{2\pi})$\\
       \bottomrule 
        	\end{longtable}\normalsize
Note that to achieve a marginal improvement in delta-v, integer multiples of linear combinations of $\nu_{opt,1}$ \textit{and} $\nu_{opt,2}$ can be used for reconfigurations in the disconnected region. 

After computing the optimal times and maneuvers using Table \ref{table:dvstars_CF_ecc}, the nested reachable set is generated according to Step 3 of the general methodology described in Sec. \ref{sec:gen_meth_cf_solutions}. Then, according to Step 4, a subset of maneuvers is chosen from $S_n$ so that it satisfies a set of constraints.
Table \ref{table:CFconstraints_ecc} gives the specific constraints that must be satisfied (column 3) and the corresponding linear system (column 4) to solve in Steps 4-5 of the general methodology in Sec. \ref{sec:gen_meth_cf_solutions}. 
In Table \ref{table:CFconstraints_ecc}, the subscript max (min) refers to the largest (smallest) value of that ROE in the nested reachable set for the chosen set of 3 maneuvers. 
\fontsize{8.5}{11}\selectfont
\begin{longtable}{l K{1.8cm} K{4.55cm} K{3.7cm}}
		\caption{Closed-form maneuver scheme constraints in eccentric orbits}\label{table:CFconstraints_ecc} \\ \toprule\toprule
		
		\textbf{{Dominant...}}& \textbf{{\# of Maneuvers} }& \textbf{{Constraints (Type 1)}} &\textbf{ {Linear System}}\\ \midrule
        \textit{... $\delta\pmb{e}$} & 3 & $\begin{matrix}a\Delta\delta  {a}_{max} \geq a\Delta\delta  {a}_{des} \geq a\Delta\delta a_{min}\\
a\Delta\delta \lambda_{max} \geq a\Delta\delta  {\lambda}_{des} \geq a\Delta\delta \lambda_{min} \end{matrix}$ & $\begin{matrix*}[l]\sum_{i=1}^3 c_i = 1 \\
\sum_{i=1}^3 c_ia\Delta\delta  {a}_i = a\Delta\delta  {a}_{des} \\ 
\sum_{i=1}^3 c_ia\Delta\delta \lambda_i = a\Delta\delta  {\lambda}_{des}\end{matrix*}$ \\ \midrule
\textit{... $\delta \pmb{i}$}\\
      \smalltab$\nu^*$ $\in (\nu_{re},\nu_{dis})$ & 1 & N/A & $c_1 = 1$\\ \hdashline
      \smalltab$\nu^*+\pi$ $\in (\nu_{re},\nu_{dis})$ & 1& N/A & $c_1 = 1$\\  \hdashline
      \smalltab Else & 2 & N/A & $\begin{matrix*}[l]\sum_{i=1}^2 c_i a\Delta\delta \tilde{i}_{x} = a\Delta\delta\tilde{i}_{x,des} \\
      \sum_{i=1}^2 c_i a\Delta\delta \tilde{i}_{y} = a\Delta\delta\tilde{i}_{y,des}\end{matrix*}$\\
\bottomrule\end{longtable}\normalsize

\section{Sub-optimal Solution Handling}\label{sec:subopt_solns}
% As summarized in Sec. \ref{sec:cf_soln_summary}, 
An optimal solution cannot always be found in closed-form. Fig. \ref{fig:Sn_Ecc} shows that the nested reachable set is so restricted in dominant $\delta a$ and $\delta\lambda$ reconfigurations that only a dominant plane reconfiguration can be achieved in closed-form. The same can occur (but less frequently) in dominant $\delta\pmb{e}$ reconfigurations where the desired pseudo-state lies outside of the nested reachable set in the non-dominant plane. This section will describe the method by which a closed-form sub-optimal solution can be found for an unreachable pseudo-state.
To generate a sub-optimal solution for a given reconfiguration, the methodology in Sec. \ref{sec:cf_maneuver_schemes} is modified to relax the requirement in Step 4 that the desired pseudo-state must lie inside or on the boundary of the nested reachable set in the non-dominant plane. Then the reconfiguration problem is solved as if the reconfiguration were dominant in the relative eccentricity vector. The increase in delta-v required to achieve the desired pseudo-state can be quantified by comparison to the reachable delta-v minimum that was calculated in Step 1 of the closed-form solution methodology. The ``best'' closed-form maneuver scheme is the sub-optimal scheme with delta-v closest to the reachable delta-v minimum. The first example in the Validation section, Sec. \ref{sec:validation}, will demonstrate application of this method to a case where the desired pseudo-state lies outside of the nested reachable set.

This methodology applies for many types of reconfigurations where a closed-form optimal solution cannot be found. However, as with the closed-form solutions in Sec. \ref{sec:cf_maneuver_schemes}, the reconfiguration span must be long enough to satisfy the other internal constraints of the system. In addition, for some missions, the delta-v budget may be so strict that the sub-optimal solution delta-v is too high to be usable. To find closed-form solutions for these restrictive cases, other sub-optimal solution methods can be applied, such as extending the reconfiguration time or breaking the reconfiguration up into multiple reconfigurations with boundary conditions. 

\section{Error Analysis}
In an ideal world, the algorithms above produce maneuver schemes that will achieve a desired reconfiguration exactly. In reality however, uncertainty in the dynamics model, state knowledge, and maneuver execution can propagate into significant errors in the ROE achieved at the end of a reconfiguration. Therefore, it is pertinent to assess the effect of common uncertainties  on achieving a desired reconfiguration. 

Recall, in the linearized framework, the ROE achieved in the presence of $p$ maneuvers with magnitudes $\delta \pmb{v}_k$ and times $t_k$ for $k=1,...,p$ are given in Eq. \eqref{eqn:prob_statement}. This section analyzes the effect on the achieved ROE $\delta\pmb{\alpha}_f$ of errors from four sources: the magnitude of the maneuver vector $\delta \pmb{v}_k$, maneuver execution time $t_k$, initial absolute state $\pmb{\alpha}_{c,0}$, and initial relative state $\delta\pmb{\alpha}_0$. The error in each of the four cases is assumed to be normally distributed with mean $\pmb{\mu}_{var}$ and covariance $\pmb{V}_{var}$, where the subscript $var$ will be replaced with the name of the error source. Error will be represented as $\epsilon$ for scalars and as $\pmb{\epsilon}$ for vectors. First, each error source's effect on the reachable sets $S(c,T)$ and $S^*(c,T)$ is analyzed graphically to gain intuition into the abstract 6D space. Then, the effect of the error on the ROE reached at the end of the reconfiguration is quantified by including the error mean and covariance in the equations in place of the nominal values.

\subsection{Effect of errors on the reachable sets}
\subsubsection{Maneuver execution errors}
\begin{figure}[H]
    \centering
    \begin{subfigure}[t]{0.49\textwidth}
    \centering 
        \includegraphics[height = 1.2in]{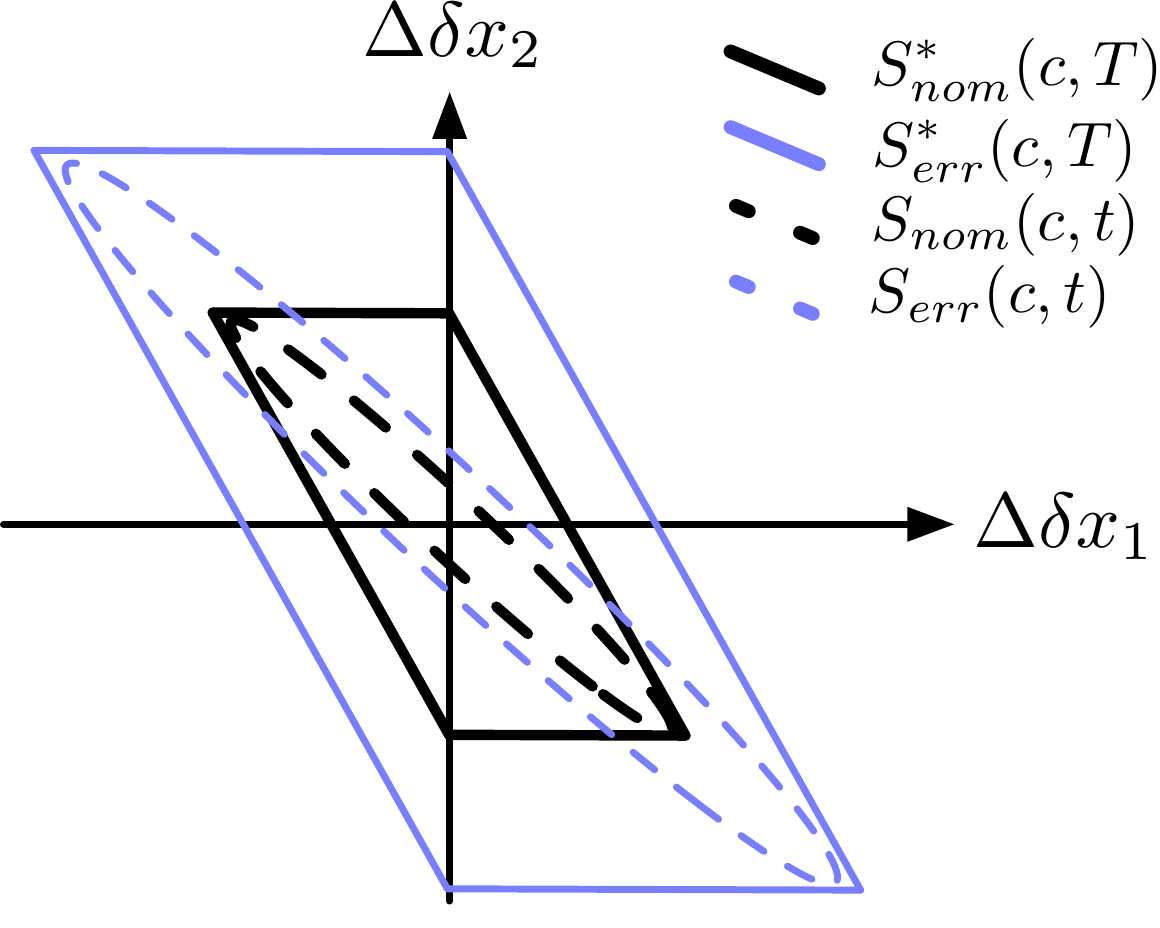}
        \caption{An error in the magnitude of the maneuver vector expands or contracts the nominal reachable sets $S_{nom}$ and $S^*_{nom}$ (black) linearly. The scaled reachable sets $S_{err}$ and $S^*_{err}$ are shown in blue.}
        \label{fig:maneuver_mag_err}
    \end{subfigure}\hfill~\begin{subfigure}[t]{0.49\textwidth}
    \centering 
        \includegraphics[height = 1.2in]{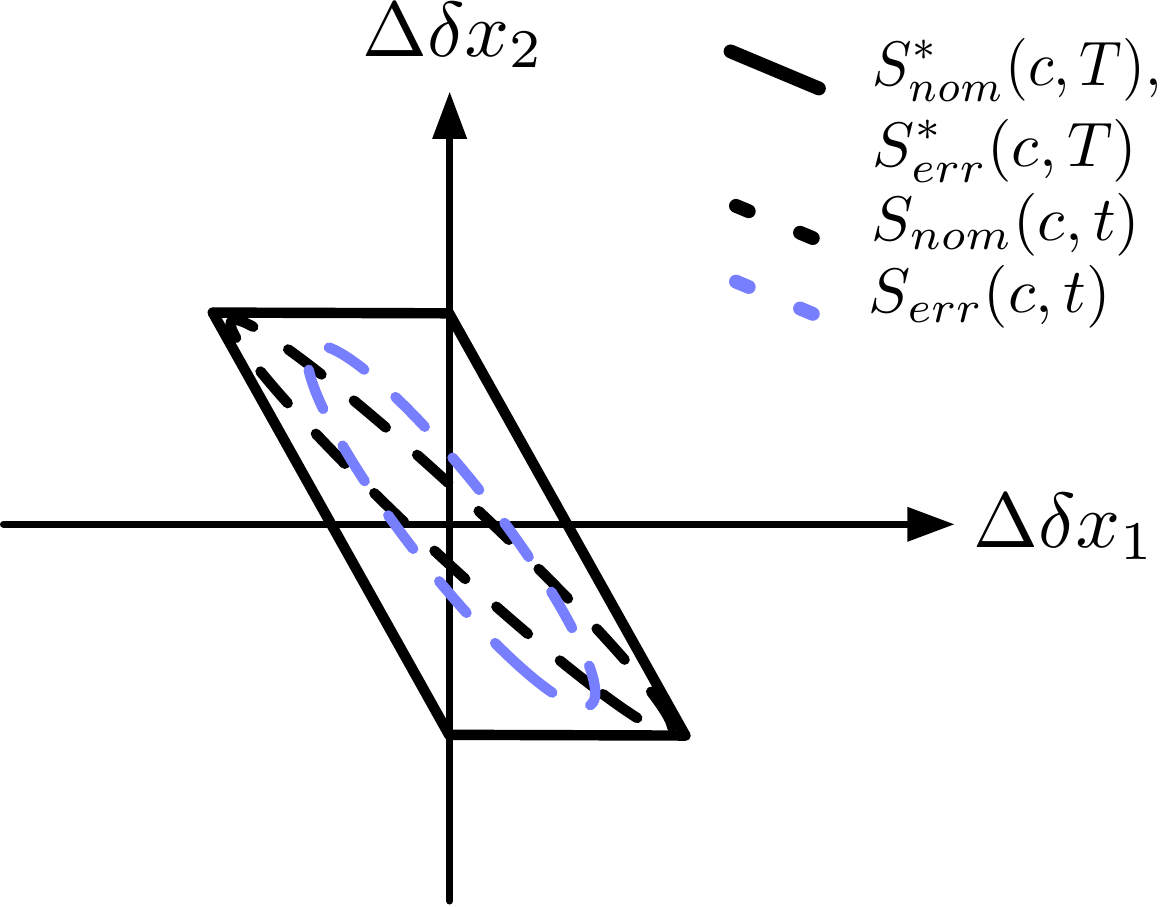}
        \caption{An error in the maneuver execution time warps $S_{nom}(c,t)$ from a nominal configuration (black dashed line) to an error configuration $S_{err}(c,t)$ (blue dashed line), but has no effect on the convex hull of the reachable set $S_{nom}^*(c,T)$.}
        \label{fig:maneuver_time_err}
    \end{subfigure}
    \caption{Effect of errors in maneuver magnitude and execution time}
    \label{fig:mag_error_time_err}
\end{figure}

Maneuver execution errors result from multiple factors including time delays, thruster misalignment, and thruster imperfections. An error in the magnitude of the maneuver vector translates to an expansion or contraction of the reachable sets $S(c,t)$, $S(c,T)$, and $S^*(c,T)$ by the same amount. This is shown in Fig. \ref{fig:maneuver_mag_err} for an arbitrary 2D case and can be easily deduced by substituting $(1+\epsilon)\delta\pmb{u}$ in place of $\delta\pmb{u}$ in the definitions of the reachable sets given in Eqs. \eqref{eqn:Sdvt}, \eqref{eqn:Sdv}, and \eqref{eqn:Sstar}, where the pseudo-state is linear in delta-v. In all figures in this section, the solid lines indicate $S^*(c,T)$, the convex hull of the reachable set, and the dashed lines indicate $S(c,t)$, the reachable set at an arbitrary time in the reconfiguration span. The subscript $nom$ denotes the shapes of the reachable sets if there were no errors, while the subscript $err$ denotes the shapes of the reachable sets in the presence of errors.
The effect of an error in the maneuver time is more complicated than the effect of an error in the vector itself. Substituting $t_j+\epsilon$ into the equations of the reachable sets in place of $t_j$ shows that the effect is nonlinear. Luckily, the effect is simple to visualize. If the maneuver time is executed earlier or later than expected, $S(c,t_j)$ becomes $S(c,t_j + \epsilon)$, which, as shown in Fig. \ref{fig:maneuver_time_err}, represents a rotation and possible deformation. However, the reachable sets $S(c,T)$ and $S^*(c,T)$ do not change at all, as long as the new maneuver time is within the reconfiguration span because by definition if $t_j+\epsilon \in \Delta t_f$, then the $S(c,t_j+\epsilon)$ was already used to construct $S(c,T)$ and $S^*(c,T)$. In this arbitrary example, the maneuver time error makes the maneuver time less effective, because the maneuver is no longer occurring at the optimal time. As will be shown in the next subsection, luckily the time error has to be very large to cause a significant error in the achieved ROE.

\subsubsection{Navigation errors}
\begin{figure}[H]
    \centering
    \begin{subfigure}[t]{0.49\textwidth}
    \centering 
        \includegraphics[height = 1.2in]{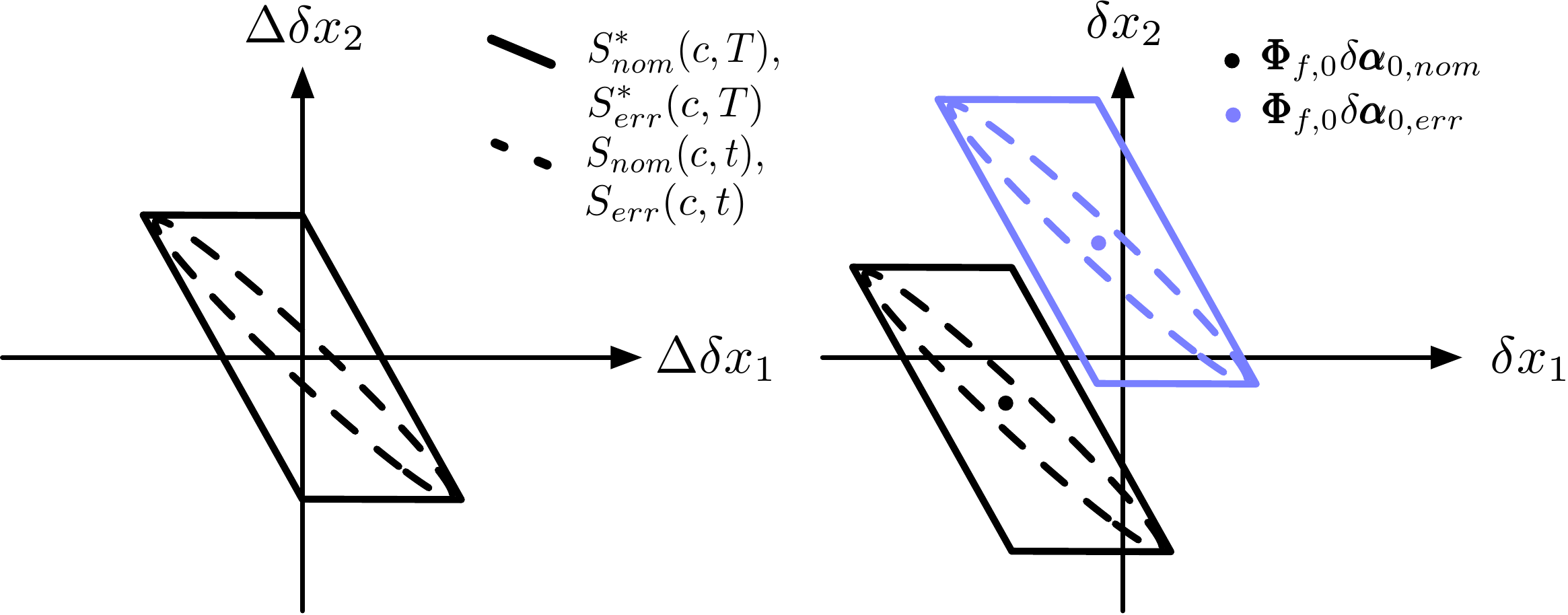}
        \caption{An error in the initial ROE causes no change in the pseudo-state space (left) but shifts the nominal reachable sets (black to blue) around linearly in the state space (right).}
        \label{fig:init_roe_err}
    \end{subfigure}\hfill~\begin{subfigure}[t]{0.49\textwidth}
    \centering 
        \includegraphics[height = 1.2in]{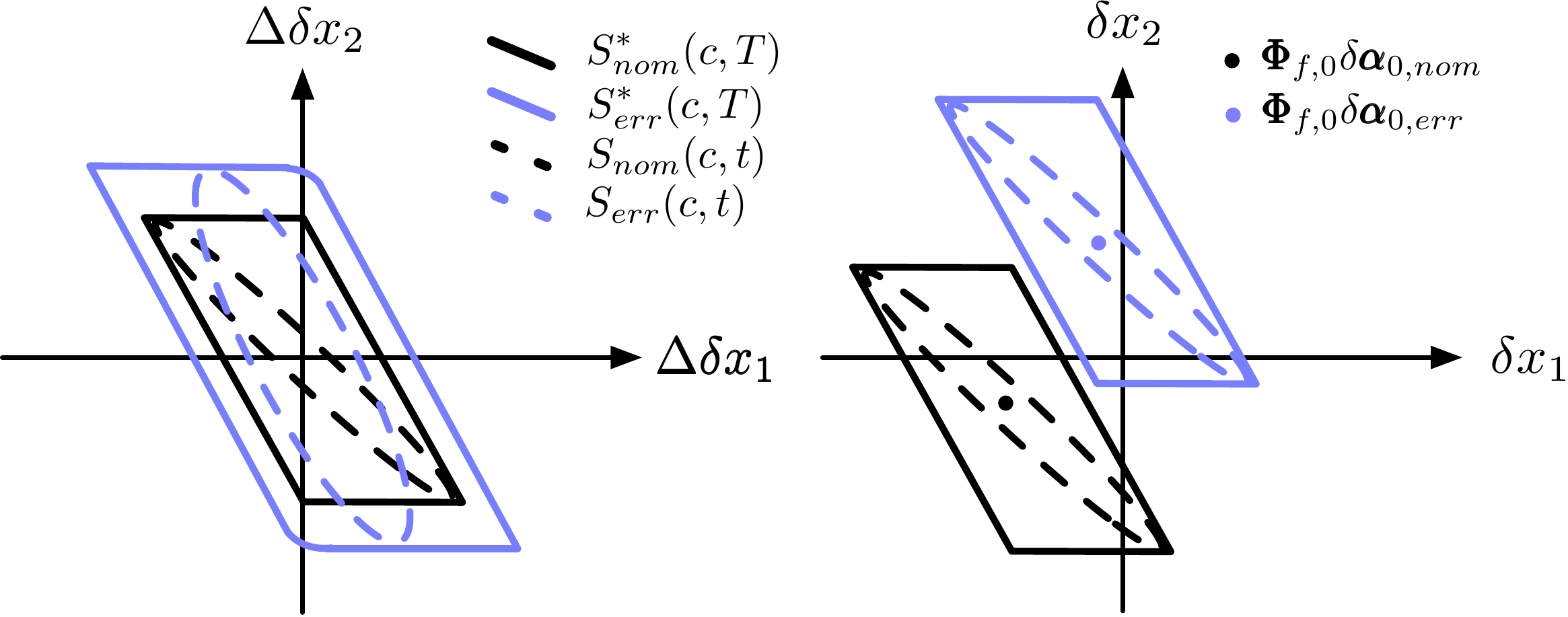}
        \caption{An error in the initial chief orbit elements causes a distortion of the shape of the nominal reachable sets $S_{nom}$ and $S^*_{nom}$ (left) and can shift the reachable sets around nonlinearly in the state space (right).}
        \label{fig:init_oe_err}
    \end{subfigure}
    \caption{Effect of relative and absolute navigation errors}
    \label{fig:nav_err}
\end{figure}

Navigation errors represent uncertainty in the knowledge of the state due to sensor errors or estimation algorithms. This section will look at the effect on the final achieved ROE due to errors in the initial relative state of the formation and in the initial absolute orbit elements state of the chief satellite.
An error in estimating the initial relative state does not change the reachable sets. This is because the definitions of the reachable sets are not a function of the initial ROE. Recall that the initial ROE only appears in precompensation of the ROE.  Therefore, in the state space $\delta\pmb{\alpha}$, which can be written simply as a function of the pseudo-state $\Delta\delta {\pmb{\alpha}}$ as $\delta\pmb{\alpha} = \pmb{\Phi}_{f,0}\delta\pmb{\alpha}_0 + \Delta\delta {\pmb{\alpha}}$, the origin of the reachable sets is actually translating. Figure \ref{fig:init_roe_err} shows this graphically. 

The plot in Fig. \ref{fig:init_roe_err} on the left is pseudo-state space, where the reachable sets remain unchanged. In contrast, in the plot on the right in Fig. \ref{fig:init_roe_err}, the origin of the reachable sets (indicated by a dot) is translating (linearly) based on the error in the initial ROE. 

An error in the absolute state is harder to quantify mathematically than an error in the relative state, but as in the case of a maneuver execution time error, is easy to visualize geometrically in the reachable sets. The chief absolute state appears in both the pseudo-state and state equations. Therefore, an error in the absolute state may distort the shapes of reachable sets and move them around in the phase plane; the overall effect is a composite of the effects shown in Fig. \ref{fig:init_oe_err}.  The effect of error in the initial chief orbit elements is nonlinear, but very small, as will be demonstrated in the next subsection with an analysis that includes realistic error.

Note that this paper does not consider the effect of maneuver vector orientation error. Orientation error can be represented with an azimuth and elevation angle from nominal RTN. In that case, a normal distribution for each angle would be required to describe the full effect of orientation error on the final ROE. Another option is to consider three planar cases (rotation about R, rotation about T, rotation about N) with a single angle to represent orientation error. In either case, orientation has no effect on the reachable sets as long as it is decoupled from the maneuver magnitude change, because the set of \textit{all} control actions of norm one is used to generate the reachable sets in time, and rotating the maneuver vector does not change its norm.

\subsection{Quantifying the effect of error on the achieved ROE}\label{sec:quantify_error}
The accuracy of the achieved final ROE is a linear function of error in the maneuver vector or in the initial ROE and a nonlinear function of the error in the initial chief OE or in the maneuver execution time. This section will show that if the final ROE are a linear function of a given error source, the mean and covariance of the effect of that source can be found analytically. For the cases where the effect of an error source is nonlinear, numerical realistic mean and covariance data will be used to demonstrate that these two sources are less impactful on the achieved ROE at the end of a reconfiguration. 

\subsubsection{Linear error effects}
This section shows how a known mean and covariance of a linear error source can be transformed analytically to a mean and covariance of the error in the achieved ROE. Suppose the error in the magnitude of the $\delta \pmb{v}$ vector, $\epsilon_{magn.}$, is a  normal distribution with known mean $\mu_{magn.} \in \mathbb{R}$ and covariance $V_{\text{magn.}}\in\mathbb{R}$. The ROE achieved in the presence of errors is found by simply substituting $(1+\epsilon_{magn.})\pmb{I}_3\delta \pmb{v}_k$ into Eq. \eqref{eqn:prob_statement} in place of $\delta \pmb{v}_k$. Performing this substitution and manipulating the equation yields the following form
\fontsize{9}{11}\selectfont\begin{equation}\label{eqn:ROE_actual_man_vec_err}
    \delta\pmb{\alpha}_{actual} = \pmb{A}_{\text{magn.}} + \pmb{B}_{\text{magn.}}\epsilon_{\text{magn.}},
\end{equation}\normalsize
where 
\fontsize{9}{11}\selectfont\begin{equation}\label{eqn:coeff_define_man_vec_err}
    \begin{matrix*}[l]
    \pmb{A}_{\text{magn.}} = \pmb{\Phi}_{f,0}\delta\pmb{\alpha}_0 + \sum_{k=1}^p\pmb{\Phi}_{f,k}\pmb{\Gamma}_k\delta\pmb{v}_k
 \in \mathbf{R}^6 \text{ and}\\
    \pmb{B}_{\text{magn.}} = \sum_{k=1}^p\pmb{\Phi}_{f,k}\pmb{\Gamma}_k\delta\pmb{v}_k \in \mathbf{R}^{6}.
    \end{matrix*}
\end{equation}\normalsize
Because the set of all normal distributions is closed under linear operations, a linear combination of multiple multivariate normal distributions is also normally distributed, and the mean and covariance of the achieved ROE in the presence of maneuver vector magnitude error are given simply by
\fontsize{9}{11}\selectfont\begin{equation}\label{eqn:mean_covar_man_vec_err}
    \begin{matrix*}
        \pmb{\mu}_{\delta\pmb{\alpha}_{actual}} = \pmb{A}_{\text{magn.}}+\pmb{B}_{\text{magn.}}\mu_{\text{magn.}}\\
        \pmb{V}_{\delta\pmb{\alpha}_{actual}} = \pmb{B}_{\text{magn.}}V_{\text{magn.}}\pmb{B}_{\text{magn.}}^{\text{T}}.
    \end{matrix*}
\end{equation}\normalsize
If $\mu_{magn.}$, the mean of the maneuver vector magnitude error, is 0, then the mean of the actual achieved ROE is the final desired ROE for the reconfiguration.
The same methodology can be applied to the final ROE achieved in the
presence of errors in the initial ROE. Substituting $\delta\pmb{\alpha}_0 + \pmb{\epsilon}_{\text{init roe}}$ into Eq. \eqref{eqn:prob_statement} in place of $\delta\pmb{\alpha}_0$ and manipulating as in Eq. \eqref{eqn:ROE_actual_man_vec_err} yields
%\fontsize{9}{11}\selectfont\begin{equation}\label{eqn:ROE_actual_init_ROE_err}
%    \delta\pmb{\alpha}_{actual} = \pmb{A}_{\text{init roe}} + \pmb{B}_{\text{init roe}}\pmb{\epsilon}_{\text{init roe}},
%\end{equation}\normalsize
%where
\fontsize{9}{11}\selectfont\begin{equation}\label{eqn:coeff_define_init_roe_err}
    \begin{matrix*}[l]
    \pmb{A}_{\text{init roe}} = \pmb{\Phi}_{f,0}\delta\pmb{\alpha}_0 + \sum_{k=1}^p\pmb{\Phi}_{f,k}\pmb{\Gamma}_k\delta \pmb{v}_k \in \mathbf{R}^6 \text{ and}\\
    \pmb{B}_{\text{init roe}} = \pmb{\Phi}_{f,0} \in \mathbf{R}^{6\times6}. 
    \end{matrix*}
\end{equation}\normalsize
%Following the same procedure as the previous section, the mean and covariance of the achieved ROE in the presence of error in the initial ROE is given by 
%\fontsize{9}{11}\selectfont\begin{equation}\label{eqn:mean_covar_init_roe_err}
%    \begin{matrix*}
%        \pmb{\mu}_{\delta\pmb{\alpha}_{actual}} = \pmb{A}_{\text{init roe}}+\pmb{B}_{\text{init roe}}\pmb{\mu}_{\text{init roe}}\\
%        \pmb{V}_{\delta\pmb{\alpha}_{actual}} = \pmb{B}_{\text{init roe}}\pmb{V}_{\text{init roe}}\pmb{B}_{\text{init roe}}^{\text{T}}.
%    \end{matrix*}
%\end{equation}\normalsize
%which takes the same form as Eq. \eqref{eqn:mean_covar_man_vec_err}.
The mean and covariance of the achieved ROE in the presence of error in the initial ROE takes the same form as Eq. \eqref{eqn:mean_covar_man_vec_err}, but with $\pmb{A},\pmb{B}_{\text{init roe}}$ in place of $\pmb{A},\pmb{B}_{\text{magn}}$.
\subsubsection{Nonlinear error effects}
The mean and covariance of a nonlinear error effect cannot be directly translated to an analytic expression for the mean and covariance of the error in the final ROE without linearization. Instead, a large number of data points of the errored variable are generated, and each is transformed nonlinearly to its effect on the final ROE. By the law of large numbers, the mean and covariance of the result of the many transformed data points will closely approximate the actual mean and covariance. Test 2 in Sec. \ref{sec:validation} will apply the methods to determine linear and nonlinear error effects to a realistic relative orbit reconfiguration.

The newly-found standard deviation of the error in each ROE, given by the square root of the on-diagonal entries of the resulting covariance matrices, can be used to determine the likelihood that the desired end-state lies within a certain range about the expected value. However, this common method neglects the off-diagonal covariances, which, as will be shown in the example in Sec. \ref{sec:validation}, are significant. Therefore, the cross-coupling cannot be discounted, and a better method is necessary to understand if a given maneuver scheme satisfies performance requirements in the presence of known errors. The next section demonstrates how to use the full 6D covariance matrix to assess performance. 

\subsection{Performance}\label{sec:performance}
As formation flying missions grow more ambitious, the requirements for delta-v budgeting and maneuvering performance grow more strict. This section shows how an error ellipsoid can be used to assess the statistical likelihood that the final ROE (in the presence of errors and uncertainty) lies within a given performance envelope. 

An error ellipsoid is an $n$-D generalization of an error ellipse, centered at the mean $\pmb{\mu}$ with shape, size, and orientation defined by a covariance matrix $\pmb{V}\in\mathbb{R}^{n\times n}$ and a desired confidence level. For an $n$-D state, the performance envelope is defined as a polygon in nD, or, equivalently the intersection of hyperplanes of dimension $n$-1, centered at the mean $\pmb{\mu}\in\mathbb{R}^n$. Because the polygon is essentially just a desired range that the final state must lie in in each coordinate direction, its axes align with the coordinate vectors of the $n$-D state space. To determine if the error ellipsoid lies entirely inside the performance envelope, it is enough to show that the smallest bounding box that contains the error ellipsoid lies inside of the polygon. This section shows how to find the ellipsoid maxima which define the edges of the smallest bounding box for an $n$-D state.

Given the mean $\pmb{\mu}\in \mathbb{R}^n$ and covariance $\pmb{V}\in \mathbb{R}^{n\times n}$, an error ellipsoid can be represented as a transformation matrix that stretches and rotates the unit spheroid in $\mathbb{R}^n$ to the size, shape, and orientation of the error ellipsoid. The covariance matrix is positive semi-definite by definition, and is therefore diagonalizable and has distinct (linearly independent) eigenvectors. Let $\pmb{\lambda}$, $\pmb{v}$ represent the eigendecompensation of the covariance matrix where $\pmb{\lambda}\in\mathbb{R}^{n\times n}$ is a diagonal matrix with $\left(i,i\right)$ entry eigenvalue $\lambda_i$ and associated eigenvector $\pmb{v}_i\in\mathbb{R}^n$ column i of the eigenvector matrix $\pmb{v}\in\mathbb{R}^{n\times n}$.  The error ellipsoid is centered at $\pmb{\mu}$ with semi-axis i of length $2\chi\sqrt{\lambda_i}$ in the direction of $\pmb{v}_i$ for $i=1,..,n$ \cite{bib:statistical_ellipsoids}. $\chi^2$ is the $\chi^2$ statistic value for a given confidence range and degrees of freedom equal to the size of the state (i.e. $n$). The transformation matrix $\pmb{T}\in\mathbb{R}^{n\times n}$ that maps the unit spheroid to the error ellipsoid is given by $\pmb{T} = \pmb{v}\text{diag}\left(2\chi\sqrt{\lambda_1}, ..., 2\chi\sqrt{\lambda_n}\right)$
which, when applied the unit axes of the spheroid, simply represents first a stretching along each axis $i$ to be of length $2\chi\sqrt{\lambda_i}$, and then a rotation of the ellipsoid such that the axes align with the semi-axes directions. 

\begin{figure}[H]
    \centering
    \begin{subfigure}[t]{0.31\textwidth}
    \centering 
        \includegraphics[height = 1.2in]{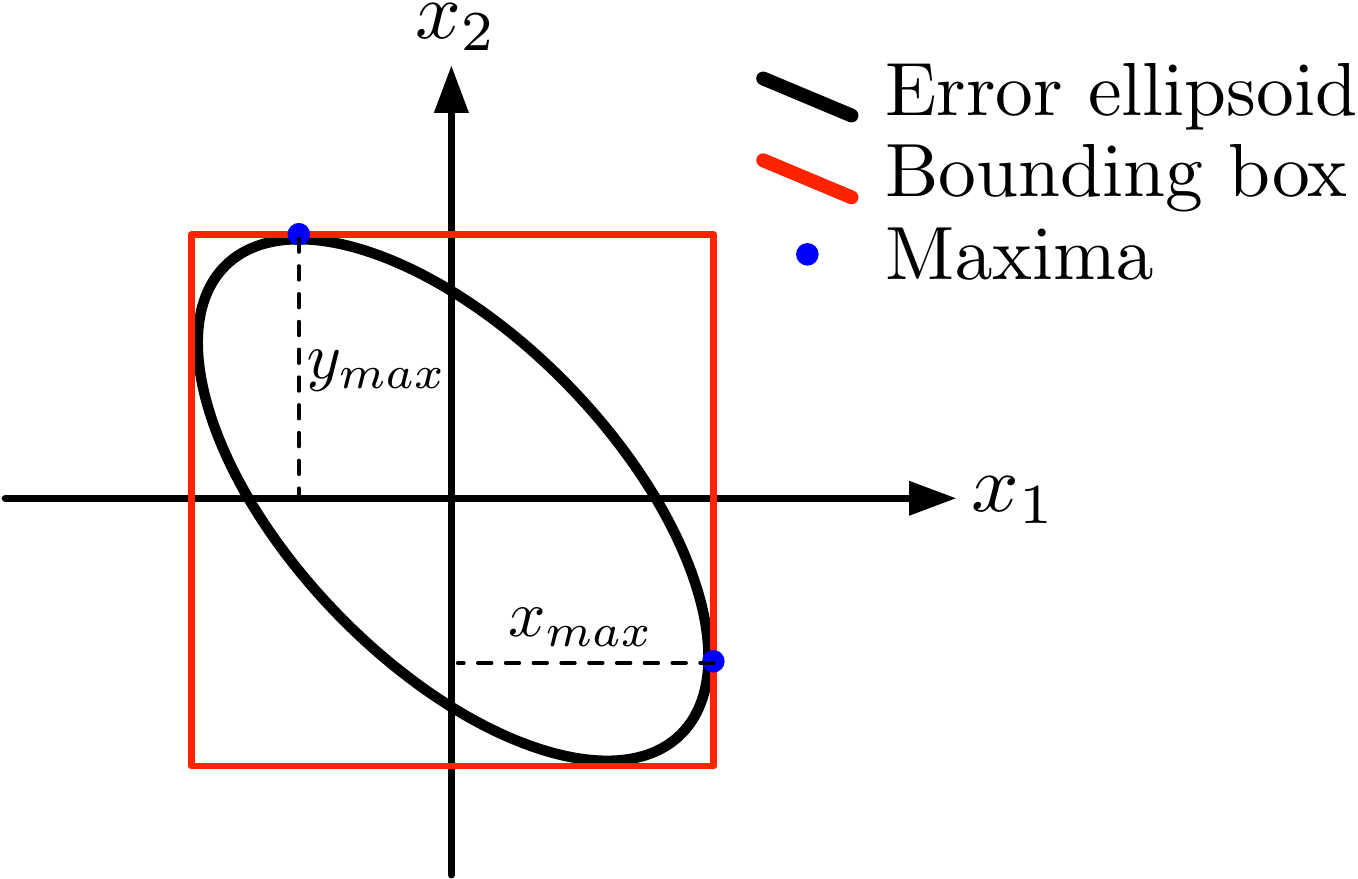}
        \caption{The normal vectors of the sides of the minimum bounding box are the basis vectors $\hat{\pmb{e}}_i$}
        \label{fig:error_ellipsoid_step1}
    \end{subfigure}\hfill~\begin{subfigure}[t]{0.31\textwidth}
    \centering 
        \includegraphics[height = 1.2in]{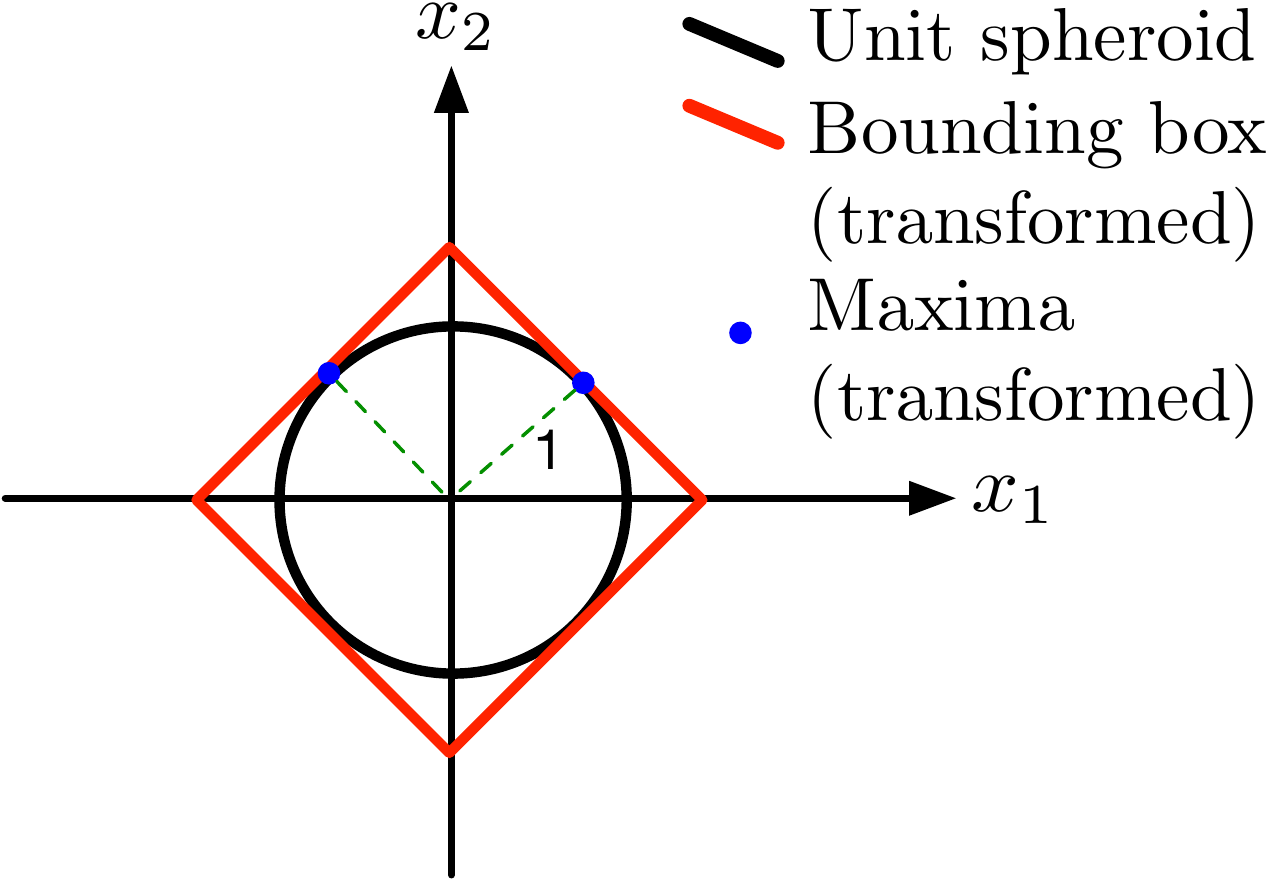}
        \caption{The transformed bounding box remains tangent to the ellipsoid when it is transformed into a unit sphere.}
        \label{fig:error_ellipsoid_step2}
    \end{subfigure}\hfill~\begin{subfigure}[t]{0.31\textwidth}
    \centering 
        \includegraphics[height = 1.2in]{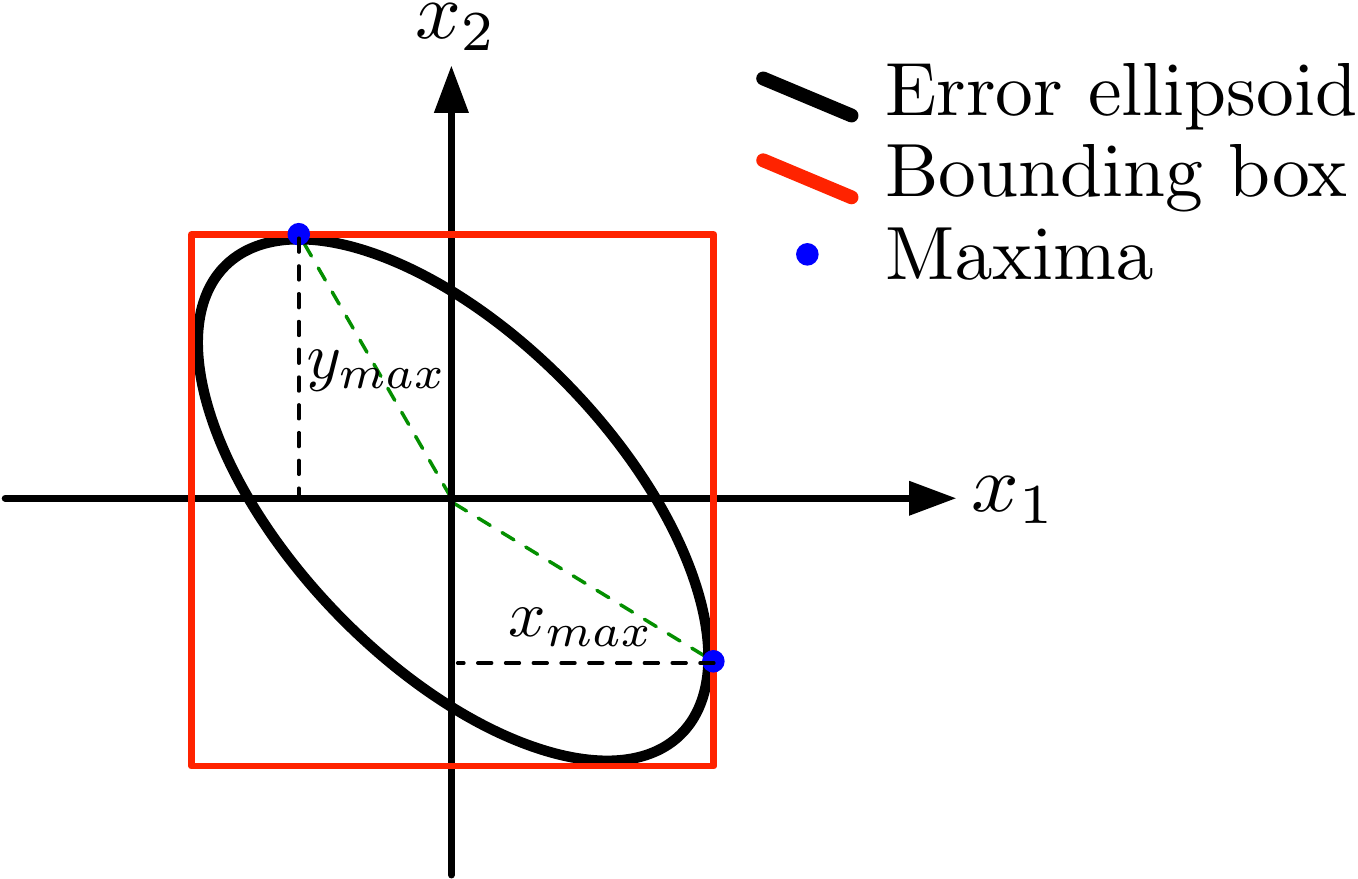}
        \caption{The normal vectors (green) to the hyperplanes on the unit sphere are the maxima of the error ellipsoid}
        \label{fig:error_ellipsoid_step3}
    \end{subfigure}
    \caption{To find the maxima of an error ellipsoid in nD, the ellipsoid is transformed into a unit sphere using the eigendecomposition of its covariance matrix.}
    \label{fig:error_ellipsoid_steps}
\end{figure}

The ellipsoid maxima are found by exploiting the simplicity of the transformation matrix and the positive semi-definiteness of the covariance matrix. This method is outlined in Fig. \ref{fig:error_ellipsoid_steps} for a 2D demonstrative example. The goal of this methodology is to find the minimum bounding ``box,'' or n-dimensional abstraction of a box, that encloses the error ellipsoid completely, for any $n$-D state. The minimum bounding box that fits around the error ellipsoid has unknown dimensions defined by the maxima of the ellipsoid. As shown in Fig. \ref{fig:error_ellipsoid_step1}, the ``sides'' of the bounding box are aligned with the coordinate directions of the space, i.e. the basis vectors $\hat{\pmb{e}}_1$, $\hat{\pmb{e}}_2$, ..., $\hat{\pmb{e}}_n$ are each normal vectors of the hyperplanes. Because it is the smallest possible box, the hyperplanes are tangent to the error ellipsoid at its maxima in each coordinate direction and remain tangent to the unit sphere after transformation because $\pmb{T}$ is linear, as shown in Fig. \ref{fig:error_ellipsoid_step2}. However, the benefit of looking at the error ellipsoid as a transformed unit sphere comes from the fact that the transformed normal vectors that define these tangent planes are now by definition radius vectors of the sphere, and have a known length of 1, as shown in green in Fig. \ref{fig:error_ellipsoid_step2}. If the directions of these normal vectors are found and transformed back, they will define the $n$ maxima, the green vectors in Fig. \ref{fig:error_ellipsoid_step3}. The basis vectors $\hat{\pmb{e}}_i$ define the hyperplanes both before and after linear transformation, so any vectors perpendicular to these transformed basic vectors can be used as direction vectors. The transformed basis vectors are calculated as $\pmb{T}^{-1}\hat{\pmb{e}}_i = \text{column i of } \pmb{T}^{-1}.$ Recall $\pmb{T}$ is just the product of a diagonal matrix and distinct eigenvectors, so $\pmb{T}$ and $\pmb{T}^{-1}$ have the same eigenvectors. Therefore  row $j$ of $\pmb{T}$ is perpendicular to column $i$ of $\pmb{T}^{-1}$ as long as $i\neq j$. So using the rows of $\pmb{T}$ as normal vectors and transforming them back to ellipsoid space yields the maxima. 
A ROE reconfiguration example can be found in Sec. \ref{sec:validation}. 
The methodology is general to an $n$-D state, and takes into account the actual shape and cross coupling of the $n$-D covariance matrix to check for violations of more complicated deadbands without loss of information or generality. Using this methodology and the error analysis in the previous section, it is now possible to determine with a given confidence level the accuracy of a given maneuver scheme in achieving a desired set of ROE.

\section{Validation}\label{sec:validation}
\begin{figure}[H]
\centering 
\includegraphics[width = .8\textwidth]{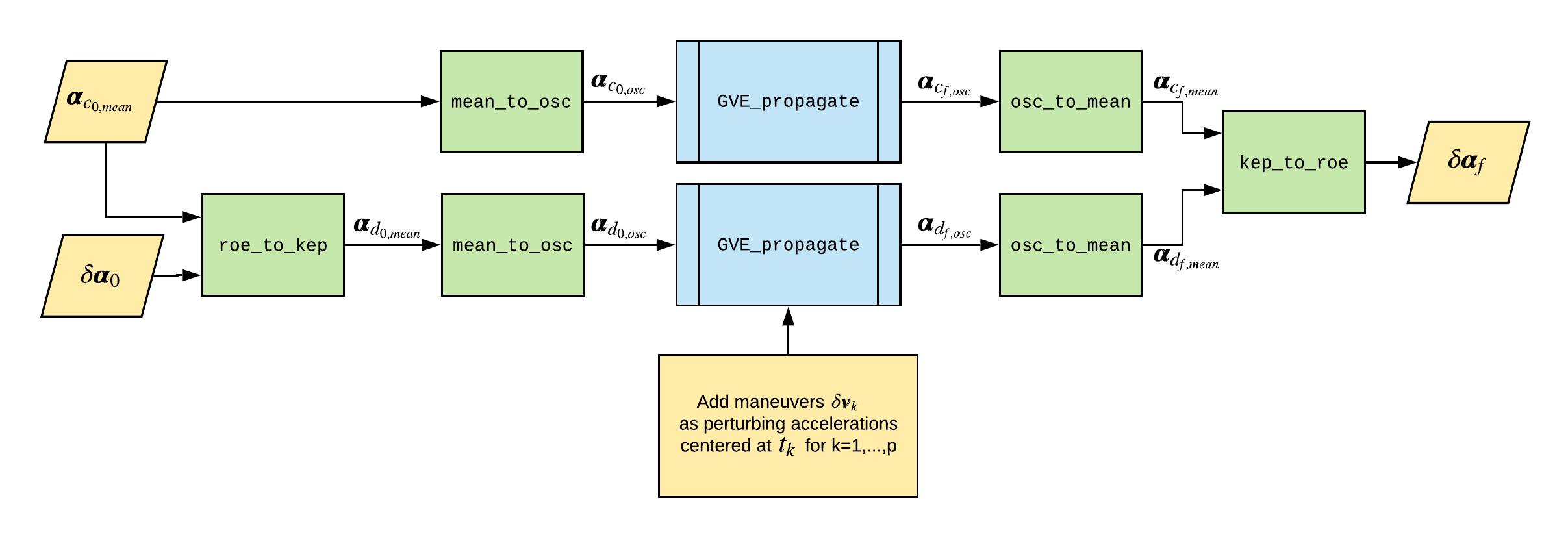}
\caption{Simulation architecture for numerical integration of the GVE including maneuvers.}
\label{fig:blockdiagram}
\end{figure}

\begin{figure}[H]
\centering 
\includegraphics[width = .5\textwidth]{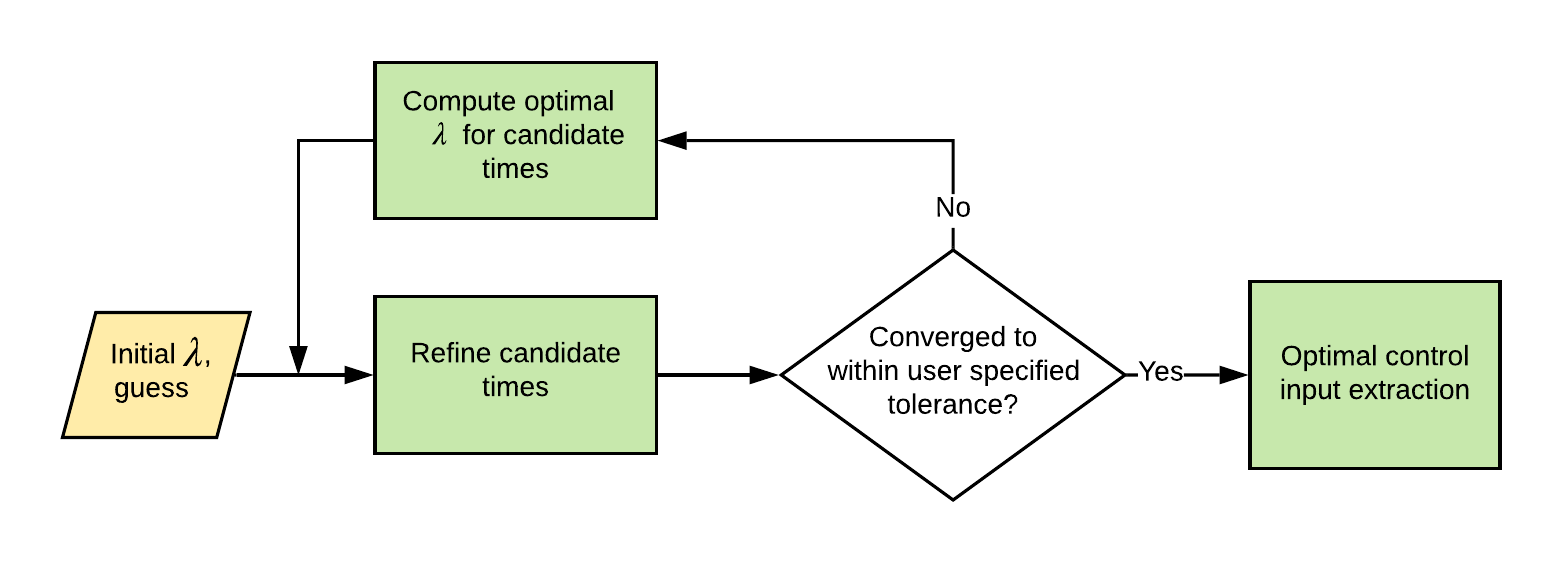}
\caption{Algorithm to iteratively compute lower bound on minimum reconfiguration cost.\cite{bib:KDAutomatica}}
\label{fig:koenig_alg}
\end{figure}

The new closed-form maneuver schemes are validated in this section. Numerical integration of the GVE including a full-force dynamics model is used to verify that the control solutions achieve the desired reconfiguration. In addition, the optimality of the maneuver schemes is confirmed by comparing the computed reachable delta-v to the output of Koenig et al.'s numerical algorithm \cite{bib:KDAutomatica}. 
The GVE-based propagator in this paper uses a subset of the capabilities of the Stanford Space Rendezvous Laboratory's state of the art propagator, including a 40$\times$40 gravity field model, atmospheric drag, solar radiation pressure, geomagnetic and solar-flux data, and third-body effects from both the Sun and moon. The numerical integration architecture is as follows. Reconfiguration parameters ($\delta\pmb{\alpha}_0$, $\pmb{\alpha}_{c,0}$, and $\Delta t_f$) and maneuver scheme ($t_k$, $\delta \pmb{v}_k$ for $k=1,...,p$) are provided as inputs. The maneuver schemes and times are represented as perturbing accelerations over a number of time steps centered on the actual maneuver time. First, the initial mean deputy OE, $\pmb{\alpha}_{d,0}$, are extracted from $\delta\pmb{\alpha}_0$ and $\pmb{\alpha}_{c,0}$ using the nonlinear relationships in Eq. \eqref{eqn:ROE}. The chief and deputy OE are both transformed to osculating OE using Brouwer's transformation \cite{bib:Brouwer} and simultaneously propagated forward by numerical integration of the GVE including perturbations. The final osculating OE are converted first back into mean OE, then back to ROE for comparison to the desired final ROE. This process is shown in compact block diagram form in Fig. \ref{fig:blockdiagram}. 
The second validation method uses Koenig et al.'s algorithm to confirm optimality. The optimization problem in Eq. \eqref{eqn:optcontrolproblem} is solved in order to maximize the delta-v cost subject to the constraint that the desired reconfiguration cannot be reached at a lesser cost. This is equivalent to maximizing the cost to reach a supporting hyperplane $L$ which contains the target. The hyperplane is defined by a normal vector $\pmb{\lambda}$ tangent to the boundary of the convex hull at the desired pseudo-state. As Koenig shows\cite{bib:KDAutomatica}, a lower bound on the minimum reconfiguration cost can be found using an algorithm which iteratively refines $\pmb{\lambda}$ by adding and removing candidate maneuver times in order to improve the cost. Figure \ref{fig:koenig_alg} is a graphical overview of the algorithm. Once the lower bound, $\delta v_{lb}$, has been computed numerically using Koenig's algorithm, it will be compared to the reachable minimum delta-v, $\delta v_{min}$, derived in this paper to confirm global optimality.
Two example reconfigurations are defined to demonstrate the new closed-form maneuver schemes in realistic mission scenarios. The first test demonstrates application of the methodology presented in this paper to derive optimal and quantifiably sub-optimal maneuver schemes for a 6D reconfiguration in an eccentric chief orbit. The second test demonstrates the effects of four different error sources on the ROE achieved at the end of the reconfiguration span for a given reconfiguration.

\subsection{Test 1: 6D reconfiguration, eccentric chief orbit}
The reconfiguration parameters for Test 1 are given by
\fontsize{9}{11}\selectfont\begin{equation}\label{eqn:reconfig_params_test1}
    \begin{matrix*}[l]
    \pmb{\alpha}_{c,0} = \begin{bmatrix}15000\text{ km} & 0.5 & 10^o & 0^o & 20^o & 0^o\end{bmatrix}\\
    \delta\pmb{\alpha}'_0 = \begin{bmatrix}30 & -10500 & 0 & -50 & 0 & -30\end{bmatrix}\text{ m}\\
    \delta\pmb{\alpha}'_f = \begin{bmatrix}100 & -12500 & 200 & 300 & 20 & 0\end{bmatrix}\text{ m}\\
    \Delta t_f = 2.2 \text{ orbits.}\\
    \end{matrix*}
\end{equation}\normalsize 
As discussed, using the $\delta\pmb{e}'$ state allows for the full 6D reconfiguration to be treated as a 4D in-plane reconfiguration and a 2D out-of-plane reconfiguration. The desired decoupled pseudo-states are given by
\fontsize{9}{11}\selectfont
\begin{align}
    \Delta\delta {\pmb{\alpha}}_{IP} = \begin{bmatrix}a\Delta\delta  {a}_{des} & a\Delta\delta {\lambda}_{des} & a\Delta\delta\tilde{e}_{x,des} & a\Delta\delta \tilde{e}_{y,des}\end{bmatrix}^{\text{T}} = \begin{bmatrix}
        70 & -1377.965 & 307.646 & 260.488
    \end{bmatrix}^{\text{T}} \text{ m} \label{eqn:Ddroe_IP_reconfig_1}\\
    \Delta\delta {\pmb{\alpha}}_{OOP} = \begin{bmatrix}a\Delta\delta\tilde{i}_{x,des} & a\Delta\delta \tilde{i}_{y,des}\end{bmatrix}^{\text{T}} = \begin{bmatrix}29.0545 & 21.350
    \end{bmatrix}^{\text{T}} \text{ m.} \label{eqn:Ddroe_OOP_reconfig_1}
\end{align}\normalsize
Note that $a\Delta\delta  {\pmb{e}}'_{des}$ has already been transformed to the equivalent desired pseudo-state $a\Delta\delta\tilde{e}_{des}$ using Eq. \eqref{eqn:Ddeprime_to_Dde}. The next two subsections treat the reconfiguration problems separately to derive the in-plane and out-of-plane nested reachable sets. This yields all possible optimal maneuvers. Then, in the \textit{Results} section, appropriately sized sets of maneuvers are chosen from each nested reachable set to satisfy inherent system constraints and achieve the desired reconfiguration. 

\subsubsection{Out-of-plane reconfiguration}
The reachable delta-v minimum for the reconfiguration in Eq. \eqref{eqn:Ddroe_OOP_reconfig_1} is found using Table \ref{table:dvmin_ecc_oop} to be $\delta v_{min,OOP} = 0.00854$ m/s. The total maneuver scheme cost matches the output of Koenig's algorithm, $\delta v_{lb} = 0.00854$ m/s. $\nu_{opt}$ are found using column 3 of Table \ref{table:dvstars_ecc} to be $\nu_{opt}$ = [$3.7753,   10.0585$]. The $\nu$ values are converted to times by first calculating the equivalent mean anomalies and then scaling by the mean motion to be 
 $T_{opt}$ = [$13397.11, 
   31680.13$] s. The points in the corresponding nested reachable set $S_n^*(\delta v_{min,\delta\pmb{i}},T_{opt})$, calculated using Eq. \eqref{eqn:Sdv}, are given by
\fontsize{9}{11}\selectfont\begin{equation}\label{eqn:reconfig_1_Sn_OOP}
S_{n,OOP} = \begin{bmatrix}
   20 & 20 \\
   30 & 30
   \end{bmatrix} \text{ m}.
\end{equation}\normalsize
After the in-plane maneuver schemes are calculated, one of the columns in Eq. \eqref{eqn:reconfig_1_Sn_OOP}, which represents a possible maneuver that satisfies the Type 1 constraints in column 3 of Table \ref{table:CFconstraints_ecc}, will be chosen for use along with the three maneuvers from the in-plane calculations. Typically this choice is based on satisfying mission constraints, attitude control timing, or science operations. 

\subsubsection{In-plane reconfiguration}
The reachable delta-v minimum for the reconfiguration in Eq. \eqref{eqn:Ddroe_IP_reconfig_1} is $\delta v_{min,IP} = 0.07801$ m/s, which is equal to $\delta v_{min,\delta\pmb{e}}$ and calculated using Table \ref{table:dvmin_de}. Applying Koenig's algorithm gives a global minimum delta-v $\delta v_{lb} = 0.07815$ m/s. The reachable delta-v minimum matches up to $0.14$ mm/s (0.18\%) which is approximately the stopping tolerance set by the user, therefore the closed-form maneuver scheme is globally optimal. From there, $\nu_{opt}$ are found using column 3 of Table \ref{table:dvstars_ecc} to be $\nu_{opt}$ = [$0.8967,    3.5907,    7.1799,    9.8738,   13.4631$]. The $\nu$ values are converted to times, $T_{opt}$, given by [$826.28,   12328.94,   19109.30,   30611.95,   37392.32$] s.   The points in the corresponding nested reachable set $S_n^*(\delta v_{min,\delta\pmb{e}},T_{opt})$, calculated using Eq. \eqref{eqn:Sdv}, are given by
\fontsize{9}{11}\selectfont\begin{equation}\label{eqn:reconfig_1_Sn_IP}
S_{n,IP} = \begin{bmatrix}
   704.20 &  -299.86 &  704.20 &  -299.86 &  704.20\\
  -14326.00 &   4235.94 & -7689.06 &  1409.81 & -1052.11\\
   197.91 &  200.00 &  197.91 &  200.00 &  197.91\\
   346.34 &  350.00 &  346.34 &  350.00 &  346.34
   \end{bmatrix} \text{ m},
\end{equation}\normalsize
where each column represents a 4D vector of the in-plane change in ROE ($a\Delta\delta {\pmb{\alpha}}_{des}$) achieved using the optimal maneuver $\delta \pmb{v}^*$, given in Eq. \eqref{eqn:dvtstar_example}. 

\subsubsection{Results}
\begin{figure}[H]
    \begin{subfigure}[h]{0.32\textwidth}
    \centering 
        \includegraphics[width=.95\textwidth]{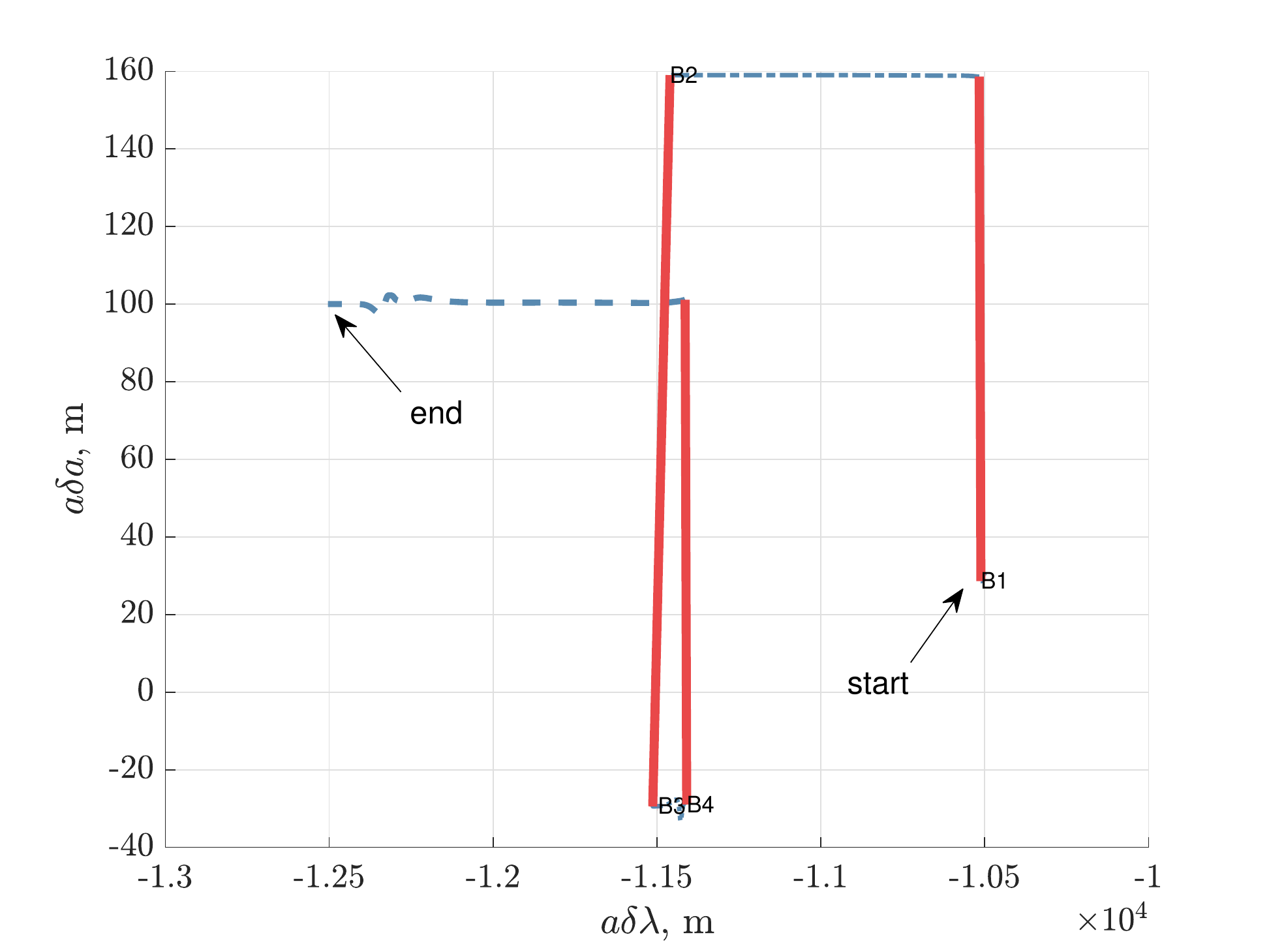}
        \caption{$\delta a, \delta\lambda$ plane}
        \label{fig:ROE_evolution_reconfig_1_dadl}
    \end{subfigure}\begin{subfigure}[h]{0.32\textwidth}
    \centering 
        \includegraphics[width=.95\textwidth]{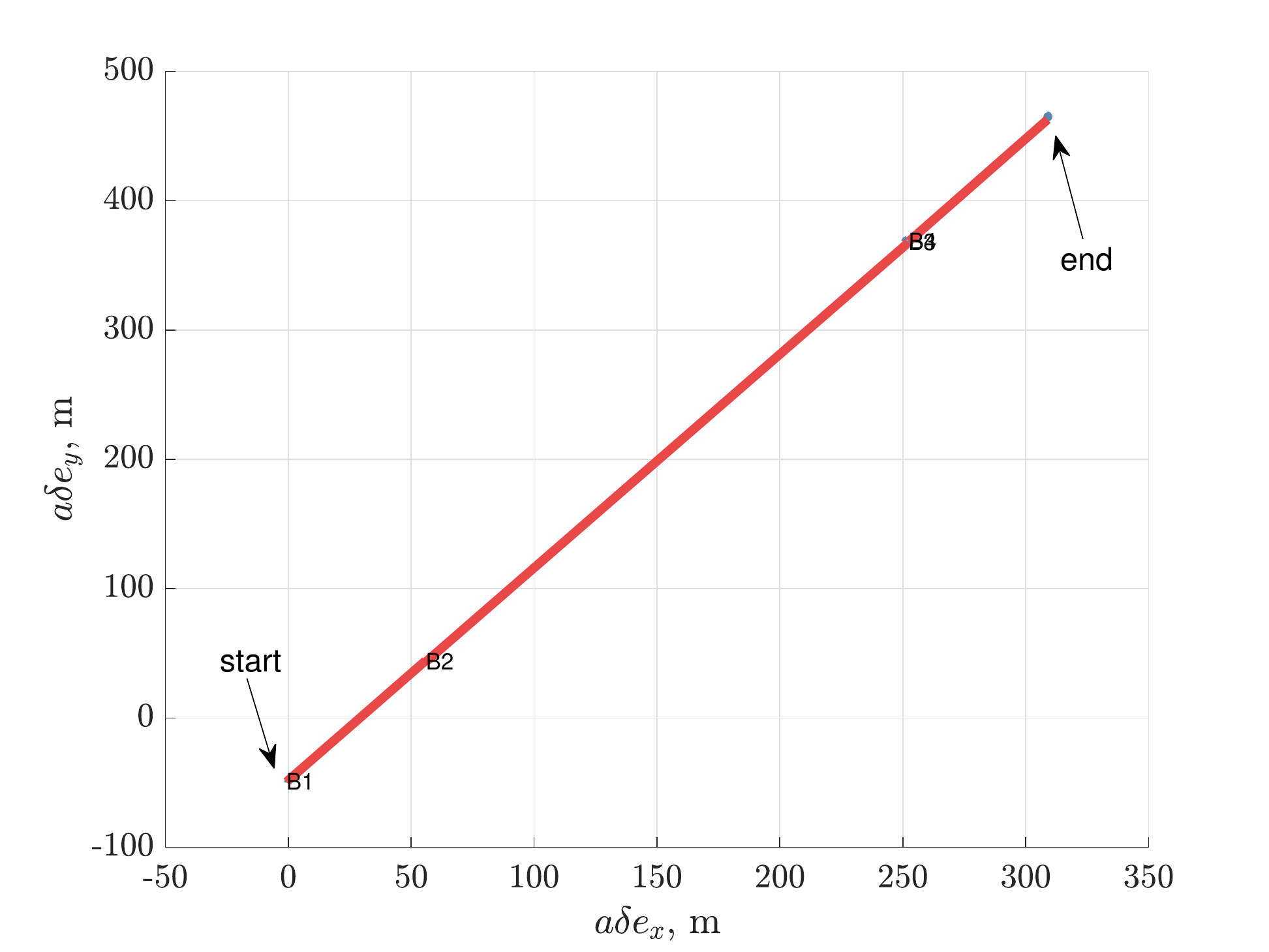}
        \caption{$\delta \pmb{e}'$ plane}
        \label{fig:ROE_evolution_reconfig_1_de}
    \end{subfigure}\begin{subfigure}[h]{0.32\textwidth}
    \centering 
        \includegraphics[width=.95\textwidth]{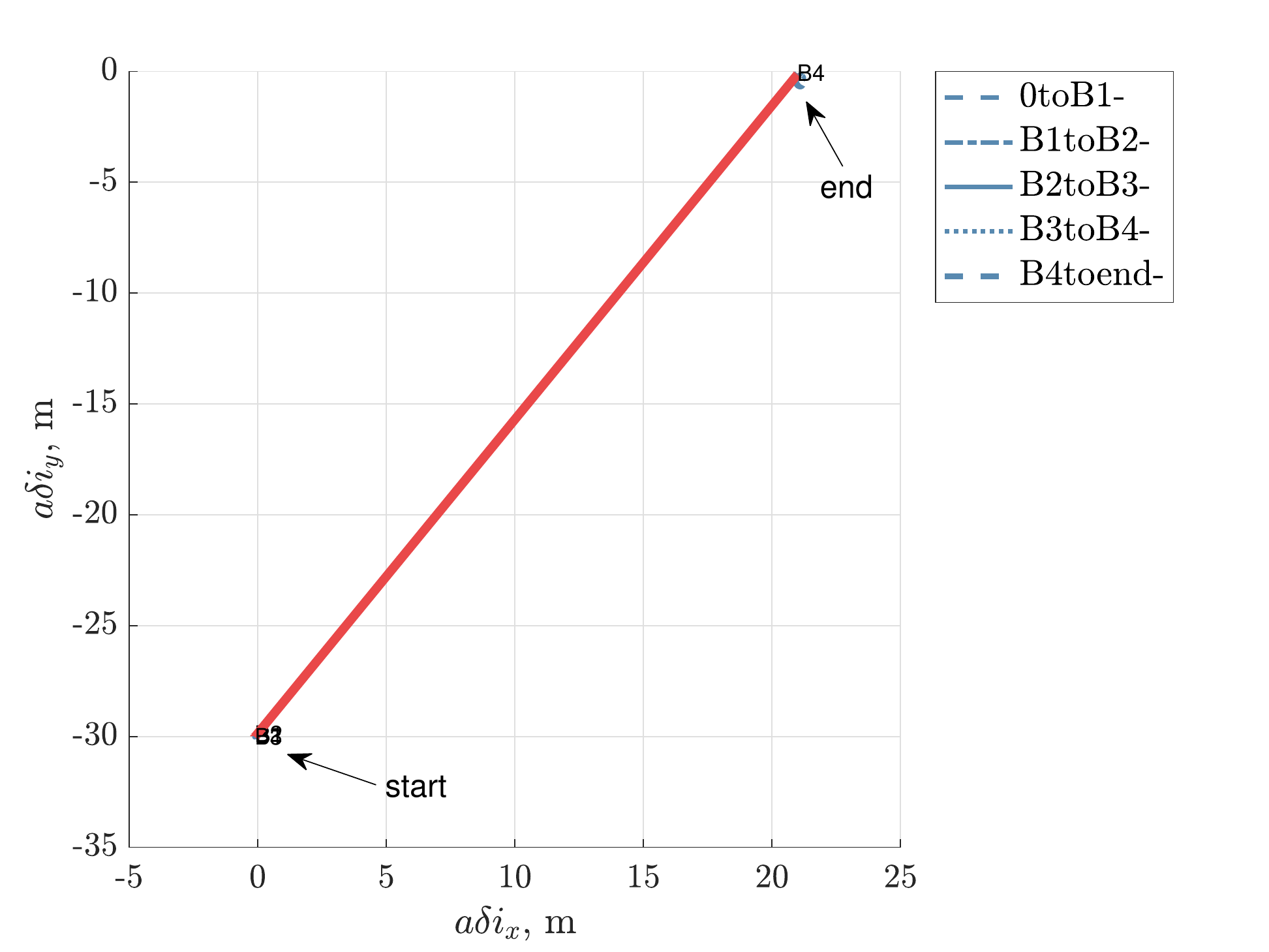}
        \caption{$\delta \pmb{i}$ plane}
        \label{fig:ROE_evolution_reconfig_1_di}
    \end{subfigure}
    \caption{Evolution of the ROE in Test 1 with full-force dynamics model}
    \label{fig:ROE_evolution_reconfig_1}
    \end{figure}

There are ten sets of three in-plane maneuvers and one out-of-plane maneuver that can be chosen from the five maneuvers in $S_{n,IP}$ (Eq. \eqref{eqn:reconfig_1_Sn_IP})  and the two maneuvers in $S_{n,OOP}$ (Eq. \eqref{eqn:reconfig_1_Sn_OOP}) to satisfy the inherent system constraints. One of the combined schemes is given by 
\fontsize{9}{11}\selectfont\begin{equation}\label{eqn:dvsreconfig_1}
\begin{matrix*}[l]
\delta \pmb{v}_1 = \begin{bmatrix}0.00136 &  0.0143 & 0\end{bmatrix}^{\text{T}} \text{ m/s at } t_1 =826.28\text{ s}\\
\delta \pmb{v}_2 = \begin{bmatrix}0.00603 &  -0.0490 & 0\end{bmatrix}^{\text{T}}\text{ m/s at } t_2 = 12328.94\text{ s} \\
\delta \pmb{v}_3 = \begin{bmatrix}0 & 0 & -0.008543\end{bmatrix}^{\text{T}}\text{ m/s at } t_3 = 13397.11\text{ s} \\ 
\delta \pmb{v}_4 = \begin{bmatrix}0.00137 &  0.0143 & 0\end{bmatrix}^{\text{T}}\text{ m/s at } t_4 = 19109.30\text{ s}
\end{matrix*}
\end{equation}\normalsize
which corresponds to columns 1, 2, and 3 of $S_{n,IP}$ and column 1 of $S_{n,OOP}$.
Fig. \ref{fig:ROE_evolution_reconfig_1} shows the evolution of the ROE when the maneuvers in Eq. \eqref{eqn:dvsreconfig_1} are applied. 

In addition, the reconfiguration error, defined as the 1-norm of the difference between the desired mean ROE and the achieved mean ROE, scaled by the desired mean ROE, is given in Table \ref{table:reconfig_1_results}.

\fontsize{9}{11}\selectfont
\begin{longtable}{l K{1.5cm} K{1.5cm} K{1.5cm} K{1.5cm} K{1.5cm} K{1.5cm}}
		\caption{Reconfiguration accuracy, Test 1}\label{table:reconfig_1_results} \\ \toprule\toprule
		
	    {}                          & \multicolumn{6}{c}{\textbf{Relative orbit element}} \\ 
	    {}                          & $a\delta a$   & $a\delta\lambda$  & $a\delta e_x'$    & $a\delta e_y'$    & $a\delta i_x$ & $a\delta i_y$ \\\midrule 
	    \textbf{Desired value, m}   & 100.00        & -12500.00         & 307.65            & 470.98            & 20.00            & 0.00\\
	    \textbf{Achieved value, m}  & 100.02        & -12504.33         & 308.33            & 464.12            & 20.94         & -0.48\\
	    \textbf{\% Error}           & 0.02\%        & 0.03\%            & 0.22\%            & 1.46\%            & 4.69\%        & -\% \\\bottomrule    
	\end{longtable}   \normalsize

As shown in the third and fourth columns of Table \ref{table:reconfig_1_results}, the maneuver scheme calculated in the $\delta\tilde{\pmb{e}}$ representation achieves the desired reconfiguration in the $\delta\pmb{e}'$ with high accuracy. In fact, even though the open-loop maneuver scheme was calculated excluding perturbations, the desired reconfiguration is achieved with less than 10 meters (<5\%) of error.
For the first time in literature, this example demonstrates the ability of the new closed-form maneuver schemes to reconfigure the relative motion in an eccentric chief orbit with globally optimal delta-v.

\begin{figure}[H]
    \centering
    \begin{subfigure}[t]{0.32\textwidth}
    \centering 
        \includegraphics[width=.9\textwidth]{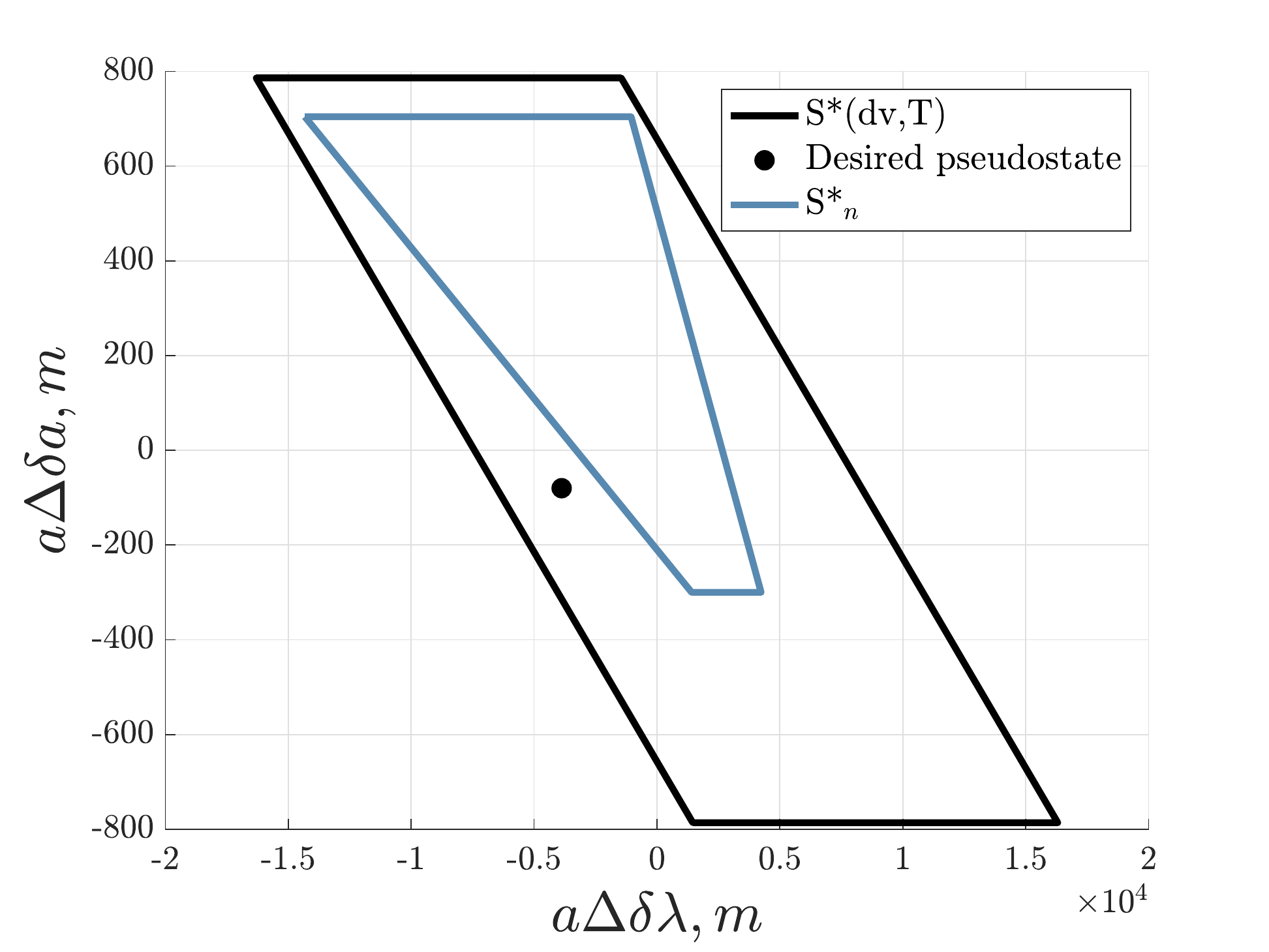}
        \caption{The desired pseudo-state lies outside of the nested reachable set}
        \label{fig:NRS_subopt_test1}
    \end{subfigure}\hfill\begin{subfigure}[t]{0.32\textwidth}
    \centering 
        \includegraphics[width=.9\textwidth]{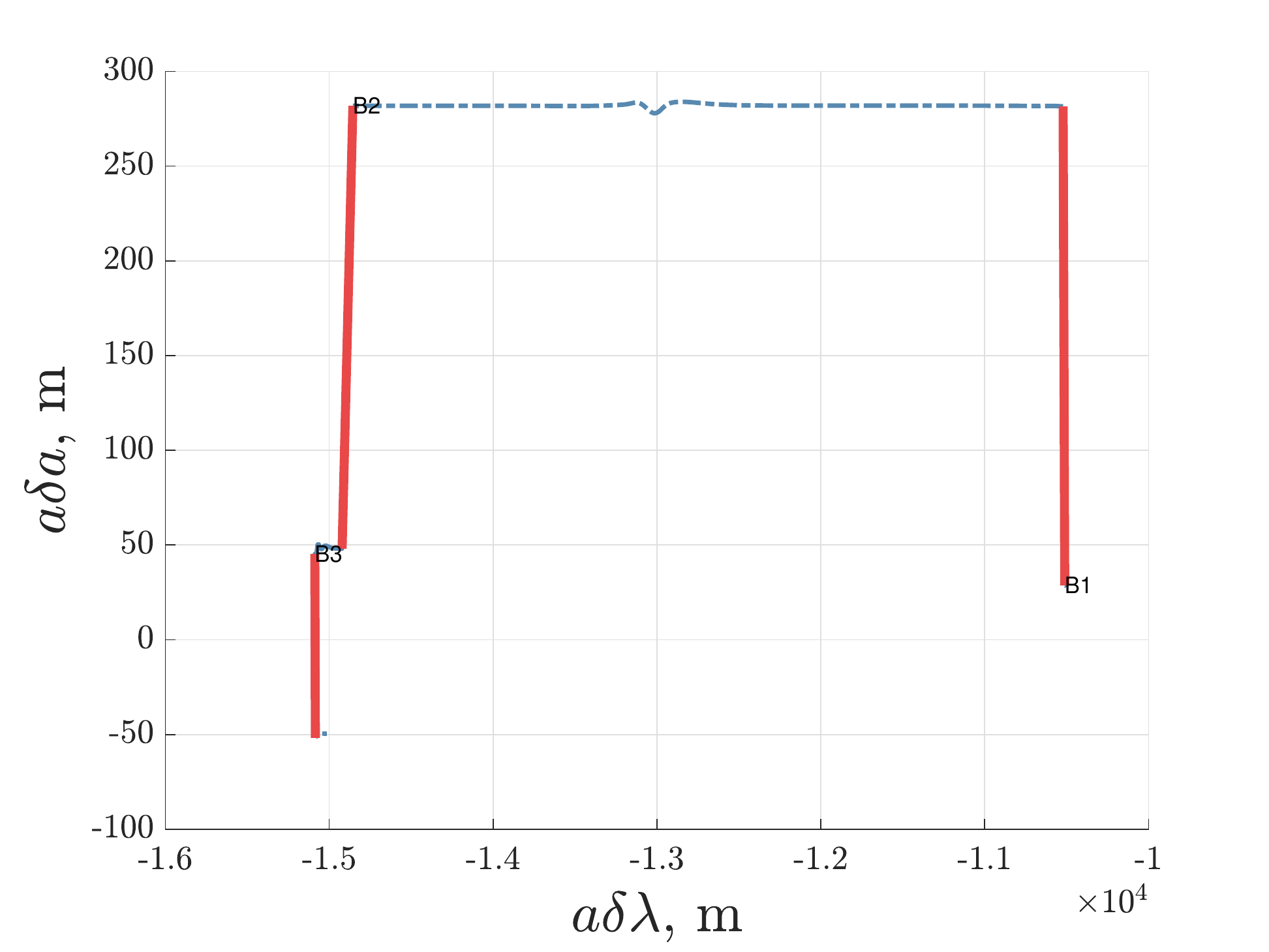}
        \caption{Evolution of the ROE in $\delta a, \delta\lambda$ plane with full-force dynamics model}
        \label{fig:ROE_evolution_reconfig_1_subopt_dadl}
    \end{subfigure}\hfill\begin{subfigure}[t]{0.32\textwidth}
    \centering 
        \includegraphics[width=.9\textwidth]{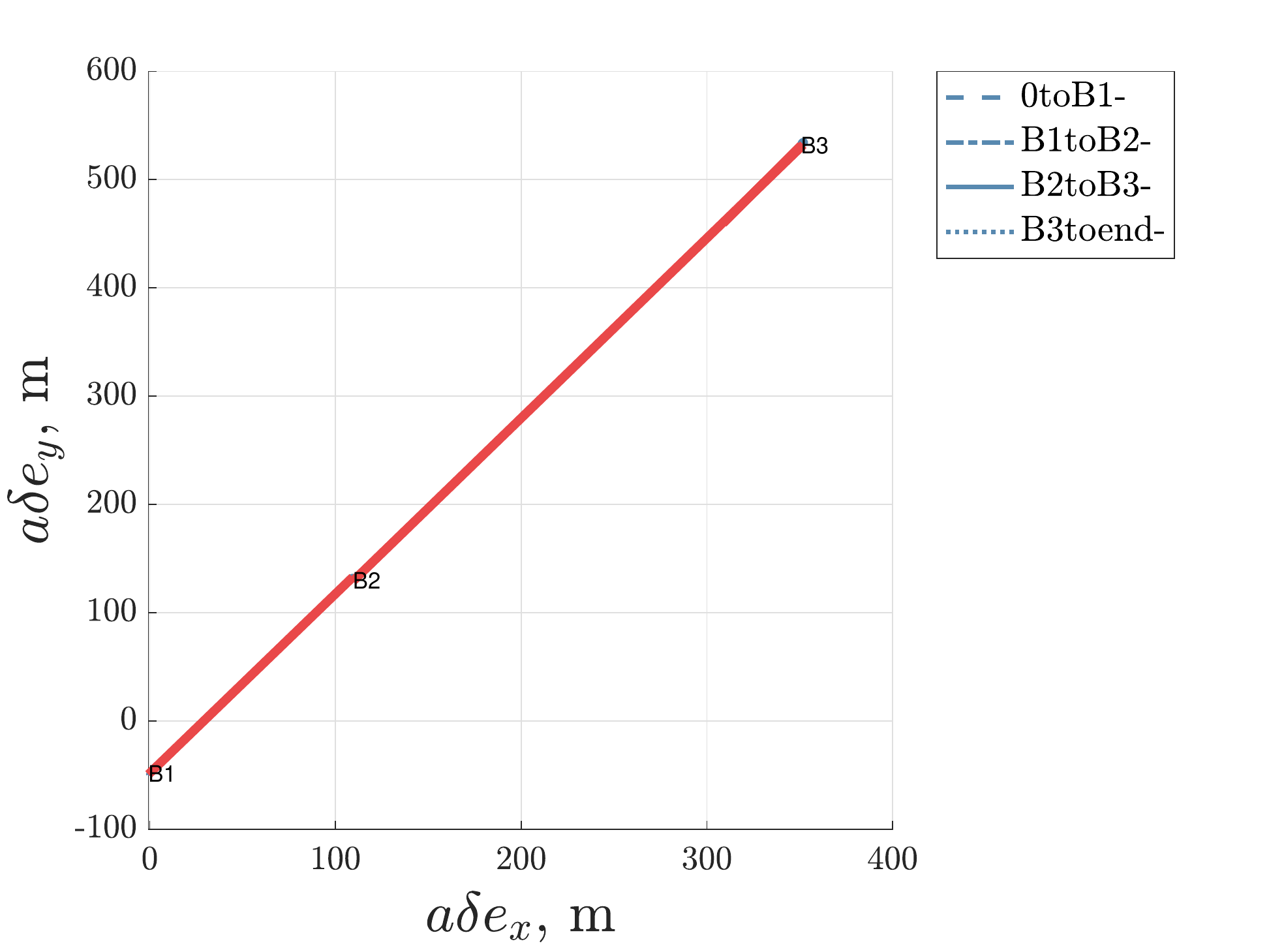}
        \caption{Evolution of the ROE in $\delta \pmb{e}'$ plane with full-force dynamics model}
        \label{fig:ROE_evolution_reconfig_1_subopt_de}
    \end{subfigure}
    \caption{Application of sub-optimal solution handling methodology}
    \label{fig:subopt_results_test1}
\end{figure}

Now suppose $a\delta a_f = -50$ m and $a\delta \lambda_f = -15000$ m. Then $a\Delta\delta {a}_{des} = -80$ m and $a\Delta\delta {\lambda}_{des} = -3877.96$ m. The desired psuedo state lies outside of the nested reachable set in the non-dominant plane as shown in Fig. \ref{fig:NRS_subopt_test1}, so the sub-optimal solution methodology presented in Sec. \ref{sec:subopt_solns} must be used. Applying the methodology in Sec. \ref{sec:subopt_solns} yields nine possible maneuver schemes for the in-plane reconfiguration. %The scheme with minimum delta-v is 
%\fontsize{9}{11}\selectfont\begin{equation}\label{eqn:dvsreconfig_1_subopt}
%\begin{matrix*}[l]
%\delta \pmb{v}_1 = \begin{bmatrix}0.00265 & 0.02779 & 0\end{bmatrix}^{\text{T}} \text{ m/s at } t_1 =826.28\text{ s}\\
%\delta \pmb{v}_2 = \begin{bmatrix}0.00747  &  -0.0606 & 0\end{bmatrix}^{\text{T}}\text{ m/s at } t_2 = 30611.95\text{ s} \\
%\delta \pmb{v}_3 = \begin{bmatrix}-0.00103 & -0.0107 & 0\end{bmatrix}^{\text{T}}\text{ m/s at } t_3 = 37392.32\text{ s} \\ 
%\end{matrix*}
%\end{equation}\normalsize
%which has a total delta-v of $0.0998$ m/s. 
The scheme with minimum delta-v has a total delta-v of $0.0998$ m/s. This a $27\%$ increase over the optimal delta-v. Figs. \ref{fig:ROE_evolution_reconfig_1_subopt_dadl} and \ref{fig:ROE_evolution_reconfig_1_subopt_de} show the evolution of the ROE.
Unlike Fig. \ref{fig:ROE_evolution_reconfig_1_de}, where the maneuvers follow a straight line path towards the desired pseudo-state, which indicates optimality, the path in Fig. \ref{fig:ROE_evolution_reconfig_1_subopt_de} doubles back on itself after the third burn. However, the desired pseudo-state in 6D is still achieved. 

\begin{figure}[H]
    \centering
    \begin{subfigure}[t]{0.32\textwidth}
    \centering 
        \includegraphics[width=.9\textwidth]{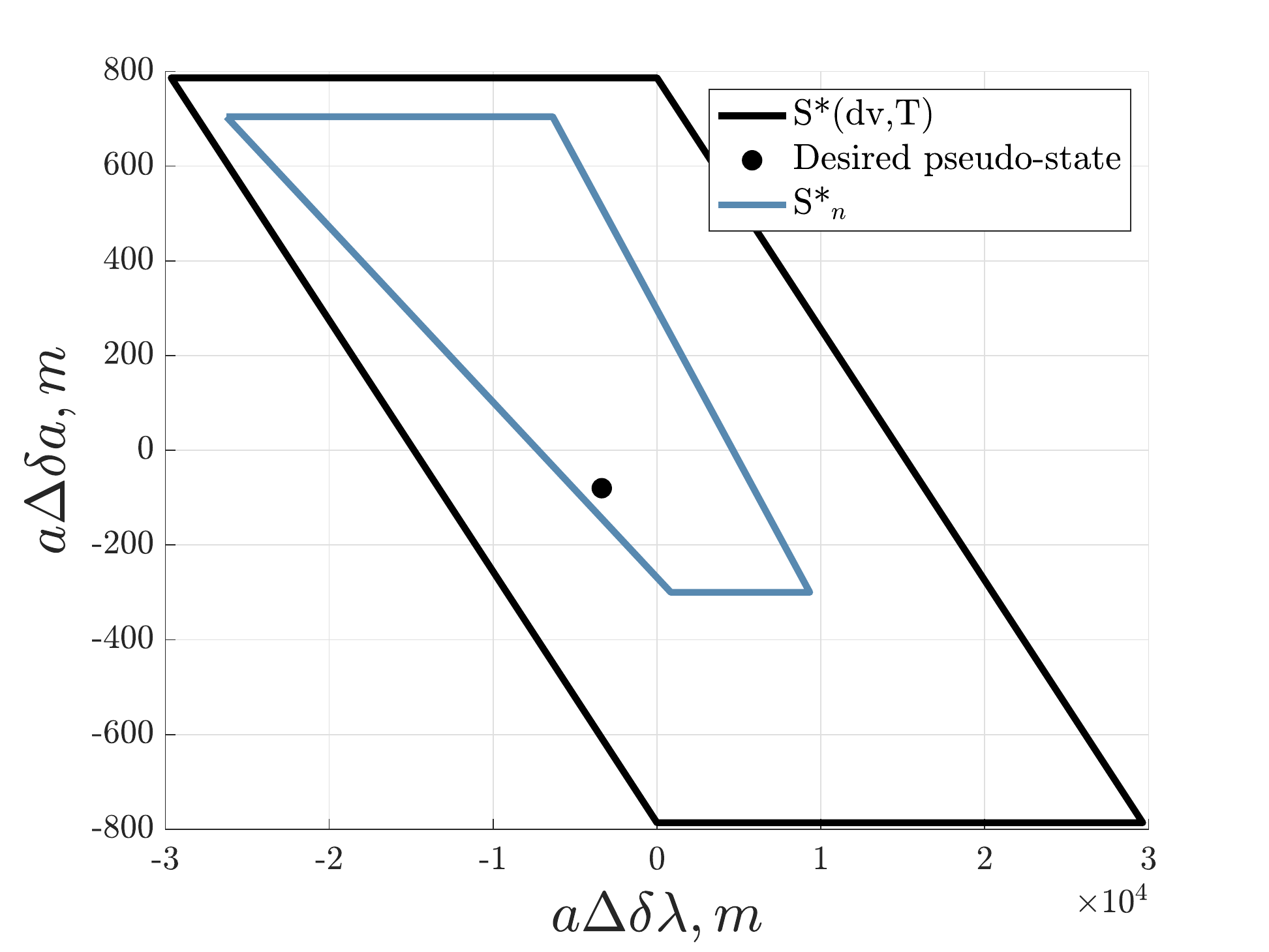}
        \caption{The desired pseudo-state lies inside of the nested reachable set when the reconfiguration time is extended}
        \label{fig:NRS_longer_test1}
    \end{subfigure}\hfill\begin{subfigure}[t]{0.32\textwidth}
    \centering 
        \includegraphics[width=.9\textwidth]{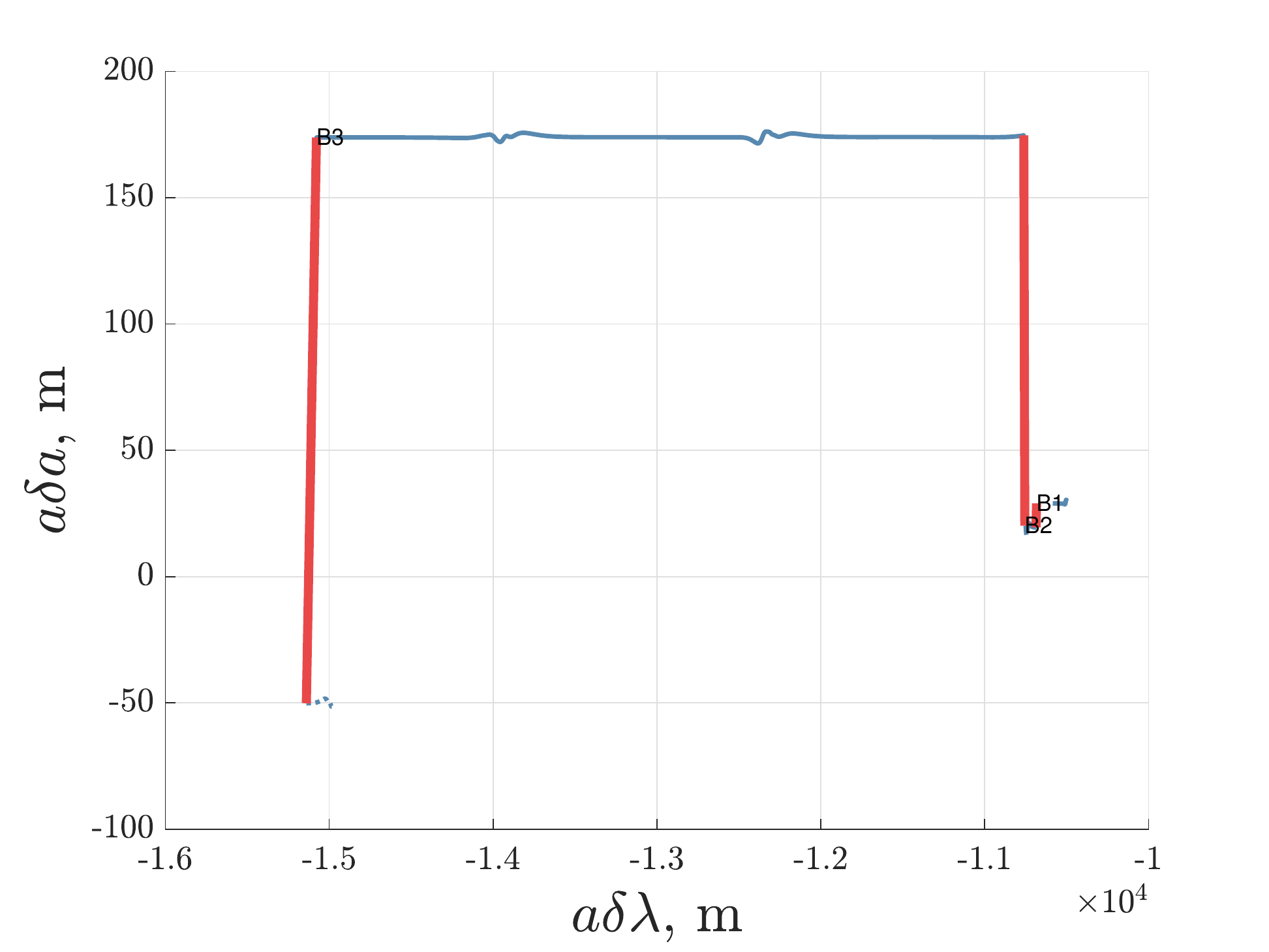}
        \caption{Evolution of the ROE in $\delta a, \delta\lambda$ plane with full-force dynamics model}
        \label{fig:ROE_evolution_reconfig_1_longer_dadl}
    \end{subfigure}\hfill\begin{subfigure}[t]{0.32\textwidth}
    \centering 
        \includegraphics[width=.9\textwidth]{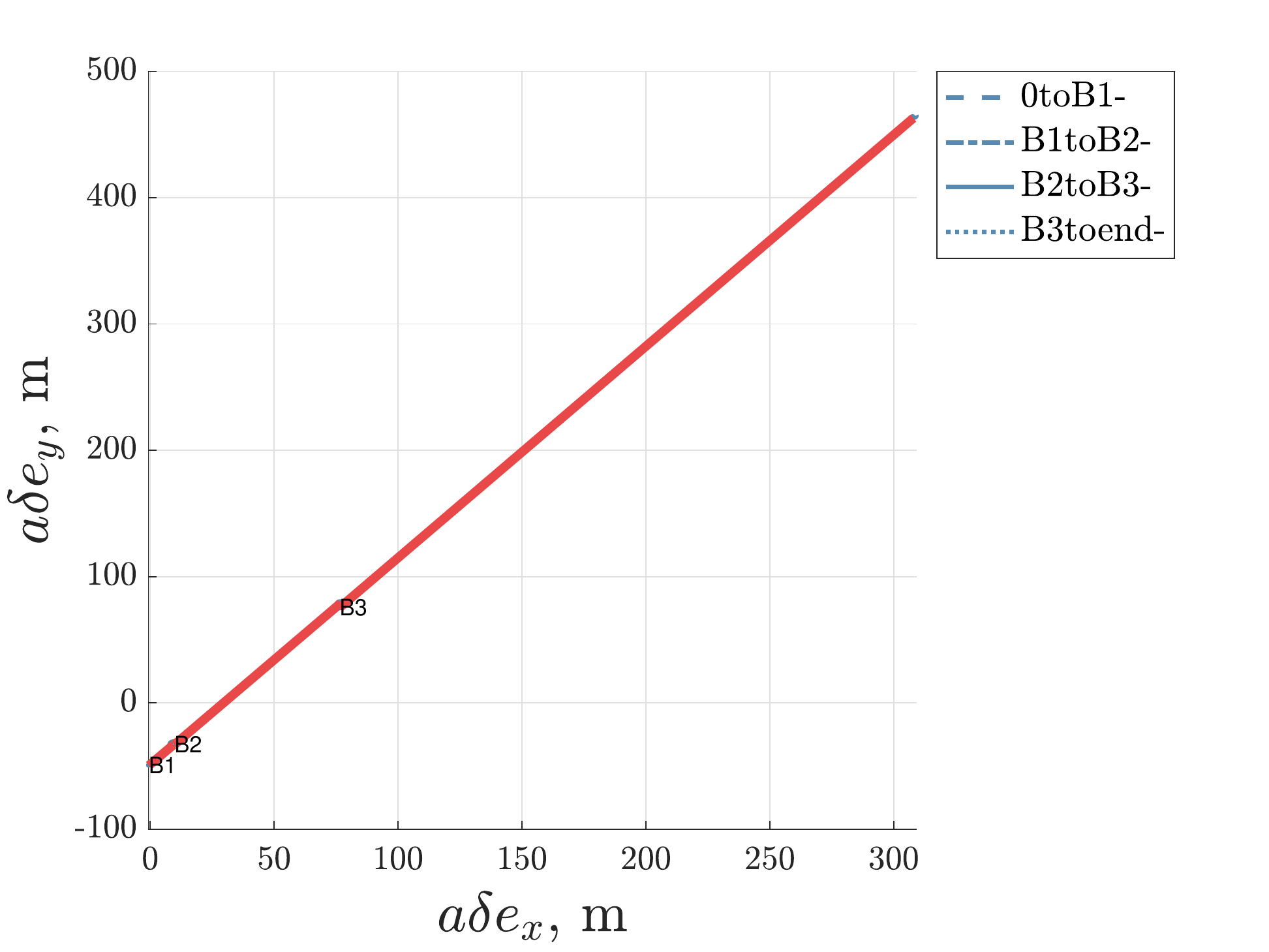}
        \caption{Evolution of the ROE in $\delta \pmb{e}'$ plane with full-force dynamics model}
        \label{fig:ROE_evolution_reconfig_1_longer_de}
    \end{subfigure}
    \caption{Evolution of the ROE in Test 1 with full-force dynamics model for a longer reconfiguration time}
    \label{fig:longer_results_test1}
\end{figure}

In missions where the reconfiguration time is not a strict constraint, an alternative approach to solving the sub-optimal problem is to increase the reconfiguration time. This allows $\delta\lambda$ to drift for longer, thus decreasing the required change in $\delta\lambda$ by control. Suppose the reconfiguration time is increased to 4 orbits. Then with the same reconfiguration parameters and $\delta a_f = -50$ m and $\delta\lambda_f = -15000$ m, the desired pseudo-states in the $\Delta\pmb{a}$ plane are $a\Delta\delta {a}_{des} = -80$ m and $a\Delta\delta {\lambda}_{des} = -3369.0$ m. The free dynamics take care of about 500m of the desired change in $\delta\lambda$. In addition, because the reconfiguration time is extended, there are also more times at which $S(c,t)$ aligns with the desired pseudo-state in the relative eccentricity vector plane. For this example, these two changes 1) increased the size of the nested reachable set and 2) moved the desired pseudo-state to the interior of the nested reachable set, as shown in Fig. \ref{fig:NRS_longer_test1}.

The desired pseudo-state now lies inside of the nested reachable set, therefore the reconfiguration can be achieved with the optimal delta-v. Also, $\delta v_{min}$ is still $0.07801$ m/s because the desired pseudo-state $a\Delta\delta\pmb{e}_{des}$ has not changed. Figs. \ref{fig:ROE_evolution_reconfig_1_longer_dadl} and \ref{fig:ROE_evolution_reconfig_1_longer_de} show the evolution of the ROE using an optimal maneuver scheme generated using the method in Sec. \ref{sec:gen_meth_cf_solutions}. The maneuvers follow a straight line path towards the desired pseudo-state in Fig. \ref{fig:ROE_evolution_reconfig_1_longer_de}, indicating optimality. 

\subsection{Test 2: Performance analysis for 6D near-circular reconfiguration}
This example will look at how realistic error sources - $\delta \pmb{v}_k$, $t_k$, $\delta\pmb{\alpha}_0$, $\pmb{\alpha}_{c,0}$ - affect the final achieved ROE for a 6D reconfiguration in an eccentric chief orbit. The reconfiguration parameters are given by
\fontsize{9}{11}\selectfont\begin{equation}\label{eqn:reconfig_params_test2}
    \begin{matrix*}[l]
    \pmb{\alpha}_{c,0} = \begin{bmatrix}9000\text{ km} & 0.2 &   0.1 &    0.1 &  0.444 &   0\end{bmatrix}^{\text{T}} \\ 
    \delta\pmb{\alpha}'_0 = \begin{bmatrix}-99.998 &  -43.193 &   247.690 &   107.851 &   63.000 &  0 \end{bmatrix}^{\text{T}} \text{ m} \\ 
    \delta\pmb{\alpha}'_f = \begin{bmatrix}20 & 2000 & 205.64 & -943.0 & 56.6 & 56.6\end{bmatrix}\text{ m}\\
    \Delta t_f = 2.5 \text{ orbits.}\\
    \end{matrix*}
\end{equation}\normalsize 
The desired decoupled pseudo-states are given by $
    a\Delta\delta\pmb{\alpha}_{des} = \begin{bmatrix}
        120.0 & -313.0 & -42.0 & -210.2 & -6.43 & 56.6
    \end{bmatrix}^{\text{T}} \text{ m}.$ Note that $a\Delta\delta  {\pmb{e}}'_{des}$ has already been transformed to the equivalent desired pseudo-state $a\Delta\delta\tilde{e}_{des}$ using Eq. \eqref{eqn:Ddeprime_to_Dde}. Table \ref{table:dvstars_CF_ecc} is used to obtain an optimal 6D maneuver scheme as given by
\fontsize{9}{11}\selectfont\begin{equation}\label{eqn:dvs_reconfig_2}
\begin{matrix*}[l]
\delta \pmb{v}_1 = \begin{bmatrix}-0.0006 & -0.0095 & 0\end{bmatrix}^{\text{T}} \text{ m/s at } t_1 =1502.30\text{ s}\\
\delta \pmb{v}_2 = \begin{bmatrix}0        &           0  & -0.0402\end{bmatrix}^{\text{T}}\text{ m/s at } t_2 = 6459.91\text{ s} \\
\delta \pmb{v}_3 = \begin{bmatrix}-0.0005 &  -0.0070 &              0\end{bmatrix}^{\text{T}}\text{ m/s at } t_3 = 999.95\text{ s} \\
\delta \pmb{v}_4 = \begin{bmatrix}-0.0042 &   0.0637 &            0\end{bmatrix}^{\text{T}}\text{ m/s at } t_3 = 14956.9\text{ s}, 
\end{matrix*}
\end{equation}\normalsize
where $\delta v_{min,total} = \delta v_{min,IP} + \delta v_{min,OOP} = 0.1205$ m/s. 
After finding a maneuver scheme, the next step is to add in realistic error. The mean of the error is $\pmb{0}$ for each source.
To capture the uncertainty in maneuver execution errors, the covariances in the maneuver magnitude $\delta\pmb{v}$ and maneuver execution time $t_k$ are given by 
\fontsize{9}{11}\selectfont\begin{equation}\label{eqn:stats_dv_t_ex}
    \sigma_{\delta\pmb{v}_k} = 0.03^2 \text{( m/s)}^2 \text{ and } \sigma_{t_k} = 60^2 \text{ s}^2
\end{equation}\normalsize respectively.
%and in the maneuver execution time is given by
%\fontsize{9}{11}\selectfont\begin{equation}\label{eqn:stats_t_ex}
%        \sigma_{t_k} = 60^2 \text{ s}^2.
%\end{equation}\normalsize 
The same distribution of error is applied to each of the four maneuver vectors and times in Eq. \eqref{eqn:dvs_reconfig_2} for consistency. For the error in the initial absolute and relative states, this example will use the typical errors generated by the state of the art in GNSS. Precision GNSS yields meter-level accuracy in the absolute Cartesian state and centimeter-level accuracy in the relative Cartesian state \cite{bib:GiraloDamico}. To be consistent with the state representations in this example, the mean and standard deviations must be transformed into the quasi-nonsingular ROE and Keplerian orbit element state representations using the well-known nonlinear transformations.
Given a covariance in the absolute Cartesian state, the equivalent covariance for the absolute OE state is calculated by applying the nonlinear Cartesian to Keplerian transform to a large set of simulated absolute Cartesian state data points. Similarly, an equivalent covariance for the ROE state can be found from a covariance in the relative Cartesian state by first rewriting the ROE as quasi-nonsingular orbit element differences and then using Eq. (B14) in Ref. \cite{bib:SullivanGrimberg}. Suppose the error covariances are given by \fontsize{9}{11}\selectfont
\begin{align}
     \pmb{V}_{\pmb{x}_0} = \text{diag}\left(1^2,1^2,1^2,0.1^2,0.1^2,0.1^2\right) \label{eqn:stats_OE_ex}\\
    \pmb{V}_{\delta\pmb{x}_0} = \text{diag}\left(0.01^2,0.01^2,0.01^2,0.001^2,0.001^2,0.001^2\right) \label{eqn:stats_ROE_ex}  
\end{align} \normalsize
for the absolute and relative Cartesian states, respectively.
After calculating the equivalent OE/ROE covariances using the Law of Large Numbers and Eqs. \ref{eqn:stats_OE_ex}-\ref{eqn:stats_ROE_ex}, each error source's respective effect on the achieved final ROE can now be calculated by either the linear or nonlinear methods described in Sec. \ref{sec:quantify_error}. Because the mean of the error in each of the four sources was zero, the mean of the error in the final achieved ROE is also zero. Error is only included in one variable at a time. In the resulting $\pmb{V}_{\delta\pmb{\alpha}_f}$, the the on-diagonal terms are the variances in each final achieved ROE due to a given error source, and the off-diagonal terms are the covariances. Table \ref{table:covariances_result_ex} gives the minimum and maximum variance and covariance in the terms of $\delta\pmb{\alpha}_f$ for each error source. 
\fontsize{8.5}{11}\selectfont
\begin{longtable}{l K{4cm} K{4cm} K{4.5cm}} 
	\caption{Values of variances and covariances in $\delta \pmb{\alpha}_{f,actual}$ due to four error sources }\label{table:covariances_result_ex} \\ \toprule\toprule
	\textbf{Error source}  & \textbf{Equations used} & [Min |variance|, Max |variance|] & [Min |covariance|, Max |covariance|] \\ \midrule
	$\delta\pmb{\alpha}_0$ & \eqref{eqn:mean_covar_man_vec_err} with \eqref{eqn:stats_ROE_ex}, Linear & [0.2251, 6093.6] & [0.0009,258.43]\\ \midrule 
	$\pmb{\alpha}_0$       & Law of large numbers and \eqref{eqn:stats_OE_ex}             & [0.0008, 0.0914] & [0.002, 0.0628]\\ \midrule
	$\delta\pmb{v}_k$      & \eqref{eqn:mean_covar_man_vec_err} with \eqref{eqn:stats_dv_t_ex}, Linear  & [0.0372, 86.302] & [0.327, 87.233] \\ \midrule
	$t_k$                  & Law of large numbers and \eqref{eqn:stats_dv_t_ex}             & [0.0091, 1.5677] & [0.023, 1.342]\\ 
    \bottomrule
\end{longtable}    \normalsize
The variances and covariances from the two linear sources ($\delta\pmb{\alpha}_0$ in row 1 and $\delta \pmb{v}_k$ in row 3) are consistently about $10^3$ times larger in magnitude than the variances and covariances from the two nonlinear sources  ($\pmb{\alpha}_0$ in row 2 and $t_k$ in row 4). Because their effect is so small, the nonlinear error sources can typically be neglected in a performance analysis; it is not necessary to run a law of large numbers simulation on-board.
As discussed previously and confirmed in Table \ref{table:covariances_result_ex} above, however, the off-diagonal covariances are non-zero, and neglecting them in a performance analysis would be incorrect. Therefore, the method in Sec. \ref{sec:performance} is used to determine the bounds that include the achieved ROE for a given confidence level by finding the smallest bounding ``box'' in 6D that would contain the error ellipsoid due to each error source. This example uses a confidence level of 95\%, which corresponds to a $\chi^2$ value of 12.59.

\fontsize{8.5}{11}\selectfont
\begin{longtable}{P{1cm} c c c c c c} 
	\caption{Performance bounds in each ROE in the presence of four independent error sources}\label{table:performance_bounds_example_2} \\ \toprule\toprule
	\textbf{Error}   & \multicolumn{6}{c}{\textbf{95\% confidence bounds on each ROE}}\\ \midrule
	{}                      & $a\delta a$, m   & $a\delta\lambda$, m & $a\delta e_x$, m   & $a\delta e_y$, m    & $a\delta i_x$, m   & $a \delta i_y$, m \\\midrule
	$\delta\pmb{\alpha}_0$  & [-3.503,  43.503] & [1446.0,   2553.9] & [187.96,  223.33] & [-956.2,  -929.8] & [49.492,   63.645] & [53.202,   59.935]\\\midrule
    $\pmb{\alpha}_0$        & [19.618,  20.382] & [1998.8,  2001.2] & [203.50,   207.79] & [-944.5,  -941.5] & [56.195,   56.942] & [56.372,   56.765]\\\midrule
    $\delta\pmb{v}_m$       & [-5.547, 45.547] & [1933.4, 2066.6] & [184.20,   227.09] & [-1008, -877.1] & [55.199,   57.938] & [44.525,   68.612] \\ \midrule
    $t_m$                   & [18.313,  21.687] & [1991.1, 2008.9] & [198.04,   213.25] & [-946.3,  -939.7] & [54.479,   58.658] & [55.893,   57.244] \\ \bottomrule   
\end{longtable}    \normalsize    
It is clear in Table \ref{table:performance_bounds_example_2} that the confidence interval is large for the linear error sources and extremely small for the nonlinear error sources. In other words, even in the presence of error in the initial absolute OE of the chief and the maneuver times, the final ROE are achieved with high accuracy. This is consistent with earlier claims that the error due to the nonlinear sources can be ignored. However, the linear error sources reduce the accuracy of the maneuver scheme significantly, and therefore must be mitigated. One way to reduce the effect of error is to use a receding time horizon and keep replanning the maneuver scheme until 1) a maneuver must occur in the current scheme, or 2) the time horizon becomes too short to replan. At each time step, the performance bounds of the previously calculated maneuver scheme and the new one are compared, and the more accurate scheme is chosen. 

\section{Conclusion}
To address the challenges of multi-spacecraft control, this paper presents new solutions to the satellite relative orbit reconfiguration problem of achieving a desired spacecraft end state in fixed-time. The relative orbit reconfiguration problem is cast in relative orbit element (ROE) space, which inherently allows for the linearization of the dynamics equations that govern relative motion and the straightforward inclusion of perturbations. 

This paper leverages the geometric advantages of reachable set theory to derive general closed-form, globally optimal impulsive maneuver schemes in orbits of arbitrary eccentricity. The reachable set is the geometric state space that can be achieved with multiple maneuvers in specified finite time with specified total cost. This paper develops a new metric for quantifying maneuver scheme optimality, and shows that optimality can be assessed without loss of generality by projecting a general $2n$-dimensional reconfiguration into $n$ 2-dimensional (2D) planes. The plane that drives the minimum delta-v is called the dominant plane. It is shown that the minimum delta-v is exactly equal to the delta-v required by the dominant plane reconfiguration if the desired end state lies in the nested reachable set, the set formed by mapping the optimal times and maneuvers for the dominant change onto the non-dominant planes. This means that a complicated higher-dimensional problem can be solved by looking at its projections into multiple 2D problems. A general methodology to derive the minimum delta-v is presented and then applied specifically to the ROE state representation for each dominance case. 

The reachability of the minimum delta-v for each dominance case is quantified by analyzing the nested reachable sets in the non-dominant planes. According to the results of this analysis, this paper presents the explicit expressions for all possible maneuver schemes that are achievable with minimum delta-v in each dominance case. The maneuver schemes are validated in realistic mission scenarios to show that they achieve a desired reconfiguration and do so optimally in eccentric chief orbits. Even with perturbations such as a full-force gravity model, drag, third body, and solar radiation pressure included in the simulation, the error in achieving the desired reconfiguration is less than 5\% in the example. 

The paper demonstrates that the same algorithms can be used to generate quantifiably sub-optimal solutions for reconfigurations that cannot be reached with delta-v driven by the dominant plane, which therefore extended the applicability of the closed-form solutions. However, there are reconfigurations that are still unachievable that must be explored, e.g. when the optimal maneuvers are linearly dependent, or when the reconfiguration time is less than one orbit. In addition, the sub-optimal delta-v, though quantifiable, may be too large for missions with strict budgets. This motivates the need to develop a new method for deriving maneuver schemes that achieve the minimum delta-v. One method, as shown in this paper, is to increase the reconfiguration time. Other possible solutions are to split a reconfiguration, add more maneuvers, or numerically solve for a maneuver scheme using the sub-optimal delta-v as an initial guess.

Finally, the paper describes a new method to assess maneuver scheme performance by fitting an $n$-dimensional bounding box to an $n$-dimensional error ellipsoid using the non-diagonal covariance matrix for a desired confidence level. An error analysis using this method shows that maneuver timing error and the high navigation error in the absolute state can usually be neglected, but that error mitigation is important for errors in the initial relative state and in the maneuver magnitudes. The same method can be applied to other potential error sources such as thruster misalignment.

\section*{Appendix A: Dynamics of Relative Motion}
The state transition matrix for the modified quasi-nonsingular ROE in Eq. \eqref{eqn:ROE} for dominant $J_2$ effects is given by
\fontsize{9}{11}\selectfont\begin{equation}\label{eqn:STMJ2}
	\pmb{\Phi}(t_f,t_0) = \begin{bmatrix}
	1 & 0 & 0 & 0 & 0 & 0 \\
	-7\kappa \eta P \tau - \frac{3}{2}n\tau & 1 & 7\kappa e_{x0}P\tau/\eta & 7\kappa e_{y0}P\tau/\eta & -7\kappa \eta S\tau & 0\\
	\frac{7}{2}\kappa e_{yf} Q\tau & 0 & \cos{(\dot{\omega}\tau)}-4\kappa e_{x0}e_{yf}GQ\tau & -\sin{(\dot{\omega}\tau)}-
    4\kappa e_{y0} e_{yf}GQ\tau & 5\kappa e_{yf}S\tau & 0 \\
	-\frac{7}{2}\kappa e_{xf} Q\tau & 0 & \sin{(\dot{\omega}\tau)}+4\kappa e_{x0}e_{xf}GQ\tau & \cos{(\dot{\omega}\tau)}+4\kappa e_{y0}e_{xf}GQ\tau & -5\kappa e_{xf}S\tau & 0 \\
	0 & 0 & 0 & 0 & 1 & 0 \\
	\frac{7}{2}\kappa S\tau & 0 & -4\kappa e_{x0}GS\tau & -4\kappa e_{y0} GS\tau & 2\kappa T\tau & 1 
	
	\end{bmatrix},
	\end{equation}\normalsize
where the constants are defined in Eq. (6) in Ref. \cite{bib:ChernickDamico}.
The subscripts 0 and f denote initial and final values of the chief’s orbit elements, respectively, and $e_x$, $e_y$ are the $x$, $y$ components of the absolute eccentricity vector. $\mu$ is the Earth’s gravitational parameter, $R_e$ is the Earth’s equatorial radius, and $n$ is the mean motion of the chief spacecraft. 
\section*{Appendix B: Proof of Decomposition of $2n$-D Reconfiguration into $n$ 2D Reconfigurations} 
\begin{figure}[H]
    \centering
    \begin{subfigure}[t]{0.46\textwidth}
    \centering 
        \includegraphics[width=.9\textwidth]{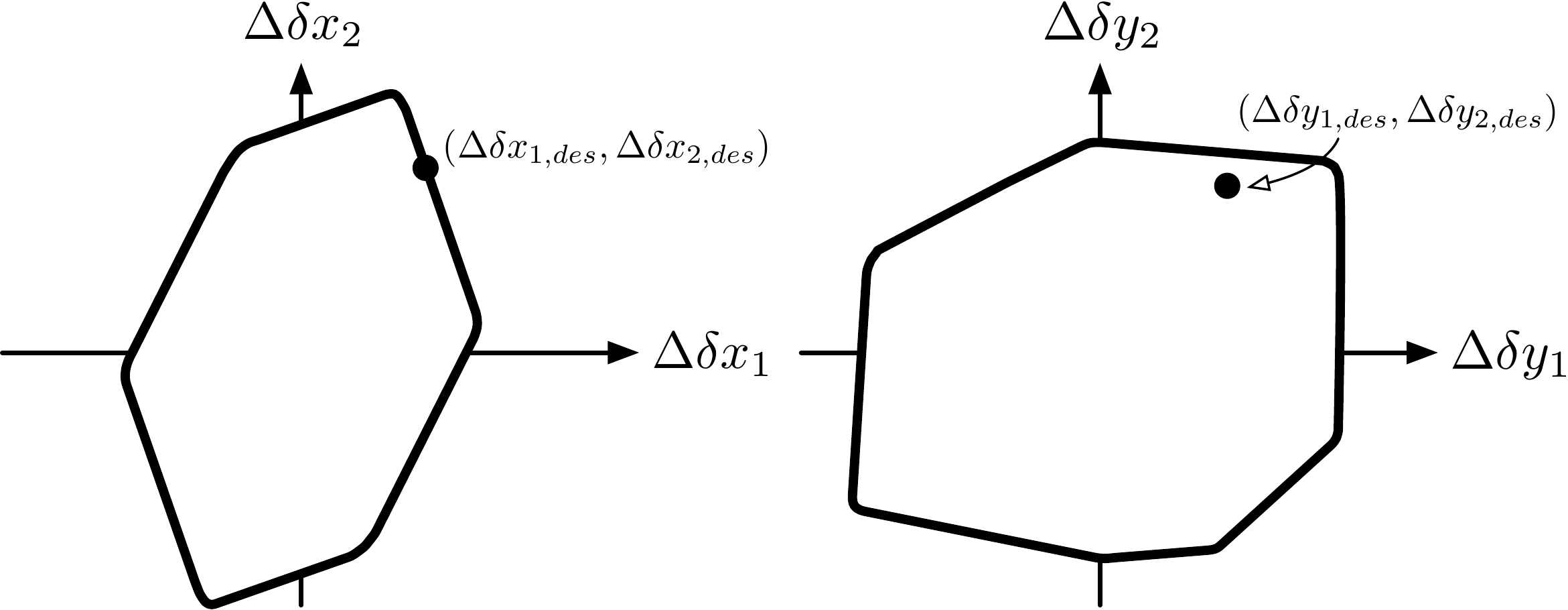}
        \caption{Reachable sets in both planes defined by $\delta v_{min,\delta\pmb{x}}$, represented by the solid line}
        \label{fig:prooftop}
    \end{subfigure}\hfill\begin{subfigure}[t]{0.46\textwidth}
    \centering 
        \includegraphics[width=.9\textwidth]{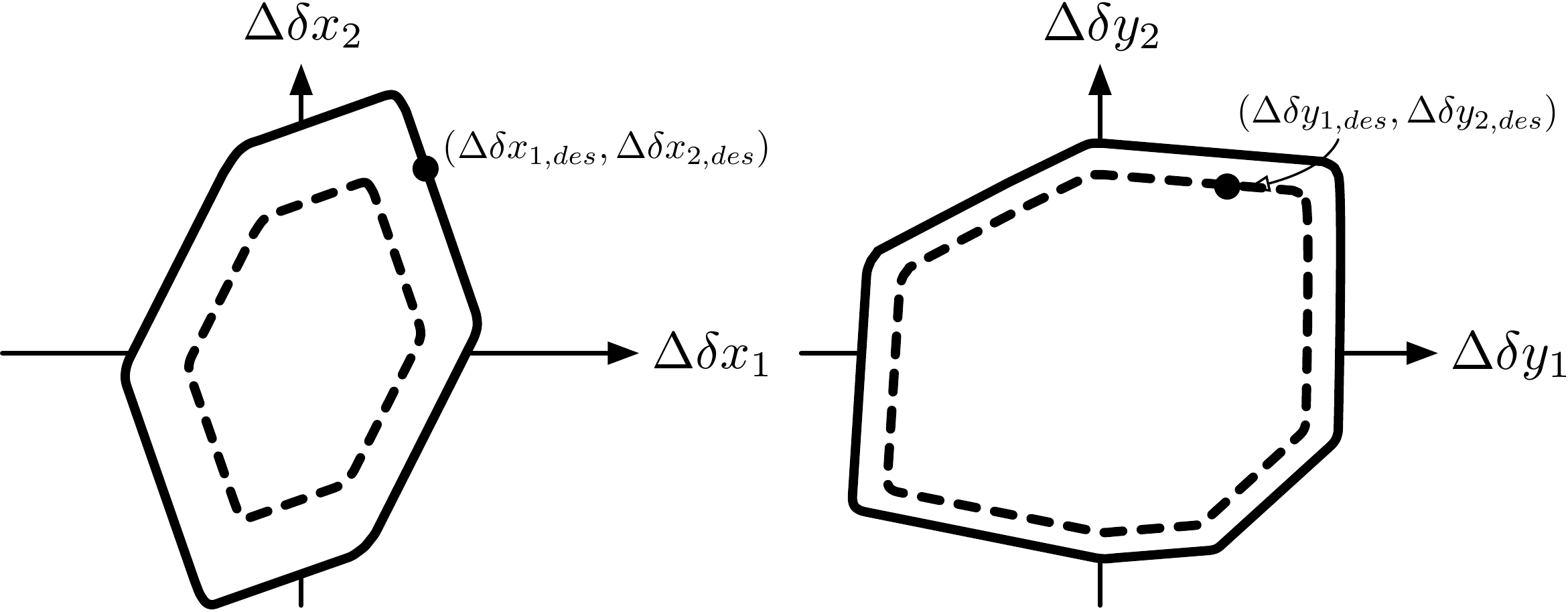}
        \caption{Reachable sets in both planes defined by $\delta v_{min,\delta\pmb{y}}$, represented by the dashed line}
        \label{fig:proofbottom}
    \end{subfigure}
    \caption{Illustration of proof that total cost of entire reconfiguration is driven by one 2D plane.}\label{fig:proof_explanation}\end{figure}

This section proves the claim in Eq. \eqref{eqn:dvminproof} by applying it to a 4D reconfiguration. 
Suppose that a desired reconfiguration given by $[\Delta\delta  {x}_{1,des},\Delta\delta  {x}_{2,des},\Delta\delta  {y}_{1,des},\Delta\delta  {y}_{2,des}]$ is split into the 2D $\Delta \delta  {\pmb{x}}_{des}$ reconfiguration and the 2D $\Delta \delta  {\pmb{y}}_{des}$ reconfiguration, and that the minimum delta-v required to achieve each 2D reconfiguration independently is known. Let the minimum delta-v required to achieve a desired pseudo-state in $\delta \pmb{x}$, $\delta\pmb{y}$ be called $\delta v_{min,\delta \pmb{x}}$, $\delta v_{min,\delta\pmb{y}}$ respectively. The minimum delta-v required to achieve the entire desired reconfiguration is $\delta v_{min} \geq \max \{ \delta v_{min,\delta\pmb{x}}, \delta v_{min,\delta\pmb{y}} \}$.
\textit{Proof.} Suppose $\delta v_{min,\delta\pmb{x}}>\delta v_{min,\delta \pmb{y}}$. If $S^*(\delta v_{min,\delta\pmb{x}},T)$, the set of pseudo-states reachable by a series of maneuvers with total magnitude $\delta v_{min,\delta\pmb{x}}$ executed within period T, is computed in both planes using Eq. \eqref{eqn:Sstar}, the desired pseudo-state will lie on its boundary in the $\Delta\delta \pmb{x}$ plane by definition. Because $\delta v_{min,\delta\pmb{x}} > \delta v_{min,\delta\pmb{y}}$, the desired pseudo-state will lie inside the boundary of $S^*(\delta v_{min,\delta\pmb{x}},T)$ in the $\Delta\delta\pmb{y}$ plane. Therefore, the desired pseudo-state in $\delta \pmb{x}$ is reachable (see Fig. \ref{fig:prooftop}). Now suppose $S^*(\delta v_{min,\delta\pmb{y}},T)$ is computed in both planes. In the $\Delta\delta\pmb{y}$ plane, the desired pseudo-state now lies on the boundary of the convex hull by definition (see Fig. \ref{fig:proofbottom}). However, in the $\Delta\delta\pmb{x}$ plane, the desired pseudo-state is outside of the reachable region, because the delta-v required to reach it is larger than the delta-v that defines the boundary of the convex hull. Therefore, the delta-v for the total reconfiguration is no less than $\delta v_{min,\delta\pmb{x}}$. $\centerdot$ 
   
\section*{Appendix C: Reachable Delta-v Minimum for Dominant $\delta\lambda$}
As discussed in Sec. \ref{sec:reachabledvmin}, a dominant $\delta\lambda$ reconfiguration is only achieved optimally in the dominant plane. Therefore, typically it suffices to check that a reconfiguration is \textit{not} dominant $\delta\lambda$. However, for completeness, the reachable delta-v minima for dominant $\delta\lambda$ reconfigurations is given in the table below.
\fontsize{9}{11}\selectfont
   \begin{longtable}{ l K{4cm} P{8cm}}
		\caption{Reachable $\delta v_{min}$, eccentric chief orbits, dominant $\delta\lambda$}\label{table:dvmin_dl} \\ \toprule\toprule
		
		{\textbf{Dominant...}} & \textbf{{Max normalized effect}} & {$\delta v_{min}$, \textbf{(m/s)}}\\ \midrule
      
\textit{... $\delta\lambda$, Transition region}& $a\Delta\delta  {a}_{max} =  \frac{2\sqrt{e^2+2e\cos(\nu^*)+1}}{\eta n} $ & $\delta v_{min,trans} = 
\frac{|a\Delta\delta  {a}_{des}|}{|a\Delta\delta  {a}_{max}|}$ \\ \midrule
        %%%%%%%%% 
        
    \textit{... $\delta \lambda$} & N/A & $\delta v_{min,\delta\lambda} = n\left|\frac{a\Delta\delta {a}_{des} - ma\Delta\delta {\lambda}_{des}}{ma\Delta\delta {\lambda}_0 - a\Delta\delta {a}_0}\right|$ where $m=\frac{a\Delta\delta  {a}_t +a\Delta \delta  {a}_0}{a\Delta\delta {\lambda}_t +a\Delta\delta {\lambda}_0}$ \\ \midrule
        
     \multirow{2}{*}{\textit{... $\delta\lambda$, Extended region}} & N/A & $\delta v_{min} = n \frac{a\Delta\delta a_{des} - m_1 a\Delta\delta\lambda_{des}}{a\Delta\delta a_t - m_1 a\Delta\delta\lambda_t}$ where $m_1 = \frac{a\Delta\delta a_t - a\Delta\delta a_f}{a\Delta\delta\lambda_t - a\Delta\delta \lambda_f}$ \smalltab (if $\Delta\delta {\lambda}_{des} < \delta v_{min,\delta\lambda}\Delta\delta {\lambda}_f$) \\
     
       & & $\delta v_{min} = n \frac{m_2 a\Delta\delta\lambda_{des} - a\Delta\delta a_{des}}{a\Delta\delta a_0 - m_2 a\Delta\delta\lambda_0}$ where $m_2 = \frac{a\Delta\delta a_f + a\Delta\delta a}{a\Delta\delta\lambda_f + a\Delta\delta\lambda_0} $ \smalltab (if $\Delta\delta {\lambda}_{des} > \delta v_{min,\delta\lambda}\Delta\delta {\lambda}_f$)  \\
      \midrule    
	\end{longtable}\normalsize  
	
where recall, Eqs. \eqref{eqn:parametric_dadl} and Eqs. \eqref{eqn:Dd0}, \eqref{eqn:Ddk2pi}-\eqref{eqn:Ddf} are used.

%%%%%%%%%%%%%%%%

\section*{Appendix D: Other Closed-form Solutions}

The tables below present closed-form globally optimal maneuvers schemes for cases where only the desired pseudo-state in the dominant plane is important. The general methodology is the same as for the other closed-form solutions presented in this paper. The optimal times and maneuvers are computed for a given dominance case using Table \ref{table:dvstars_CF_dom}.
\fontsize{8.5}{11}\selectfont
\begin{longtable}{ l K{8cm} K{3cm}}
		\caption{Optimal maneuver vectors $\delta \pmb{v}^*$ and optimal maneuver times $T_{opt}$ for dominant-plane-only reconfigurations}\label{table:dvstars_CF_dom} \\ \toprule\toprule
		
		\textbf{{Dominant...}} & {$\delta \pmb{v}^*$, (m/s)} & {$T_{opt}$, (s)}\\ \midrule
        \textit{... $\delta a$} & $\begin{matrix*}[l]\Delta\delta a>0: \text{ } \delta \pmb{v}^* = \begin{bmatrix} 0 & +\delta v_{min,\delta a} & 0\end{bmatrix}^{\text{T}}\\
        \Delta\delta a<0: \text{ } \delta \pmb{v}^* = \begin{bmatrix} 0 & -\delta v_{min,\delta a} & 0\end{bmatrix}^{\text{T}}\end{matrix*}$ for all $i$ & $k2\pi$, $k=\text{floor}(\frac{\nu_f}{2\pi})$ \\ \midrule
        
        \textit{... $\delta \lambda$} & $\begin{matrix*}[c] \begin{bmatrix}0 & \pm1 & 0\end{bmatrix}^{\text{T}} \\ 
\begin{bmatrix}\frac{e\sin(\nu_t)}{\sqrt{e^2 + 2e\cos(\nu_t) +1}} & \mp\frac{1+e\cos(\nu_t)}{\sqrt{e^2 + 2e\cos(\nu_t) +1}} &0\end{bmatrix}^{\text{T}}\end{matrix*}$ for $\mp \Delta\delta {\lambda}_{des}$ & $\begin{matrix*}[c]0 \\ \nu_t \end{matrix*}$\\  \bottomrule 
        	\end{longtable}\normalsize

Then, a subset of $S_n$, the nested reachable set, is chosen so that the maneuvers satisfy the constraints given in the third column of Table \ref{table:CFconstraints_dom}. There are fewer constraints for these cases because the effects of the maneuvers in the non-dominant planes is not considered. 
\fontsize{8.5}{11}\selectfont
\begin{longtable}{l K{1.3cm} K{7.2cm} K{3.2cm}}
		\caption{Closed-form maneuver scheme constraints in dominant-plane-only reconfigurations}\label{table:CFconstraints_dom} \\ \toprule\toprule
		
		\textbf{{Dominant}}& \textbf{{\# of Man.}} & \textbf{{Constraints (Type 1)}} & \textbf{{Linear System}}\\ \midrule
        \textit{... $\delta a$} & & & \\ \hdashline
& 1 & $a\Delta\delta n{\lambda}_{des} = \frac{\mp3}{\eta n} (M_f - k2\pi)(1+e)$ for $\pm\Delta\delta {a}_{des}$ & $c_1 = 1$ \\ \hdashline
& 2 & $a\Delta\delta  {\lambda}_{des} \neq \frac{\mp3}{\eta n} (M_f - k2\pi)(1+e)$  for $\pm\Delta\delta {a}_{des}$ & $\begin{matrix*}[l]\sum_{i=1}^2 c_i = 1 \\ 
\sum_{i=1}^2 c_ia\Delta\delta \lambda_i = a\Delta\delta  {\lambda}_{des}\\\end{matrix*}$ \\ \midrule

\textit{... $\delta\lambda$} & 2 & N/A & $\begin{matrix*}[l]\sum_{i=1}^2 c_i a\Delta\delta  {a}_i = a \Delta\delta  {a}_{des} \\ 
\sum_{i=1}^2 c_ia\Delta\delta \lambda_i = a\Delta\delta  {\lambda}_{des}\\\end{matrix*}$\\  \bottomrule
\end{longtable}\normalsize

\section*{Funding Sources}
This work was supported by a NASA Office of the Chief Technologists Space Technology Research Fellowship, NASA grant no. 80NSSC18K1187. 

\bibliography{references}

\begin{thebibliography}{30}
\newcommand{\enquote}[1]{``#1''}
\providecommand{\natexlab}[1]{#1}
\providecommand{\url}[1]{\texttt{#1}}
\providecommand{\urlprefix}{URL }
\expandafter\ifx\csname urlstyle\endcsname\relax
  \providecommand{\doi}[1]{\discretionary{}{}{}https://doi.org/#1}\else
  \providecommand{\doi}[1]{\discretionary{}{}{}\urlstyle{rm}\url{https://doi.org/#1}}\fi

\bibitem[{D'Amico et~al.(2015)D'Amico, Pavone, Saraf, Alhussien, Al-Saud,
  Buchman, Byer, and Farhat}]{bib:DamicoMDSS}
D'Amico, S., Pavone, M., Saraf, S., Alhussien, A., Al-Saud, T., Buchman, S.,
  Byer, R., and Farhat, C., \enquote{Miniaturized Autonomous Distributed Space
  System for Future Science and Exploration,} \emph{8th International Workshop
  on Satellite Constellations and Formation Flying, IWSCFF 2015}, Delft
  University of Technology, 2015.

\bibitem[{Ichikawa and Ichimura(2008)}]{bib:Ichi}
Ichikawa, A., and Ichimura, Y., \enquote{Optimal Impulsive Relative Orbit
  Transfer Along a Circular Orbit,} \emph{Journal of Guidance, Control, and
  Dynamics}, Vol.~31, No.~4, 2008, pp. 1014--1027.
\newblock \doi{https://doi.org/10.2514/1.32820}.

\bibitem[{Khalil et~al.(2016)Khalil, Larbi, and Stoll}]{bib:Khalil}
Khalil, M., Larbi, B., and Stoll, E., \enquote{Spacecraft Formation Control
  using Analytical Integration of Gauss' Variational Equations,} \emph{6th
  International inproceedings on Astrodynamics Tools and Techniques - ICATT},
  Darmstadt, Germany, 2016.

\bibitem[{D'Amico(2005)}]{bib:Damico_ROE_as_Integration_Constants}
D'Amico, S., \enquote{Relative Orbital Elements as Integration Constants of
  Hill’s Equations,} 2005.
\newblock DLR-GSOC TN 05-08; Deutsches Zentrum für Luft- und Raumfahrt,
  Oberpfaffenhofen.

\bibitem[{Gaias and D'Amico(2015)}]{bib:GaiasDamico1}
Gaias, G., and D'Amico, S., \enquote{Impulsive Maneuvers for Formation
  Reconfiguration Using Relative Orbital Elements,} \emph{Journal of Guidance,
  Control and Dynamics}, Vol.~38, No.~6, 2015, pp. 1036--1049.
\newblock \doi{10.2514/1.G000189}.

\bibitem[{Gaias et~al.(2015)Gaias, D'Amico, and Ardaens}]{bib:GaiasDamico2}
Gaias, G., D'Amico, S., and Ardaens, J.-S., \enquote{Generalized
  Multi-Impulsive Maneuvers for Optimum Spacecraft Rendezvous in Near-Circular
  Orbit,} \emph{International Journal of Space Science and Engineering},
  Vol.~3, No.~1, 2015, pp. 68--88.
\newblock \doi{10.1504/IJSPACESE.2015.069361}.

\bibitem[{Chernick and D'Amico(2018{\natexlab{a}})}]{bib:ChernickDamico}
Chernick, M., and D'Amico, S., \enquote{New Closed-Form Solutions for Optimal
  Impulsive Control of Spacecraft Relative Motion,} \emph{Journal of Guidance,
  Control and Dynamics}, Vol.~41, No.~2, 2018{\natexlab{a}}, pp. 301--319.

\bibitem[{Zhang and Mortari(2020)}]{bib:zhang_mortari}
Zhang, G., and Mortari, D., \enquote{Impulsive orbit correction using
  second-order Gauss’s variational equations,} \emph{Celestial Mechanics and
  Dynamical Astronomy}, Vol. 132, No.~2, 2020, p.~13.

\bibitem[{Vaddi et~al.(2005)Vaddi, Alfriend, Vadali, and Sengupta}]{bib:Vaddi}
Vaddi, S., Alfriend, K., Vadali, S., and Sengupta, P., \enquote{Formation
  Establishment and Reconfiguration Using Impulsive Control,} \emph{Journal of
  Guidance, Control and Dynamics}, Vol.~28, No.~2, 2005, pp. 262--268.

\bibitem[{Schaub and Alfriend(2001)}]{bib:SchaubAlf}
Schaub, H., and Alfriend, K., \enquote{Impulsive Feedback Control to Establish
  Specific Mean Orbit Elements of Spacecraft Formations,} \emph{Journal of
  Guidance Control and Dynamics}, Vol.~24, No.~4, 2001, pp. 739--745.

\bibitem[{Betts(1998)}]{bib:Betts}
Betts, J., \enquote{Survey of Numerical Methods for Trajectory Optimization,}
  \emph{Journal of Guidance, Control and Dynamics}, Vol.~21, No.~2, 1998, pp.
  193--207.

\bibitem[{Sobiesiak and Damaren(2015)}]{bib:SobiesiakDamaren}
Sobiesiak, L., and Damaren, C., \enquote{Impulsive Spacecraft Formation
  Maneuvers with Optimal Firing Times,} \emph{Journal of Guidance, Control and
  Dynamics}, Vol.~38, No.~10, 2015, pp. 1994--2000.

\bibitem[{Roscoe et~al.(2015)Roscoe, Westphal, Griesbach, and
  Schaub}]{bib:Roscoe}
Roscoe, C., Westphal, J., Griesbach, J., and Schaub, H., \enquote{Formation
  Establishment and Reconfiguration Using Differential Elements in
  {$J_2$}-Perturbed Orbits,} \emph{Journal of Guidance, Control and Dynamics},
  Vol.~38, 2015, p. 1725–1740.

\bibitem[{Prussing(2011)}]{bib:ConwayPrussing}
Prussing, J., \emph{Primer Vector Theory and Applications, In: Conway, B.A.
  (ed.) Spacecraft Trajectory Optimization}, Cambridge University Press,
  Cambridge, 2011.
\newblock Pp. 15--35.

\bibitem[{Gilbert and Harasty(1971)}]{bib:Gilbert}
Gilbert, E., and Harasty, G., \enquote{A class of fixed-time fuel-optimal
  impulsive control problems and an efficient algorithm for their solution,}
  \emph{IEEE Transactions on Automatic Control}, Vol.~16, No.~1, 1971, pp.
  1--11.
\newblock \doi{10.1109/TAC.1971.1099656}.

\bibitem[{Koenig and D'Amico(2018)}]{bib:KDAutomatica}
Koenig, A., and D'Amico, S., \enquote{Real-Time Algorithm for Globally Optimal
  Impulsive Control of Linear Time-Variant Systems,} \emph{IEEE Transactions on
  Automatic Control}, 2018.
\newblock Submitted, http://arxiv.org/abs/1804.06099.

\bibitem[{Allen et~al.(2014)Allen, Clark, Starek, and Pavone}]{bib:Allen}
Allen, R.~E., Clark, A.~A., Starek, J.~A., and Pavone, M., \enquote{A machine
  learning approach for real-time reachability analysis,} \emph{2014 IEEE/RSJ
  International inproceedings on Intelligent Robots and Systems}, 2014, pp.
  2202--2208.
\newblock \doi{10.1109/IROS.2014.6942859}.

\bibitem[{Vinh et~al.(1995)Vinh, Gilbert, Howe, Sheu, and Lu}]{bib:Vinh}
Vinh, N.~X., Gilbert, E.~G., Howe, R.~M., Sheu, D., and Lu, P.,
  \enquote{Reachable domain for interception at hyperbolic speeds,} \emph{Acta
  Astronautica}, Vol.~35, No.~1, 1995, pp. 1 -- 8.
\newblock \doi{https://doi.org/10.1016/0094-5765(94)00132-6}.

\bibitem[{Zagaris and Romano(2018)}]{bib:Zagaris_Romano}
Zagaris, C., and Romano, M., \enquote{Applied Reachability Analysis for
  Spacecraft Rendezvous and Docking with a Tumbling Object,} \emph{2018 Space
  Flight Mechanics Meeting}, Kissimmee, Fl, 2018.

\bibitem[{D'Amico and Montenbruck(2006)}]{bib:DamicoEI}
D'Amico, S., and Montenbruck, O., \enquote{Proximity Operations of Formation
  Flying Spacecraft using an Eccentricity/Inclination Vector Separation,}
  \emph{Journal of Guidance, Control and Dynamics}, Vol.~29, No.~3, 2006, pp.
  554--563.

\bibitem[{D'Amico et~al.(2009)D'Amico, Florio, Larsson, and
  Nylund}]{bib:DamicoFlorio}
D'Amico, S., Florio, S.~D., Larsson, R., and Nylund, M., \enquote{Autonomous
  Formation Keeping and Reconfiguration for Remote Sensing Spacecraft,}
  \emph{21st International Symposium on Space Flight Dynamics}, 2009.
\newblock Toulouse, France.

\bibitem[{Ardaens and Fischer(2011)}]{bib:ArdaensFischer}
Ardaens, J.-S., and Fischer, D., \enquote{TanDEM-X Autonomous Formation Flying
  System: Flight Results,} \emph{IFAC Proceedings Volumes}, Vol.~44, No.~1,
  2011, pp. 709 -- 714.
\newblock 18th IFAC World Congress.

\bibitem[{Koenig et~al.(2017)Koenig, Guffanti, and D'Amico}]{bib:KoenigDamico}
Koenig, A., Guffanti, T., and D'Amico, S., \enquote{New State Transition
  Matrices for Spacecraft Relative Motion in Perturbed Orbits,} \emph{Journal
  of Guidance, Control, and Dynamics}, Vol.~40, No.~7, 2017, pp. 1749--1768.

\bibitem[{Brouwer(1959)}]{bib:Brouwer}
Brouwer, D., \enquote{Solution of the problem of artificial satellite theory
  without drag,} Vol.~64, 1959, p. 378.

\bibitem[{Yamanaka and Ankerson(2002)}]{bib:YamanakaAnk}
Yamanaka, K., and Ankerson, F., \enquote{New State Transition Matrix for
  Relative Motion on an Arbitrary Elliptical Orbit,} \emph{Journal of Guidance,
  Control and Dynamics}, Vol.~25, 2002, pp. 60--66.

\bibitem[{Chernick and D'Amico(2018{\natexlab{b}})}]{bib:RST_conf}
Chernick, M., and D'Amico, S., \enquote{Closed-Form Optimal Impulsive Control
  of Spacecraft Formations using Reachable Set Theory,} \emph{2018 AAS/AIAA
  Astrodynamics Specialist inproceedings}, Snowbird, UT, 2018{\natexlab{b}}.

\bibitem[{Boyd and Vandenberghe(2004)}]{bib:Boyd_CVX}
Boyd, S., and Vandenberghe, L., \emph{Convex Optimization}, Cambridge
  University Press, 2004.
\newblock \doi{10.1017/CBO9780511804441}, p. 146.

\bibitem[{Friendly et~al.(2011)Friendly, Monette, and
  Fox}]{bib:statistical_ellipsoids}
Friendly, M., Monette, G., and Fox, J., \enquote{Elliptical Insights:
  Understanding Statistical Methods Through Elliptical Geometry,}
  \emph{Statistical Science}, 2011.
\newblock \urlprefix\url{http://datavis.ca/papers/ellipses.pdf}.

\bibitem[{Giralo and D'Amico(2019)}]{bib:GiraloDamico}
Giralo, V., and D'Amico, S., \enquote{Distributed multi-GNSS timing and
  localization for nanosatellites,} \emph{Navigation}, Vol.~66, No.~4, 2019,
  pp. 729--746.

\bibitem[{Sullivan et~al.(2017)Sullivan, Grimberg, and
  D'Amico}]{bib:SullivanGrimberg}
Sullivan, J., Grimberg, S., and D'Amico, S., \enquote{Comprehensive Survey and
  Assessment of Spacecraft Relative Motion Dynamics Models,} \emph{Journal of
  Guidance, Control and Dynamics}, Vol.~40, No.~8, 2017, pp. 1837--1859.

\end{thebibliography}

\end{document}